\documentclass[12pt]{article}

\pdfoutput=1

\usepackage{epsfig}
\usepackage{color,graphicx}
\usepackage{cite}
\usepackage{amssymb,amsmath}

\usepackage{a4wide}
\usepackage{amssymb}
\usepackage{amsmath}
\usepackage{graphicx}
\usepackage{mathdots}
\usepackage{youngtab}
\usepackage{caption}
\usepackage{subcaption}
\usepackage{enumerate}

\newcommand{\be}{\begin{equation}}
\newcommand{\ee}{\end{equation}}
\newcommand{\bea}{\begin{eqnarray}}
\newcommand{\eea}{\end{eqnarray}}

\def\drawbox#1#2{\hrule height#2pt 
        \hbox{\vrule width#2pt height#1pt \kern#1pt 
              \vrule width#2pt}
              \hrule height#2pt}

\def\Fund#1#2{\vcenter{\vbox{\drawbox{#1}{#2}}}}
\def\Asym#1#2{\vcenter{\vbox{\drawbox{#1}{#2}
              \kern-#2pt       
              \drawbox{#1}{#2}}}}

\def\funda{\Fund{6.5}{0.4}}
\def\asymm{\Asym{6.5}{0.4}}

\def\symm{\funda\kern-0.4pt\funda}

\numberwithin{equation}{section}

\begin{document}

\begin{center}  

\vskip 2cm 

\centerline{\Large {\bf Lifting 4d dualities to 5d}}

\vskip 1cm

\renewcommand{\thefootnote}{\fnsymbol{footnote}}

   \centerline{
    {\large \bf Oren Bergman}\footnote{bergman@physics.technion.ac.il}
    {\bf and}
    {\large \bf Gabi Zafrir}\footnote{gabizaf@techunix.technion.ac.il}}

\vspace{1cm}
\centerline{{\it Department of Physics, Technion, Israel Institute of Technology}} \centerline{{\it Haifa, 32000, Israel}}
\vspace{1cm}

\end{center}

\vskip 0.3 cm

\setcounter{footnote}{0}
\renewcommand{\thefootnote}{\arabic{footnote}}   
   
\begin{abstract}
In this paper we set out to further explore the connection between isolated ${\cal N}=2$ SCFT's in four dimensions
and ${\cal N}=1$ SCFT's in five dimensions.
Using 5-brane webs we are able to provide IR Lagrangian descriptions in terms of 5d gauge theories
for several classes of theories including the so-called $T_N$ theories.
In many of these we find multiple dual gauge theory descriptions.
The connection to 4d theories is then used to lift 4d ${\cal N}=2$ S-dualities that involve weakly-gauging
isolated theories to 5d gauge theory dualities.
The 5d description allows one to study the spectrum of BPS operators directly, using for example the superconformal index.
This provides additional non-trivial checks of enhanced global symmetries and 4d dualities.
 \end{abstract}
 
 \newpage
 
\tableofcontents

\section{Introduction}

It is well appreciated by now that there exist many isolated interacting 4d ${\cal N}=2$ superconformal field theories (SCFT's) 
that have no marginal coupling and no Lagrangian description.
These theories are described only through their Seiberg-Witten curves, 
and can be characterized by their global symmetry, and by the dimensions of their Coulomb and Higgs branches.
For example, there is a series of such theories with $E_6, E_7$ and $E_8$ global symmetry \cite{Minahan:1996fg}. 
The current ``state of the art" in our understanding of 4d ${\cal N}=2$ theories is in terms
of M5-branes wrapping punctured Riemann surfaces \cite{Gaiotto:2009we}.
In particular, 3-punctured spheres with various types of punctures give isolated SCFT's in 4d.

Although they are a-priori isolated in the space of 4d SCFT's, some of these theories can be continuously
connected to more ordinary superconformal gauge theories by gauging a subgroup of their global symmetry,
whereby they provide an S-dual description of strong coupling limits of the gauge theories \cite{AS}.
In the simplest example, the $E_6$ theory with a gauged $SU(2)\subset E_6$ and one flavor is S-dual
to the superconformal gauge theory with $SU(3)$ and 6 flavors.
The realization of 4d SCFT's in terms of punctured Riemann surfaces generalizes this idea,
by realizing different weak-coupling limits of a given SCFT as different degenerations of the corresponding
Riemann surface. This leads to many examples of S-duality between superconformal gauge theories
and ``weakly gauged" isolated SCFT's \cite{Distler}.
However the lack of an explicit Lagrangian description for these theories makes it difficult to test
the dualities in detail.

An approach which may shed more light on this is to lift the 4d theories to 5d ${\cal N}=1$ SCFT's.
Many of these admit deformations to 5d ${\cal N}=1$ gauge theories \cite{Seiberg:1996bd,Intriligator:1997pq}, thereby providing 
a Lagrangian description.
For example, there are 5d rank one SCFT's with $E_6$, $E_7$ and $E_8$ global symmetry, corresponding to the
UV fixed points of the 5d ${\cal N}=1$ gauge theories with gauge group $SU(2)$ and $N_f=5, 6$ and 7, 
respectively \cite{Seiberg:1996bd}. There are also rank $n$ versions of these theories, corresponding to $USp(2n)$
with the same number of flavor hypermultiplets plus an additional hypermultiplet in the antisymmetric representation of the gauge group.
These reduce to the corresponding 4d SCFT's by compactifying on a circle in the limit of vanishing radius.
The 5d Lagrangian description in principle allows one to determine the complete chiral ring of the theory,
although some of the BPS states involve non-perturbative instanton particles.
Indeed these states provide the necessary charges for the enhanced global symmetries,
as can be seen, for example, from their contributions to the superconformal index \cite{KKL}.

A useful way to visualize 5d SCFT's in general is by $(p,q)$ 5-brane webs in Type IIB string theory \cite{AH}.
This construction makes manifest all the mass parameters and moduli of the theory, 
realized geometrically as the relative motions of the external and internal 5-branes, respectively.
In many cases, the 5-brane web can be mass-deformed to exhibit a 5d low-energy ${\cal N}=1$ supersymmetric gauge theory,
with the 5d SCFT as its UV fixed point, in correspondence with the classification of \cite{Intriligator:1997pq}.
The mass in these cases corresponds to an inverse square gauge coupling of the gauge theory.
In fact, there may exist different mass deformations leading to different IR gauge theories.
This is somewhat analogous to Seiberg duality in 4d, except that in 5d there are two, or more, IR theories that
flow to the same CFT in the UV, whereas in 4d there are two, or more, UV theories that flow to the same CFT in the IR.
The different 5d gauge theories are in a sense continuations past infinite gauge coupling of one another,
since one has to go through a massless point in connecting them.
From the point of view of the 5-brane web this usually entails an $SL(2,\mathbb{Z})$ transformation
exchanging D5-branes and NS5-branes \cite{AH,Aharony:1997bh}.
These types of dualities were further explored and generalized in \cite{Bao:2011rc,Bergman:2013aca}.

In \cite{BBT} it was shown that a general class of isolated 4d ${\cal N}=2$ SCFT's known as the $T_N$ theories
lifts to interacting ${\cal N}=1$ SCFT's in 5d corresponding to simple 5-brane webs. 
This connection was further studied in \cite{Bao:2013pwa,Hayashi:2013qwa}.
In principle, this should allow one to identify the 5d IR gauge theory by suitably deforming the 5-brane web
\cite{Hayashi:2013qwa,Aganagic:2014oia}.
%

Our first goal, in section 2, is to find Lagrangian descriptions in terms of 5d ${\cal N}=1$
gauge theories for 5d lifts of isolated 4d ${\cal N}=2$ SCFT's.
We will begin with the $T_N$ theories.
Then, by looking at various limits on the Higgs branch of these theories, as described by the 5-brane webs,
we will also find 5d gauge theories for several other 5d SCFT's that reduce to isolated 4d SCFT's,
such as the ones considered in \cite{Distler}.
In some cases we will find dual gauge theories for the same fixed point theory.

Our second goal, in section 3, is to relate the S-dualities associated with weakly-gauging these 4d SCFT's
to dualities between 5d gauge theories associated with the same 5d SCFT in the UV.
In particular, this allows us to use localization to compute the superconformal index using either gauge theory,
and thereby obtain the explicit dictionary relating the BPS states of the two 4d theories.
We will exhibit this in a number of examples, starting with the Argyres-Seiberg duality involving the $E_6$ theory.

Section 4 contains our conclusions.
We have also included three appendices.
In Appendix A we give a brief review of the 5d superconformal index, and in particular of how various 
issues in the computation of instanton contributions are resolved.
In Appendix B we discuss the different representations of flavor degrees of freedom in 5-brane webs,
and in Appendix C we describe how to incorporate antisymmetric matter in 5-brane webs.

\medskip

{\bf A word on notation:} We will denote global symmetries associated with 
matter in the fundamental representation (``flavor") by an $F$ subscript, those associated with
matter in the bi-fundamantal representation by a $BF$ subscript, and those associated with
matter in the 2-index antisymmetric representation by an $A$ subscript.
In addition, we will use a $B$ subscript for the baryonic $U(1)$ symmetry (in the case where there is
a $U(N)_F = SU(N)_F\times U(1)_B$ flavor symmetry), and an $I$ subscript for the topological (instanton) $U(1)$ symmetries.
Subscripts on gauge symmetries will denote either the CS level or the value of the discrete $\theta$ parameter,
as appropriate.
Superscripts on gauge symmetries, in cases where there is a product of several identical groups, will denote their 
order of appearance in the product.

\section{5d gauge theories for 4d SCFT's}

\subsection{The $T_N$ theories}
\label{sec:TN} 

The 4d $T_N$ theory corresponds to M5-branes wrapping a 2-sphere with three maximal punctures, namely
punctures labelled by the fully symmetrized $N$-box Young tableau \cite{Gaiotto:2009we}.
This theory has no marginal couplings. The global symmetry is (at least) $SU(N)^3$, 
and therefore the theory has $3(N-1)$ mass parameters, corresponding to VEV's of scalars
in background vector multiplets associated with the global symmetry.
The dimensions of the Coulomb and Higgs branches are given by 
$d_C = (N-1)(N-2)/2$ and 
$d_H = (3N^2 -N -2)/2$.
The $N=2$ case is the theory of four free hypermultiplets, and 
the $N=3$ case is the $E_6$ theory.
The rank 1 $E_7$ and $E_8$ theories can be realized as particular limits on the Higgs branch of the $T_4$ and $T_6$
theories, respectively. We will mention these below.

\begin{figure}[h]
\center
\includegraphics[height=0.4\textwidth]{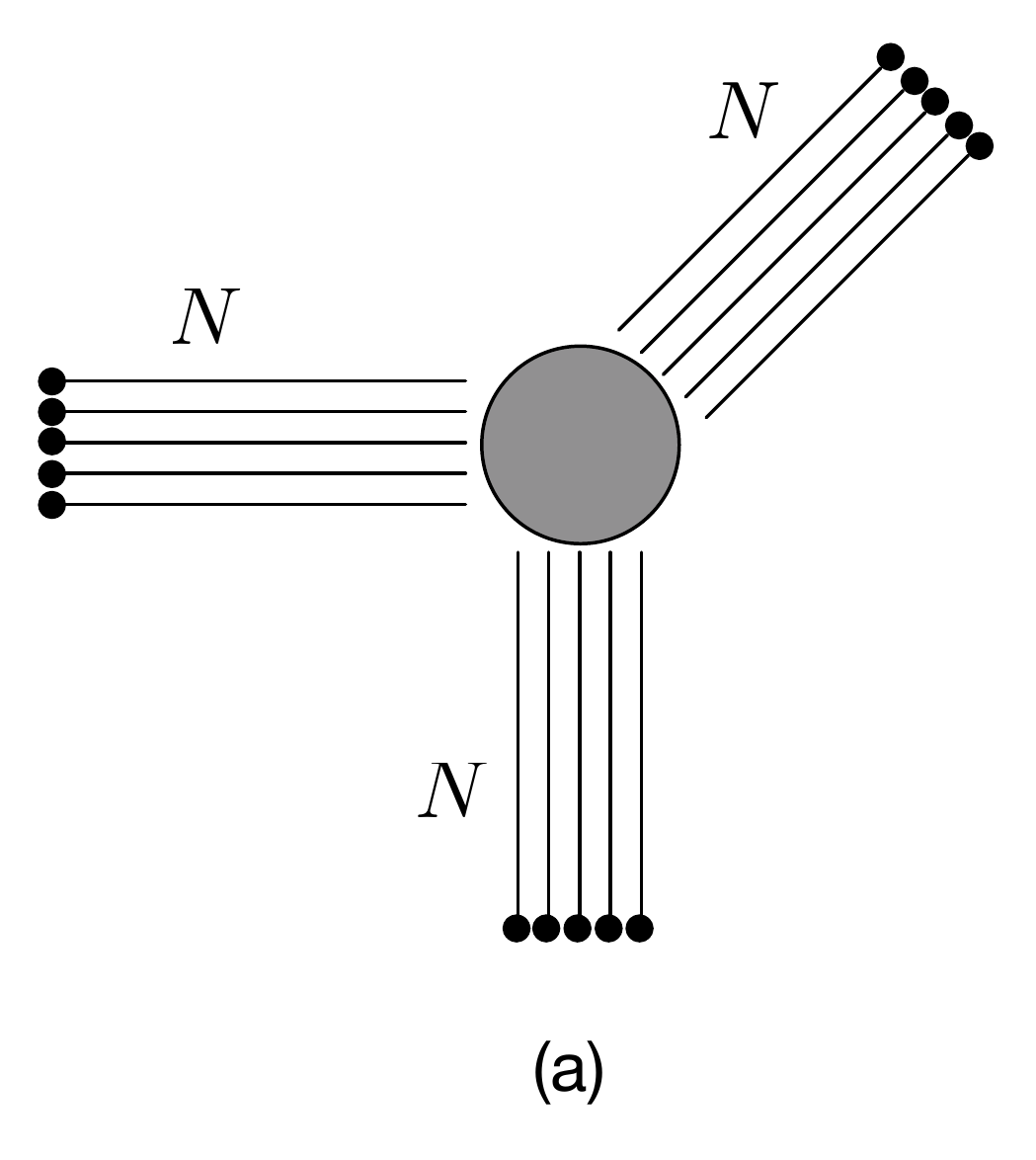} 
\hspace{1cm}
\includegraphics[height=0.45\textwidth]{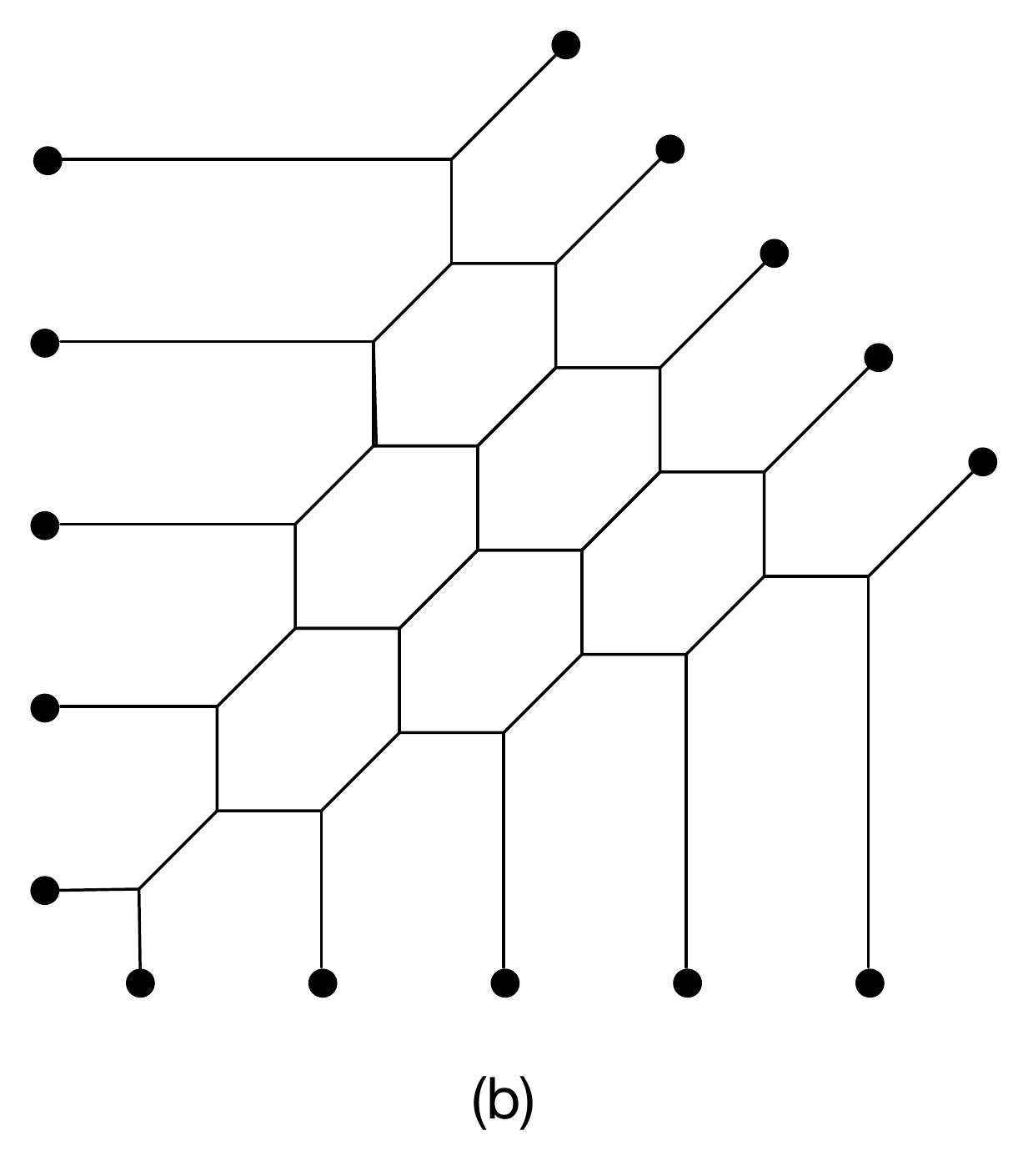} 
\caption{The 5-brane web of the $T_N$ theory (shown for $N=5$): (a) Fixed point theory (b) on the Coulomb branch.}
\label{T5web1}
\end{figure}

The 5d version of the $T_N$ theory is described by a collapsed 5-brane web, or 5-brane ``junction", with
$N$ external D5-branes, $N$ external NS5-branes, and $N$ external $(1,1)$5-branes (Fig.~\ref{T5web1}a) \cite{BBT}.
This is the 5-brane configuration resulting from the reduction of the M5-brane configuration in M theory to Type IIB string theory.
In describing the 5d theory, it is useful to have each external 5-brane end on an appropriate type of 7-brane.
One can read off the basic properties of the theory from this configuration.
The mass parameters (real in 5d) correspond to the relative positions of the 7-branes, so there are $3(N - 1)$ of them.
Indeed, the multiplicities of the external 5-branes suggest an $SU(N)^3$ global symmetry, although it may be enhanced
(as we know it should to $E_6$ for $N=3$).
The Coulomb moduli correspond to planar deformations of the web,
explicitly shown for $N=5$ in Fig.~\ref{T5web1}b, and the
Higgs moduli correspond to transverse deformations, where parts of the web separate along the 7-branes.
The counting of the web deformations reproduce the dimensions of the Coulomb and Higgs branches. 

\begin{figure}[h]
\center
\includegraphics[height=0.45\textwidth]{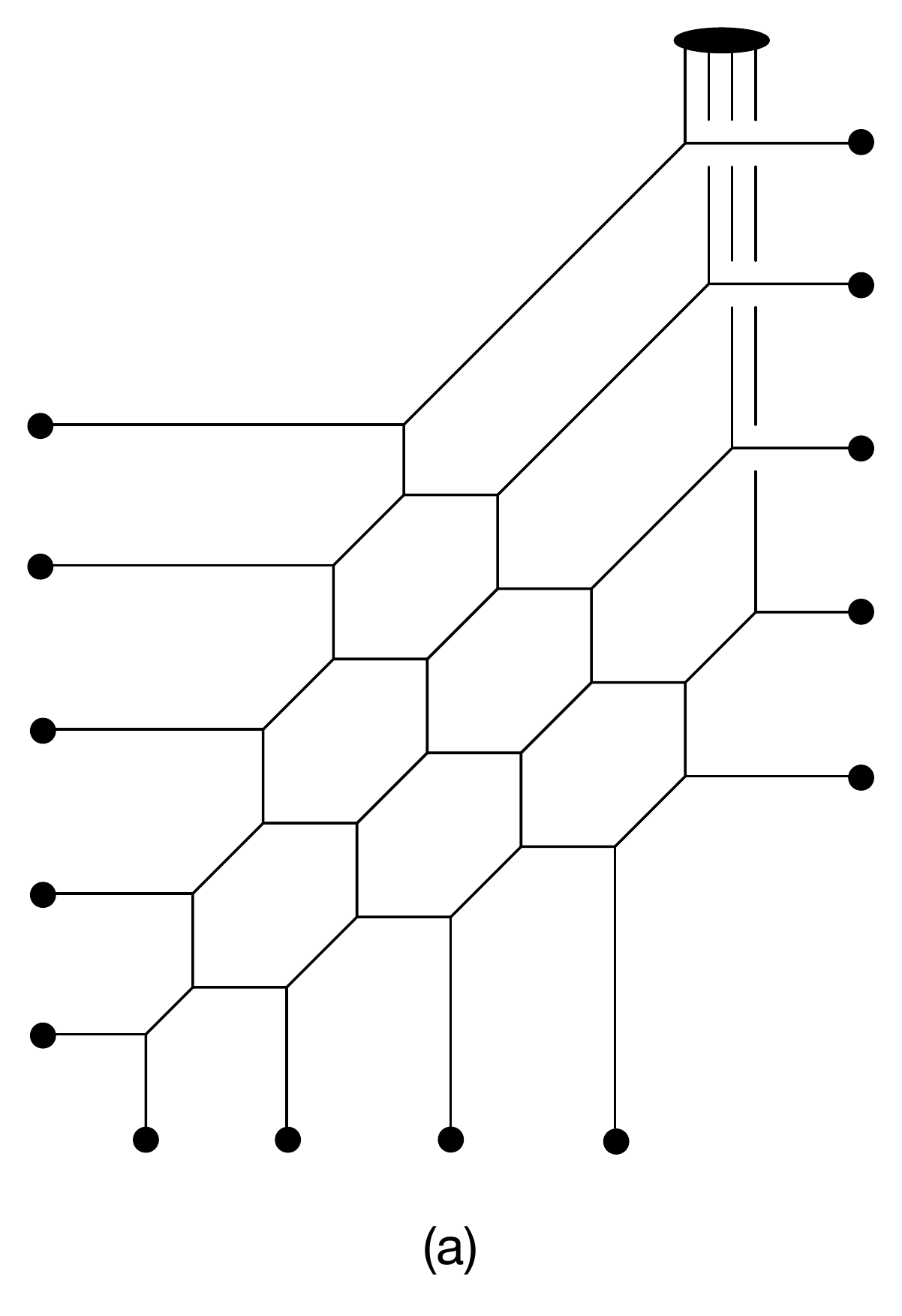}
\hspace{0.3cm}
\includegraphics[width=0.3\textwidth]{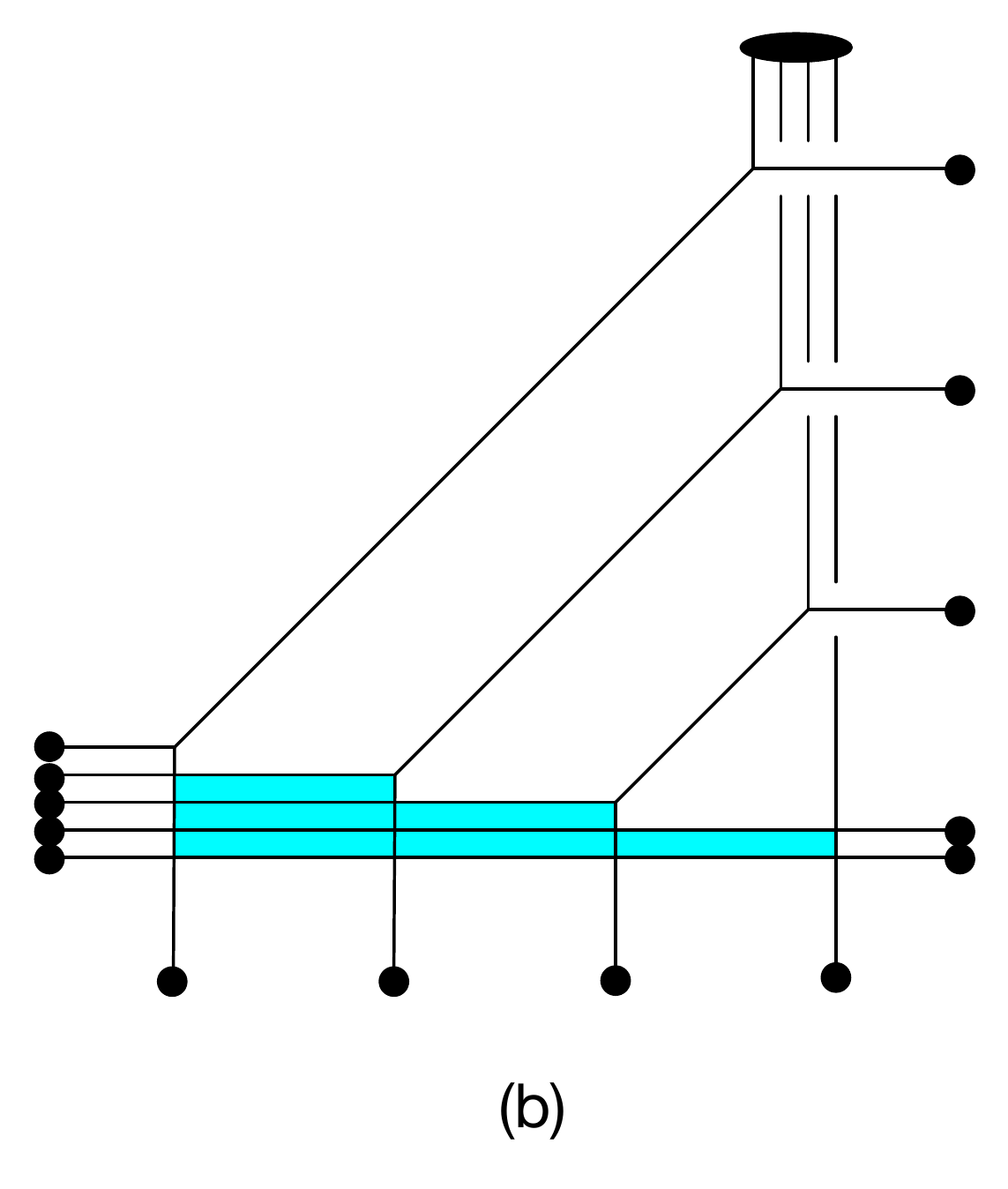} 
\hspace{0.3cm}
\includegraphics[width=0.3\textwidth]{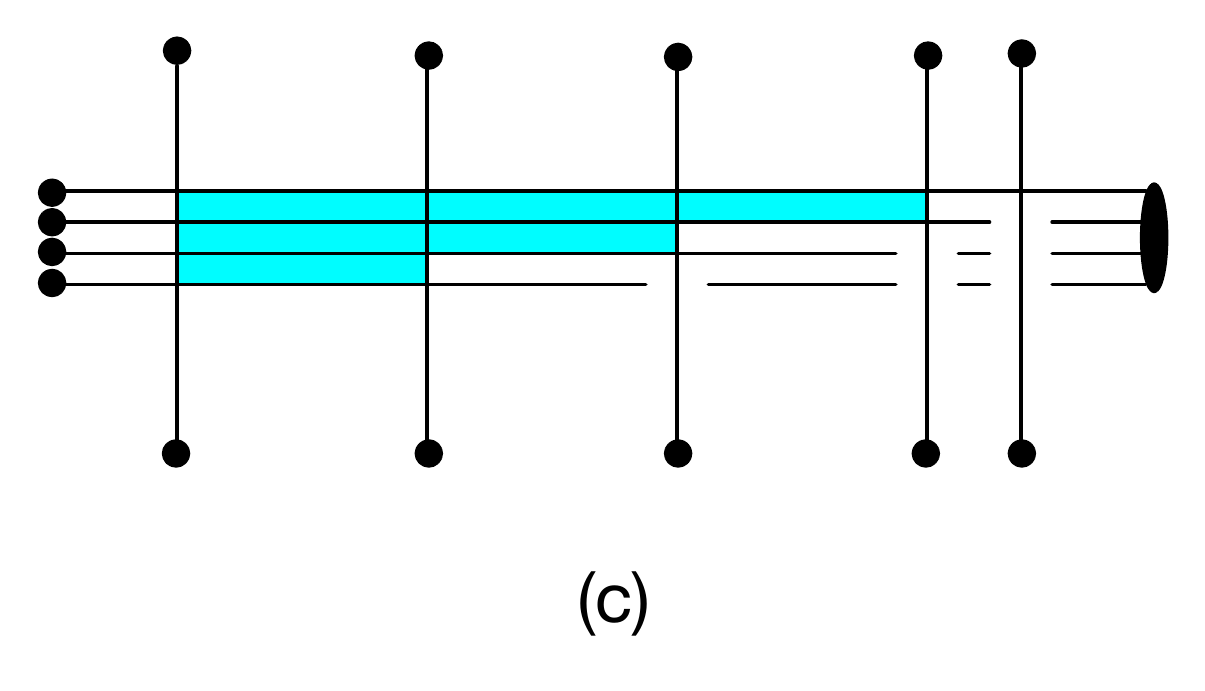} 
\caption{Another web for $T_N$ ($N=5$): (a) After an HW transition (b) Mass deforming to reveal the gauge theory 
(the shaded regions correspond to the gauge groups) (c) The S-dual web.}
\label{T5web2}
\end{figure}

\begin{figure}[h]
\center
\includegraphics[width=0.4\textwidth]{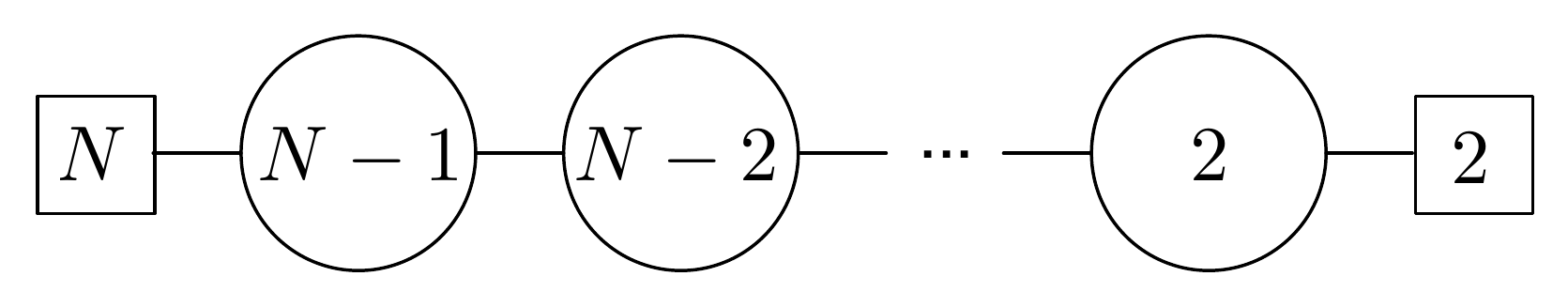} 
\caption{Quiver diagram for the IR gauge theory of $T_N$.}
\label{TNquiver}
\end{figure}

The gauge theory interpretation of the web in Fig.~\ref{T5web1} is not completely obvious. 
However we can manipulate the web so that the gauge theory becomes apparent.
Moving the 7-brane in the lower right corner upward across all the $(1,1)$5-branes we obtain, via multiple Hanany-Witten 
(brane-creation) transitions, 
the web shown in Fig.~\ref{T5web2}a. 
In this web $N-1$ NS5-branes end on the same 7-brane, leading to the 
avoided 5-brane intersections due to the s-rule \cite{Hanany:1996ie,BBT}.
The 5d IR gauge theory becomes apparent when we mass-deform the theory by
further separating the $N-1$ $(0,1)$7-branes on the bottom and going to the origin of the Coulomb branch, 
as shown in Fig.~\ref{T5web2}b.
In the limit of large mass we get a weakly interacting linear-quiver gauge theory.
For the $N=5$ case shown in the figure the gauge group is $SU(4)\times SU(3)\times SU(2)$,
and there is a single massless hypermultiplet in the bi-fundamental representation of each pair of adjacent groups,
five in the fundamental representation of $SU(4)$, and two in the
fundamental representation of $SU(2)$.
More generally, the quiver theory has the structure $N+SU(N-1)\times SU(N-2)\times \cdots \times SU(2)+2$.
The corresponding quiver diagram is shown in Fig.~\ref{TNquiver}.

S-duality gives the web shown in Fig.~\ref{T5web2}c. This actually describes the same quiver gauge theory.
This is not immediately obvious, due to the avoided intersections involving the D5-branes.
To see the flavor structure more clearly, one can go through a series of 7-brane motions,
as described in Appendix~\ref{sec:flavors}, which basically brings us back to (an S-dual of) the web of Fig.~\ref{T5web1}b.

\begin{figure}[h]
\center
\includegraphics[width=0.7\textwidth]{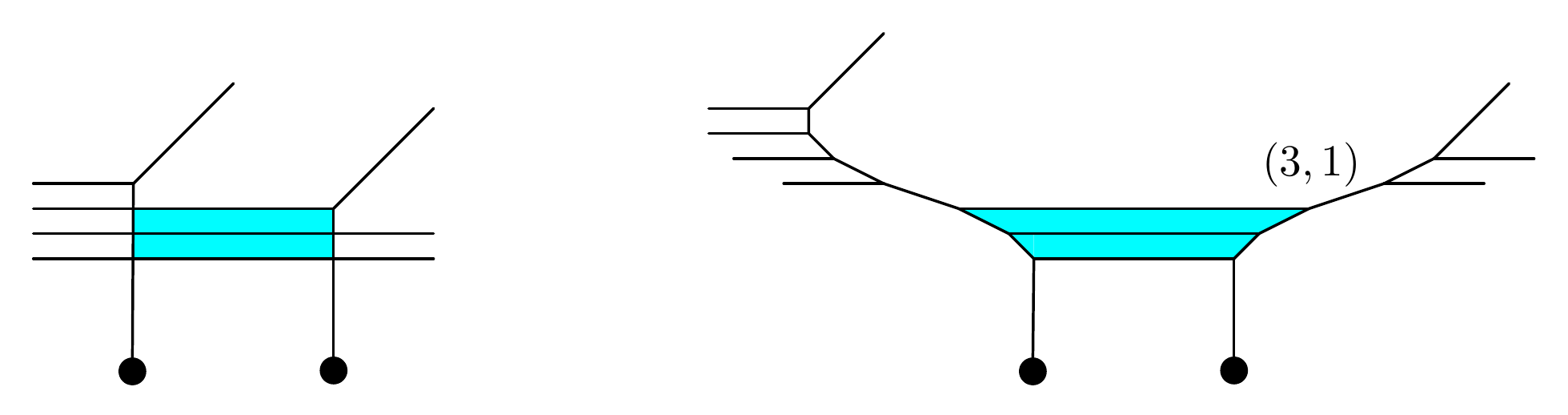}
\caption{Deforming the $SU(n)+2n$ ($n=3$ in this case) sub-web to compute the CS level.}
\label{CSweb}
\end{figure}

To completely fix the gauge theory we also need to specify the CS levels of the $SU(n)$ factors with $n\geq 3$.
These are easiest to determine by looking at each $SU(n)$ factor in the web separately.
Each such sub-web gives $SU(n)+2n$, with $n$ running from $N-1$ down to 3.
Now deform the sub-web so as to give all the flavors a mass with the same sign. 
This is shown in Fig.~\ref{CSweb}.
The CS level is renormalized (for a positive mass) as $\kappa = \kappa_0 + n$.
On the other hand, the renormalized CS level is easily read-off from the resulting pure $SU(n)$ web 
to be $\kappa = n $ (see \cite{Bergman:2013aca}). Therefore the original CS levels are all zero.

\subsubsection{Enhanced symmetry}
\label{TN_symmetry}

The global symmetry of the IR gauge theory is $U(N)_F\times SO(4)_F\times U(1)_{BF}^{N-3} \times U(1)_I^{N-2}$,
where the first two factors are associated to the flavors at the two edges, the $U(1)_{BF}$ factors to the
bi-fundamental fields, and the $U(1)_I$ factors to the instanton number currents.
This should be enhanced at the UV fixed point to $SU(N)^3$ by instantons.

The $N=3$ case is special.
The gauge theory in this case is $SU(2)+5$, so the classical global symmetry is $SO(10)_F\times U(1)_I$. 
The full global symmetry in this case is $E_6$, which
is consistent with the fact that $E_6$ is the unique rank six group which has $SU(3)^3$ and $SO(10)\times U(1)$ as subgroups.
The enhancement to $E_6$ was explicitly demonstrated in \cite{KKL} by computing the superconformal index,
including instanton contributions.

The verification of the $SU(N)^3$ symmetry in the more general case is technically harder, since it involves instantons
with charges under two gauge groups.
We did this explicitly for $T_4$ and $T_5$. 

The IR gauge theory corresponding to $T_4$ is the linear quiver $4+SU_0(3)\times SU(2)+2$.
A standard calculation of the perturbative superconformal index gives (see Appendix~\ref{appendix:index}) 
\bea
I_{pert}^{4+SU(3)\times SU(2)+2} 
& = & 1 + x^2 \left(4+\chi_{(\bold{15},\bold{1},\bold{1})}+\chi_{(\bold{1},\bold{3},\bold{1})}+\chi_{(\bold{1},\bold{1},\bold{3})}\right) 
 \nonumber \\
& + & x^3 \left( (y + \frac{1}{y}) \left(5+\chi_{(\bold{15},\bold{1},\bold{1})}+\chi_{(\bold{1},\bold{3},\bold{1})}+\chi_{(\bold{1},\bold{1},\bold{3})}\right) 
+ \frac{b}{z}\, \chi_{(\bold{4},\bold{2},\bold{2})} + \frac{z}{b}\, \chi_{(\bar{\bold{4}},\bold{2},\bold{2})} \right. \nonumber \\
& + & \left. b z^2\, \chi_{(\bold{4},\bold{1},\bold{1})} + \frac{1}{b z^2}\, \chi_{(\bar{\bold{4}},\bold{1},\bold{1})} + b^3\, \chi_{(\bar{\bold{4}},\bold{1},\bold{1})} 
+ \frac{1}{b^3}\, \chi_{(\bold{4},\bold{1},\bold{1})} \right) + {\cal O}(x^4) \,,
\eea
where $x,y$ are the superconformal fugacities, $z$ is the fugacity associated to the bi-fundamental field,
$b$ is the baryonic fugacity associated with the $U(1)_B$ subgroup of the $U(4)_F$ flavor symmetry,
and $\chi_{(\cdots)}$ denotes a character of the given representation of the non-Abelian part of the global symmetry,
in this case $SU(4)_F\times SO(4)_F = SU(4)_F\times SU(2)_F\times SU(2)_F$.

To ${\cal O}(x^3)$ there are also contributions from $(1,0), (0,1)$ and $(1,1)$ instantons.
The calculation of the instanton partition functions turns out to be simpler if we treat $SU(2)$ as $USp(2)$.
For the $SU(3)$ instanton we must use the $U(N)$ formalism and mod out the $U(1)$ part.
In general this procedure leaves some ``$U(1)$ remnants" that must be removed by hand (see Appendix~\ref{appendix:index} for a discussion).
In this case the remnant states correspond to a D1-brane between the parallel external NS5-branes in Fig.~\ref{CSweb}, and are removed 
by correcting the instanton partition function as
\be
\mathcal{Z}_{c} = PE\left[\frac{x^2 q_1 (z b^2 + \frac{1}{z b^2})}{(1-x y)(1-\frac{x}{y})}\right] \mathcal{Z} \,,
\ee 
where $q_1$ is the $SU(3)$ instanton fugacity.
Note that the correction factor is not invariant under $x\rightarrow 1/x$, which is part of the conformal symmetry.
This corrects a similar lack of invariance in the instanton partition function, due to a pole at zero in the integral over the dual 
gauge group (see Appendix~\ref{appendix:index}).

The resulting instanton contribution is given by (we present the result only to order $x^2$, although we computed to order $x^3$)
\bea
I_{(1,0)+(0,1)+(1,1)}^{4+SU(3)\times SU(2)+2} & =  & 
x^2 \Bigg[ \left(q_2 z^{\frac{3}{2}} + \frac{1}{q_2 z^{\frac{3}{2}}}\right)\chi_{(\bold{1},\bold{2},\bold{1})} 
+ \left(\frac{q_2}{z^{\frac{3}{2}}} + \frac{z^{\frac{3}{2}}}{q_2}\right)\chi_{(\bold{1},\bold{1},\bold{2})}  \nonumber\\
&+& \left(q_1 + \frac{1}{q_1}\right)\left(z b^2 + \frac{1}{z b^2}\right) 
+ \left(\frac{q_1 q_2 \sqrt{z}}{b^2} + \frac{b^2}{q_1 q_2 \sqrt{z}}\right)\chi_{(\bold{1},\bold{2},\bold{1})}
\nonumber \\
 &+&    \left(\frac{q_1 q_2 b^2}{\sqrt{z}} + \frac{\sqrt{z}}{q_1 q_2 b^2}\right)\chi_{(\bold{1},\bold{1},\bold{2})} \Bigg] + {\cal O}(x^3) \,,
%
\eea
where $q_2$ is the $SU(2)$ instanton fugacity.
Together with the perturbative contribution, the $x^2$ terms exhibit an enhancement of
$SO(4)_F \times U(1)^4  \rightarrow SU(4)^2$,
and we can express the full index in terms of characters of $SU(4)^3$ (now including the $x^3$ terms):
\bea
I^{T_4} & = & 1 + x^2(\chi_{(\bold{15},\bold{1},\bold{1})} + \chi_{(\bold{1},\bold{15},\bold{1})} 
+ \chi_{(\bold{1},\bold{1},\bold{15})}) \\
&+& x^3\left((y+\frac{1}{y})(1+\chi_{(\bold{15},\bold{1},\bold{1})} 
+ \chi_{(\bold{1},\bold{15},\bold{1})} + \chi_{(\bold{1},\bold{1},\bold{15})})+\chi_{(\bold{4},\bold{4},\bold{4})} 
+ \chi_{(\bar{\bold{4}},\bar{\bold{4}},\bar{\bold{4}})} \right) + O(x^4) \,. \nonumber
\eea

For $T_5$, the gauge theory is $5+SU_0(4)\times SU_0(3)\times SU(2)+2$, and the perturbative contribution to the superconformal index is
\bea
I_{pert}^{T_5}  
& = & 1 + x^2 (6+\chi_{(\bold{24},\bold{1},\bold{1})} 
+ \chi_{(\bold{1},\bold{3},\bold{1})} + \chi_{(\bold{1},\bold{1},\bold{3})}) \nonumber \\ 
& + & x^3 (y+\frac{1}{y}) (7+\chi_{(\bold{24},\bold{1},\bold{1})} + \chi_{(\bold{1},\bold{3},\bold{1})} + \chi_{(\bold{1},\bold{1},\bold{3})}) + O(x^4) \,.
\eea   
In this case the classical global symmetry is $SU(5)_F\times SO(4)_F \times U(1)_B\times U(1)^2_{BF}\times U(1)^3_I$.
The instanton part is again computed by treating $SU(2)$ as $USp(2)$. 
The ``$U(1)$-remnant" states, which can be read-off from the $5+SU(4)\times SU(3)+2$ sub-web, are removed by the correction:
\be
\mathcal{Z}_{c} = PE\left[\frac{x^2 \left( q_2 (z_2 z^2_1 + \frac{1}{z_2 z^2_1}) + q_1 (\frac{z^{\frac{3}{2}}_1}{b^{\frac{5}{2}}} + \frac{b^{\frac{5}{2}}}{z^{\frac{3}{2}}_1}) 
+ q_1 q_2 (\sqrt{z_1 b^5}\, z_2 + \frac{1}{\sqrt{z_1 b^5}\, z_2}) \right)}{(1-x y)(1-\frac{x}{y})}\right] \mathcal{Z} \,,
\ee 
where $q_1$ and $q_2$ are the instanton fugacities of $SU(4)$ and $SU(3)$, respectively, and 
$z_1$ and $z_2$ are the bi-fundamental fugacities for $SU(4)\times SU(3)$ and $SU(3)\times SU(2)$, respectively. 
As a consistency check, we verified that all the instanton partition functions we evaluated are $x \rightarrow 1/x$ invariant. 
To order $x^3$ there are contributions from the $(1,0,0)$, $(0,1,0)$, $(0,0,1)$, $(1,1,0)$, $(0,1,1)$ and $(1,1,1)$ instantons. 
These give 
\bea
I_{inst}^{T_5}   
&=&  x^2  \left(1 + x(y+\frac{1}{y})\right) 
\Bigg[ \left(q_3 z^{\frac{3}{2}}_2 + \frac{1}{q_3 z^{\frac{3}{2}}_2}\right) \chi_{(\bold{1},\bold{2},\bold{1})} 
+ \left(\frac{q_3}{z^{\frac{3}{2}}_2} + \frac{z^{\frac{3}{2}}_2}{q_3}\right)\chi_{(\bold{1},\bold{1},\bold{2})} \nonumber \\
&+& \left(q_2+\frac{1}{q_2}\right) \left(z_2 z^2_1 + \frac{1}{z_2 z^2_1}\right)
+ \left(q_1+\frac{1}{q_1}\right) \left(\frac{z^{\frac{3}{2}}_1}{b^{\frac{5}{2}}} + \frac{b^{\frac{5}{2}}}{z^{\frac{3}{2}}_1}\right)
\nonumber \\ 
 & + &    \left(\frac{q_3 q_2 \sqrt{z_2}}{z^2_1} + \frac{z^2_1}{q_3 q_2 \sqrt{z_2}}\right)\chi_{(\bold{1},\bold{2},\bold{1})} 
 + \left(\frac{q_3 q_2 z^2_1}{\sqrt{z_2}} + \frac{\sqrt{z_2}}{q_3 q_2 z^2_1}\right)\chi_{(\bold{1},\bold{1},\bold{2})}
 \nonumber \\  
 & + &   \left(q_2 q_1+\frac{1}{q_2 q_1}\right) \left(\sqrt{z_1 b^5}z_2 + \frac{1}{\sqrt{z_1 b^5}z_2}\right) 
 + \left(\frac{q_1 q_2 q_3 \sqrt{z_2}}{\sqrt{z_1 b^5}} + \frac{\sqrt{z_1 b^5}}{q_1 q_2 q_3 \sqrt{z_2}}\right)\chi_{(\bold{1},\bold{2},\bold{1})}
 \nonumber \\  
 & + &  \left(\frac{q_1 q_2 q_3 \sqrt{z_1 b^5}}{\sqrt{z_2}} + \frac{\sqrt{z_2}}{q_1 q_2 q_3 \sqrt{z_1 b^5}}\right)
 \chi_{(\bold{1},\bold{1},\bold{2})} 
 \Bigg]  + {\cal O}(x^3) \,.
\eea 
In this case we get an enhancement of $SO(4)_F\times U(1)^6 \rightarrow SU(5)^2$, and we can express
the full index in terms of $SU(5)^3$ characters:
\be
I^{T_5} =  1 + x^2\left(1+x(y+\frac{1}{y})\right)
\left(\chi_{(\bold{24},\bold{1},\bold{1})} + \chi_{(\bold{1},\bold{24},\bold{1})} + \chi_{(\bold{1},\bold{1},\bold{24})}\right)  +  O(x^4) \,.
\ee 

The general picture that appears to emerge is that the $SO(4)_F$ flavor symmetry of the two flavors on the $SU(2)$ end,
together with all $2N-4$ $U(1)$ symmetries, are enhanced to $SU(N)^2$, with the extra conserved currents being instantons
with charges $(0^k,1^l,0^{N-k-l-2})$ under $U(1)_I^{N-2}$, where $0\leq k \leq N-3$ and $1\leq l \leq N-k-2$.
Checking this explicitly gets hard as $N$ increases and instantons of increasingly higher orders are needed.

\subsubsection{Other theories}

\begin{figure}[h!]
\center
\includegraphics[width=0.25\textwidth]{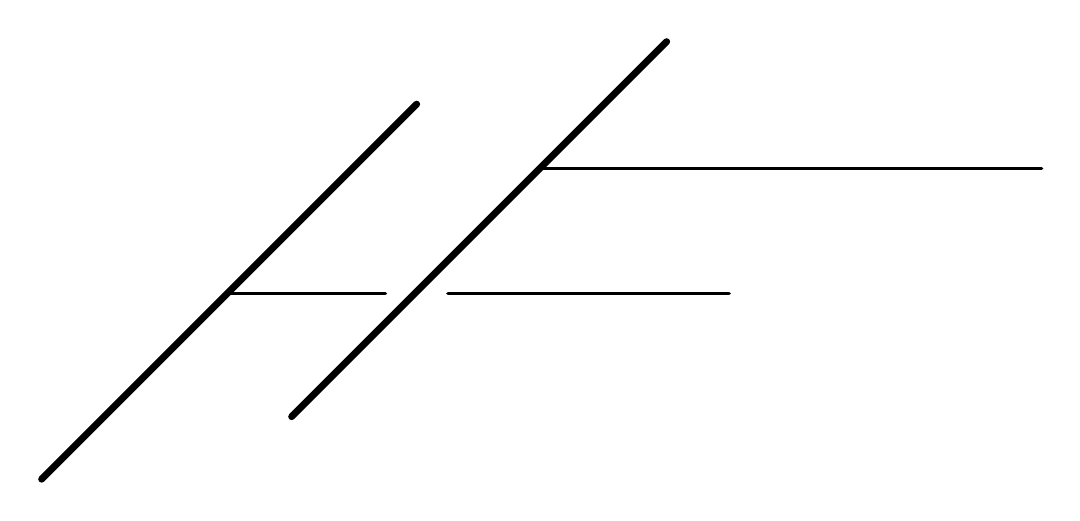} 
\hspace{0.5cm}
\includegraphics[width=0.25\textwidth]{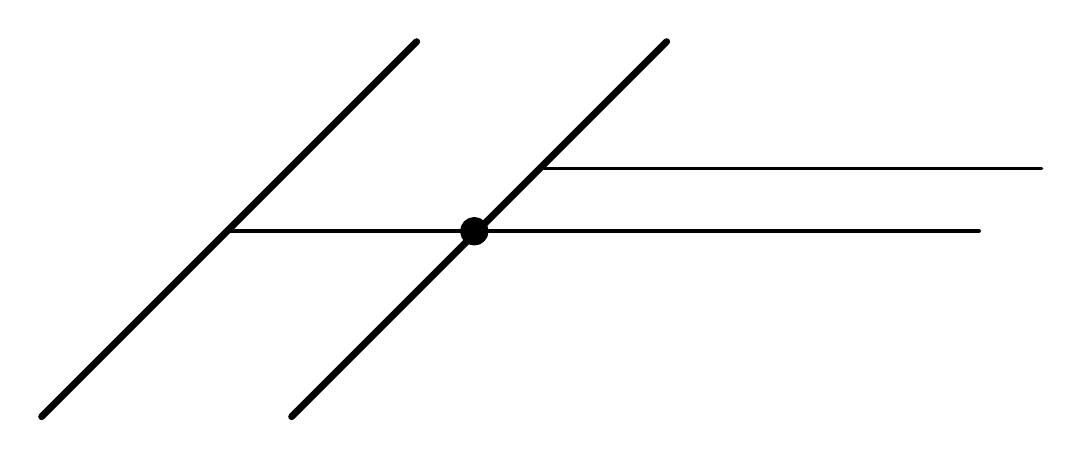} 
\hspace{0.5cm}
\includegraphics[width=0.25\textwidth]{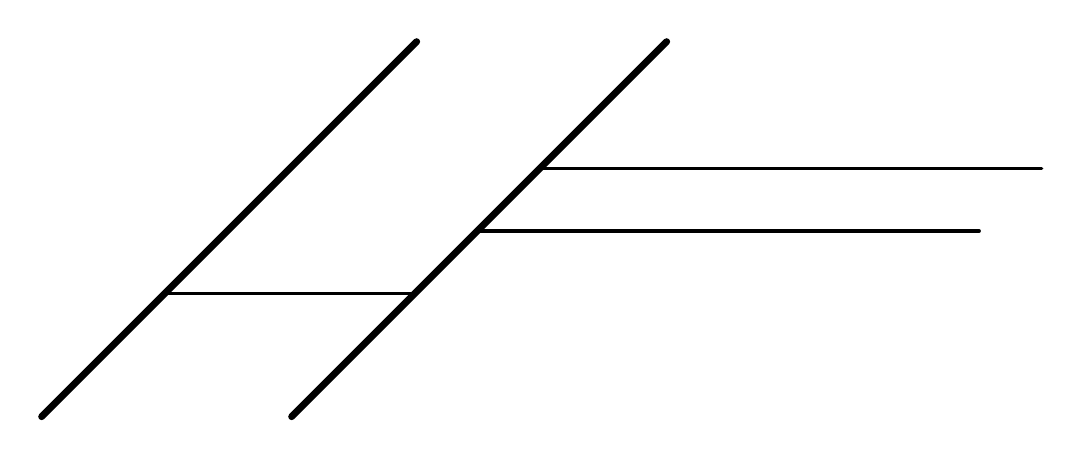} 
\caption{Higgs branch: When the positions of two 7-branes transverse to the 5-branes that end on them
coincide, one of the 5-branes can break.}
\label{break}
\end{figure}

Other isolated theories in 4d are described by 3-punctured spheres with non-maximal punctures, 
and correspond to various limits on the Higgs branch of the $T_N$ theories.
The Young diagram associated with a puncture corresponds to the pattern of symmetry breaking of the corresponding $SU(N)$.
The 5d lifts of these theories are obtained by deformations of the $T_N$ web
in which some of the 5-branes break, and subsequently share a 7-brane boundary (Fig.~\ref{break}) \cite{BBT}.
The broken 5-brane pieces can then be moved away along the 7-branes.
Each column in a Young diagram of a puncture corresponds to a 7-brane in the $N$-junction picture,
and the number of boxes in the column corresponds to the number of 5-branes that end on that 7-brane.

We can use this procedure to obtain 5d Lagrangian descriptions for many other isolated theories.
We'll concentrate on a series of isolated theories considered in \cite{Distler}.
The main properties of these theories are summarized in Table \ref{summary1}.

\begin{table}[h!]
\begin{center}
\begin{tabular}{|l|l|l|l|}
  \hline 
  SCFT & global symmetry & $d_C$ & $d_H$ \\
 \hline
  $T_N$ & $SU(N)^3$ & $\frac{1}{2}(N-1)(N-2)$ & $\frac{1}{2}(3N^2 -N -2)$ \\[3pt]
  \hline
 $R_{0,N}$ & $SU(2N)\times SU(2)$ & $N-2$ & $N^2 +2$  \\
 \hline
$R_{1,N}$ & $SU(N+2)\times SU(2)\times U(1)^2$ & $N-2$ & $\frac{1}{2}(N^2 + 3N +8)$ \\ 
 \hline
 $R_{2,2n+1}$ & $SO(4n+6)\times U(1)$ & $n$ & $2n^2+5n+4$ \\
 \hline
 $S_N$ & $SU(N+2)\times SU(3)\times U(1)$ & $N-3$ & $\frac{1}{2}(N^2 + 3N +12)$ \\[3pt]
  \hline
 $\chi_N^k$ & $SU(N)^2\times SU(k+1)\times U(1)$ &  $k(N-1) - \frac{1}{2} k(k+1)$ & $N^2 + \frac{1}{2} k(k+3) $  \\
 \hline
\end{tabular}
 \end{center}
\caption{Properties of 4d SCFT's discussed in this section} 
\label{summary1}
\end{table}

\subsection{The $R_{0,N}$ theories}
\label{sec:R0N}

Replacing one of the maximal $(1^N)$ punctures of the $T_N$ theory with an $(N-2,1,1)$ puncture gives
a series of theories known as the $R_{0,N}$ theories \cite{Distler}.
The $R_{0,N}$ theory has $d_C = N-2$, $d_H = N^2 +2$,
and a global symmetry $SU(2N)\times SU(2)$ (except for $N=3$).
Clearly $R_{0,3}$ is the same as $T_3$.

\begin{figure}[h]
\center
\includegraphics[width=0.35\textwidth]{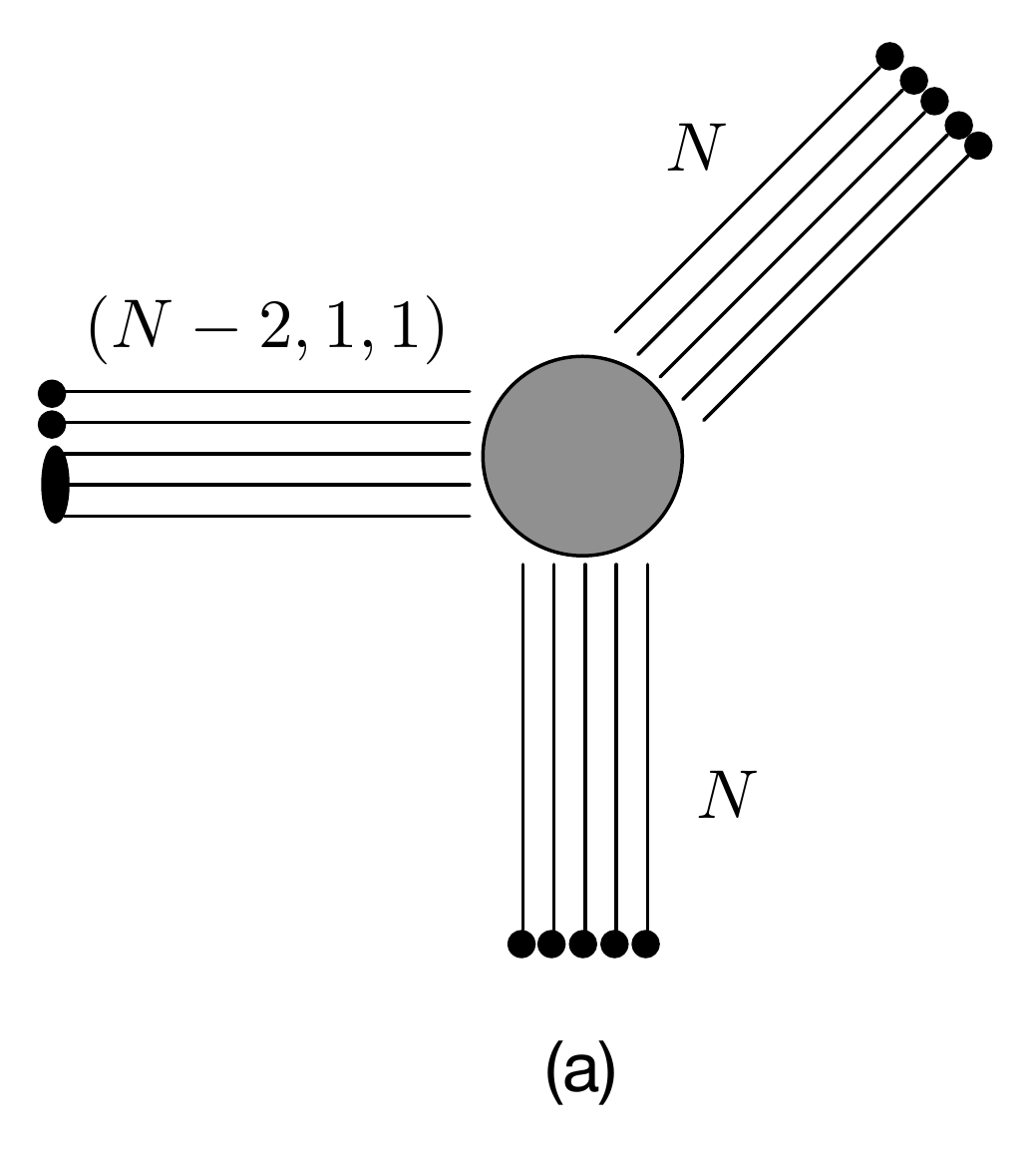} 
\hspace{0.2cm}
\includegraphics[width=0.37\textwidth]{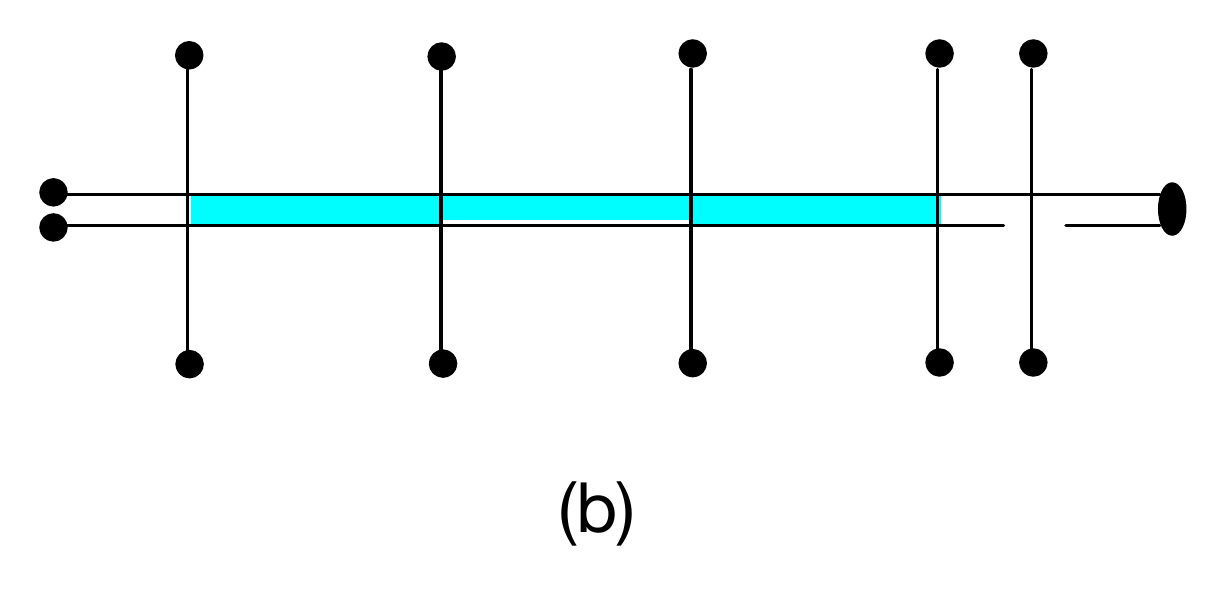} 
\hspace{0.5cm}
\includegraphics[width=0.2\textwidth]{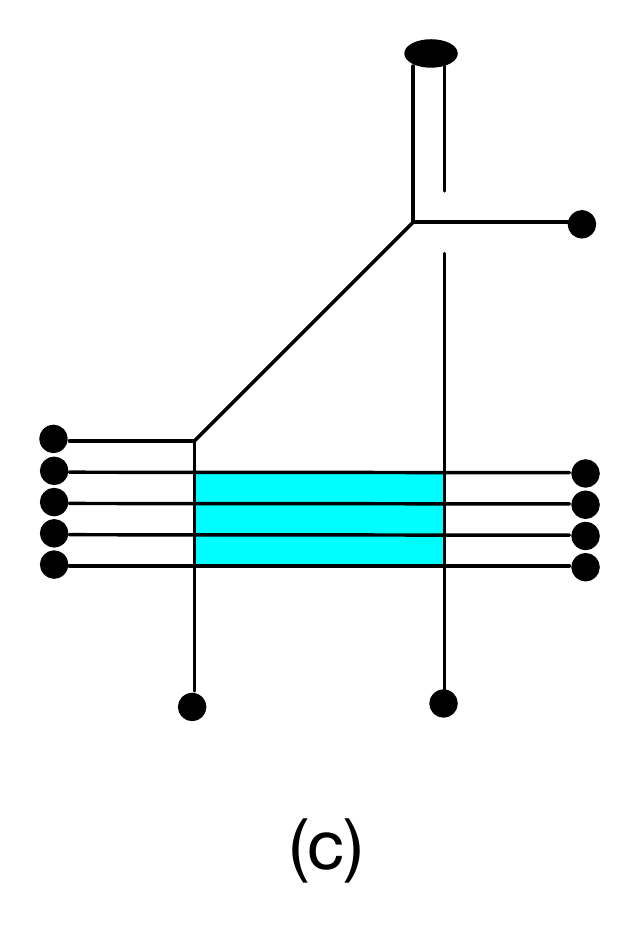}
\caption{5-brane webs for $R_{0,N}$ (shown for $N=5$).} 
\label{R0Nweb}
\end{figure}

\begin{figure}[h]
\center
\includegraphics[width=0.4\textwidth]{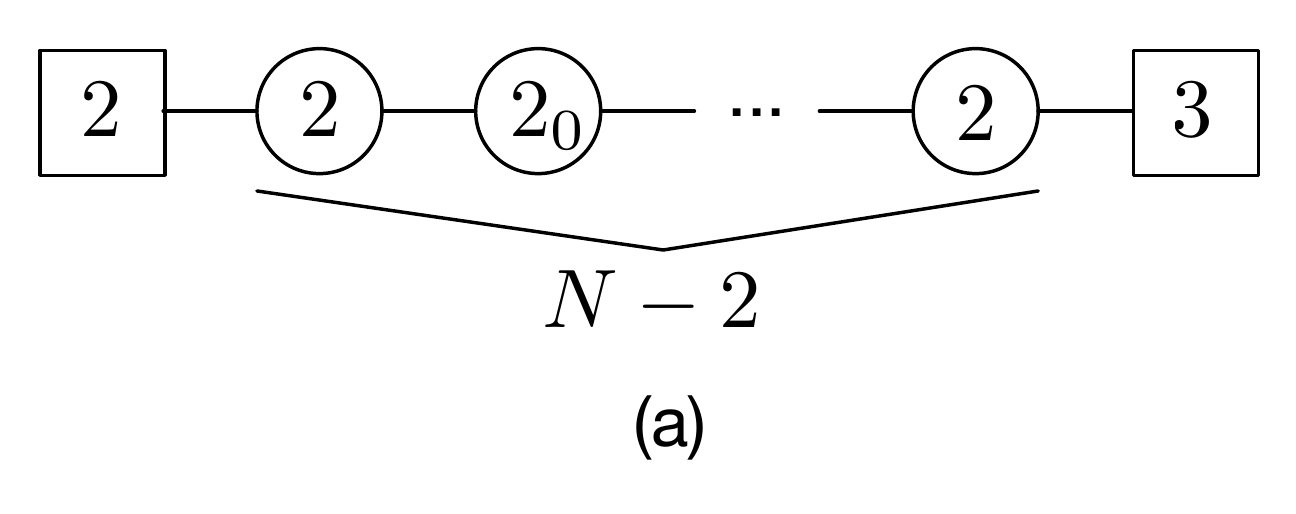}
\hspace{1cm}
\includegraphics[width=0.3\textwidth]{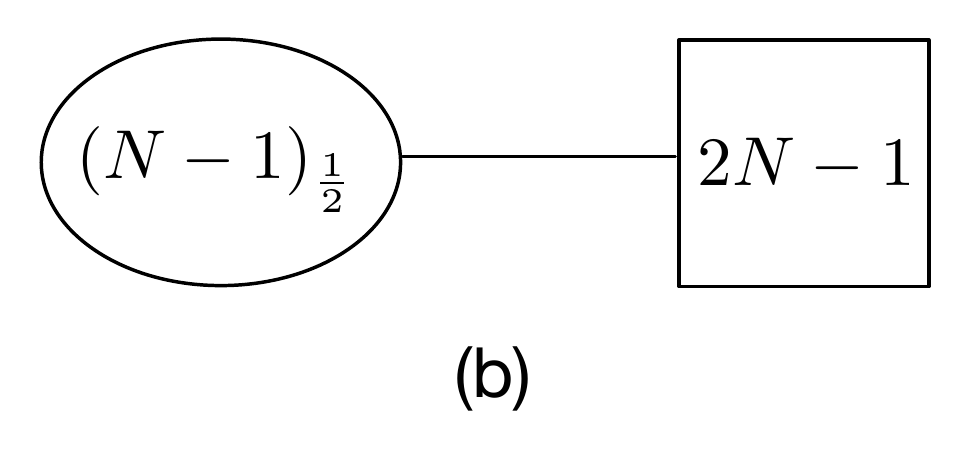}
\caption{Quiver gauge theories for $R_{0,N}$.}
\label{R0Nquivers}
\end{figure}

The corresponding 5-brane junction is shown, for $N=5$, in Fig.~\ref{R0Nweb}a, 
where now $N-2 = 3$ of the five 5-branes in one of the three prongs end on a single 7-brane.
In this case, as well as in subsequent cases, we find it useful to move 
this 7-brane to the right.
The resulting 5-brane web, after an appropriate mass-deformation, is shown in Fig.~\ref{R0Nweb}b.
This describes the quiver gauge theory with $2+SU(2)\times SU(2)\times SU(2)+3$.
The general structure is $2+SU(2)^{N-2} +3$ (Fig.~\ref{R0Nquivers}a).
Now S-duality leads to a different gauge theory, described by the S-dual 5-brane web in Fig.~\ref{R0Nweb}c.
For $R_{0,5}$ this web gives $SU(4)+9$. More generally, the dual gauge gauge theory in $SU(N-1) + (2N-1)$ (Fig.~\ref{R0Nquivers}b).

The bare CS level of second theory (for $N>3$) can be computed as before. The renormalized 
CS level after mass-deforming the flavors is $\kappa = 1- N$, 
and therefore $\kappa_0 = \kappa + \frac{1}{2}(2N-1) = \frac{1}{2}$.
No CS terms exist in the $SU(2)^{N-2}$ theory, but instead one must specify a $\mathbb{Z}_2$-valued $\theta$-parameter,
namely $\theta = 0$ or $\pi$, for each of the unflavored $SU(2)$ factors.
These can be computed in the same way as the CS level, namely by deforming the web so as to give
mass to all the matter (fundamental and bi-fundamental) fields.
Then we can read-off the value of $\theta$ for each of the separate $SU(2)$ sub-webs.
This gives $\theta =0$ for all of them.

In either gauge theory description, seeing the full global symmetry of the $R_{0,N}$ theory requires instantons.
We can however guess the form of the full symmetry by comparing 
the symmetry of the 5-brane junction, $SU(N)^2\times SU(2)\times U(1)$, with that of 
second gauge theory, $SU(2N-1)_F\times U(1)_B \times U(1)_I$.
The unique rank $2N$ group accommodating both of these as subgroups is $SU(2N)\times SU(2)$.

\subsubsection{Superconformal index}

Let us verify this in the simplest case of $R_{0,4}$ by computing the superconformal index.
We begin with the $3+SU(2)^{(1)}\times SU(2)^{(2)}+2$ description, 
which has a classical global symmetry $SO(6)_F\times SU(2)_{BF}\times SO(4)_F\times U(1)_I^2$.
The calculation is similar to the one for the $T_4$ theory, where we treat $SU(2)^{(1)}$ as $USp(2)$
and $SU(2)^{(2)}$ as $U(2)/U(1)$.
We therefore need to remove the decoupled states associated to the second $SU(2)$ factor, 
which are described by a D1-brane between the corresponding pair of external NS5-branes in the web of Fig.~\ref{R0Nweb}b.
The required correction of the instanton partition function is
\be
\mathcal{Z}_{c} = PE\left[\frac{x^2 q_2 (z f_1 + \frac{1}{z f_1})}{(1-x y)(1-\frac{x}{y})}\right]  \mathcal{Z} \,,
\ee    
where $z$ and $f_1$ are the fugacities associated to $SU(2)_{BF}$ and $SU(2)_{F_1}\subset SO(4)_F$ respectively, and $q_2$ is 
the instanton fugacity of $SU(2)^{(2)}$.
Note that in addition to the lack of $x\rightarrow 1/x$ invariance, the correction factor cannot be expressed in terms of 
characters of $SU(2)_{BF}$ and $SU(2)_{F_1}$.
This corrects another ``$U(1)$ remnant": 
for $U(2)$ the global symmetries associated with the bi-fundamental and fundamentals
are $U(1)_{BF}$ and $U(2)_F$, and the instanton partition function respects only these.
The full $SU(2)_{BF}$ and $SO(4)_F$ are recovered only after the correction factor is included.

To order $x^3$ there are contributions from the $(1,0)$, $(0,1)$, $(1,1)$ and $(2,1)$ instantons,
and the combined result for the index is (presented to order $x^2$ for conciseness)
\bea
\label{R04index1}
I^{R_{0,4}} & = & 1 + x^2\Big(2 + \chi_{(\bold{15},\bold{1},\bold{1},\bold{1})} + \chi_{(\bold{1},\bold{3},\bold{1},\bold{1})} + \chi_{(\bold{1},\bold{1},\bold{3},\bold{1})} 
+ \chi_{(\bold{1},\bold{1},\bold{1},\bold{3})} + q_1 \chi_{(\bold{4},\bold{2},\bold{1},\bold{1})} 
+ \frac{1}{q_1} \chi_{(\bar{\bold{4}},\bold{2},\bold{1},\bold{1})}
\nonumber \\  
& + &  (q_2 +\frac{1}{q_2})\chi_{(\bold{1},\bold{2},\bold{2},\bold{1})} + q_1 q_2 \chi_{(\bold{4},\bold{1},\bold{2},\bold{1})} 
+ \frac{1}{q_1 q_2} \chi_{(\bar{\bold{4}},\bold{1},\bold{2},\bold{1})}\Big) + {\cal O}(x^3) \,.
\eea 
This exhibits the enhancement of $SO(6)_F\times SU(2)_{BF}\times SU(2)_{F_1}\times U(1)_I^2 \rightarrow SU(8)$,
where the fundamental of $SU(8)$ decomposes as
${\bf 8} = ({\bf 4},{\bf 1},{\bf 1},{\bf 1})_{(\frac{1}{2},\frac{1}{4})} + ({\bf 1},{\bf 2},{\bf 1},{\bf 1})_{(-\frac{1}{2},\frac{1}{4})} 
+ ({\bf 1},{\bf 1},{\bf 2},{\bf 1})_{(-\frac{1}{2},-\frac{3}{4})}$.

\begin{figure}[h]
\center
\includegraphics[width=0.23\textwidth]{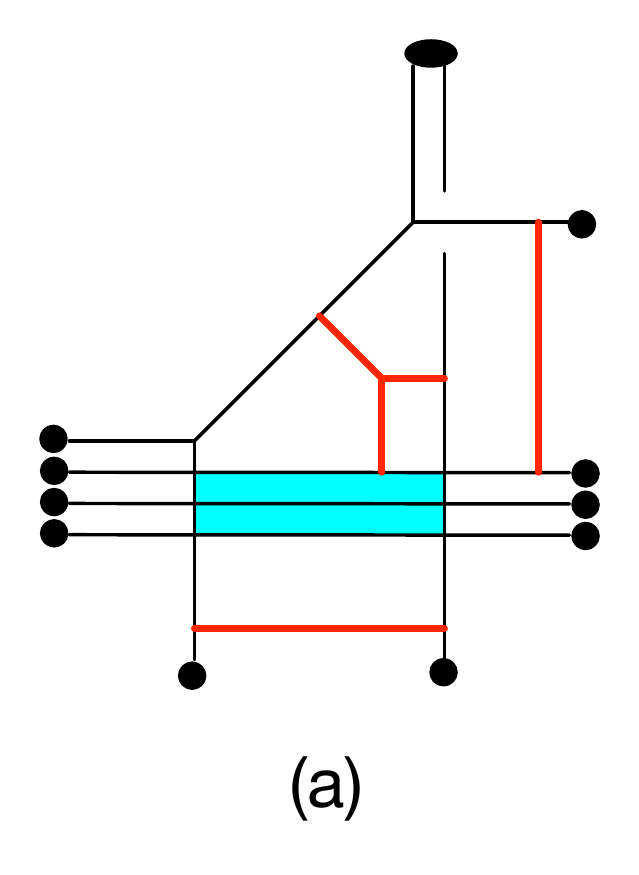} 
\hspace{1cm}
\includegraphics[width=0.25\textwidth]{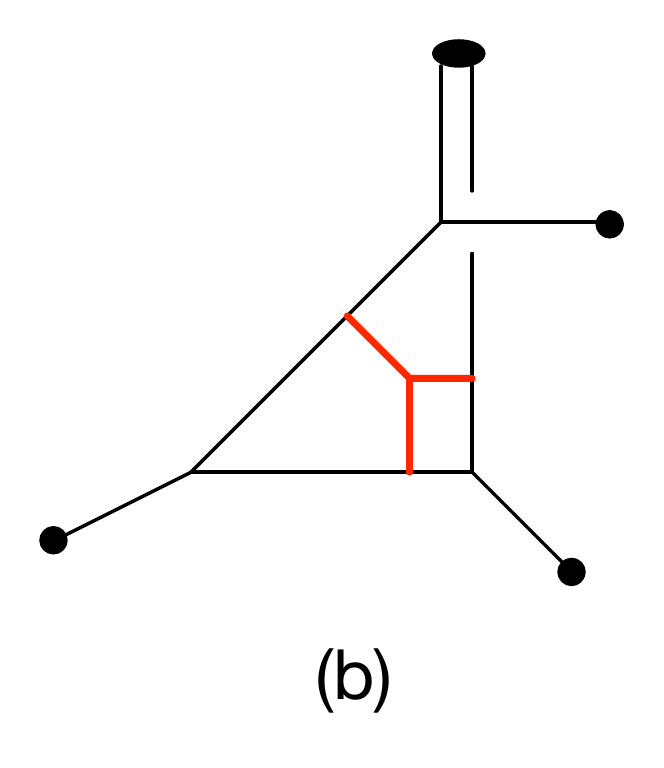}
\caption{5-brane web for the $SU(3)_{-\frac{1}{2}} + 7$ description of $R_{0,4}$ showing the ``$U(1)$ remnant" states.} 
\label{R0Nremnants}
\end{figure}

We can also demonstrate the enhancement in the dual $SU(3)_{\frac{1}{2}} +7$ description.\footnote{Note that, 
like the quiver gauge theory, this gauge theory is also outside the regime of theories classified in \cite{Intriligator:1997pq},
since the low energy effective theory has a singularity on the Coulomb branch.
This is apparent in the 5-brane web in Fig.~\ref{R0Nweb}c, where the upper triangular part shrinks to zero size
at a finite distance from the origin of the Coulomb branch.
However, as in other cases, the singularity is resolved by taking into account the additional light instantonic states. More generally, 
we would conjecture that there exists a UV fixed point for $SU(N)_{\kappa} +N_f$ 
provided $N_f + 2|\kappa| \leq 2N+4$ and $N_f<2N+4$.}
The classical global symmetry in this case is $SU(7)_F\times U(1)_B\times U(1)_I$.
As before, the instanton partition function is not invariant under $x\rightarrow 1/x$, and does not respect the full classical global symmetry.
The spectrum of ``$U(1)$ remnant" states that must be removed in this case is a bit more involved.
We can get an idea for what they are from the 5-brane web (see Fig.~\ref{R0Nremnants}).
There are three types coming from (a) the D1-brane between the separated external NS-branes, (b) fundamental strings
between the separated external D5-brane and the flavor D5-branes, and (c) a 3-string junction in the upper part of the web.
The second and third types are novel.

The state corresponding to the fundamental strings is clearly charged under $SU(7)_F$.
What is less obvious is that it also carries an instanton charge even though it is described
by a fundamental string, which is why it contributes to the instanton partition function.
This can be understood from the fact that its mass depends on the value of the gauge coupling,
which is seen geometrically by the upward motion of the separated external D5-brane as
the external NS5-branes move apart.
The 3-string junction state carries both an instanton charge and an $SU(3)$ gauge charge, so unlike the other two states,
it is not decoupled from the gauge theory.
This complicates the procedure for correcting the partition function, since this state is not expected 
to plethystically exponentiate. Unlike D1-branes or fundamental strings, one can merge multiple string junctions to make
a new state, as suggested by the fact that these states interact.
We can therefore only determine the correction to the 1-instanton partition function in this case.
Actually, the simplest setting in which this state appears is in the 5-brane web shown in Fig.~\ref{R0Nremnants}b,
which describes an empty ``$SU(1)_2$" theory. The 1-instanton partition function computed for this theory 
using $U(1)_2$ is precisely the contribution of this state,
and is given by
\be
Z_1^{SU(1)_2} = \frac{x^3}{(1-x y)(1-\frac{x}{y})} \,.
\ee
The full correction to the 1-instanton partition function for the $SU(3)+7$ theory is therefore
\be
Z^c_1 = Z_1 + \frac{x^2 q' \left(b^{-\frac{7}{2}} 
+ b^{\frac{5}{2}} \chi^{SU(7)}_{\bar{\bold{7}}} - xb^{\frac{7}{2}} \chi^{SU(3)}_{\bold{3}}\right)}
{(1-x y)(1-\frac{x}{y})} \,,
\ee   
where $q'$ is the $SU(3)$ instanton fugacity. 
The first, second and third terms in the numerator correspond respectively to the three remnant states above.
This correction restores the $x\rightarrow 1/x$ invariance to the 1-instanton partition function.
One can also check that for $N=3$, {\em i.e.}, for $SU(2)+5$ with the $SU(2)$ treated as $U(2)/U(1)$,
the analogous subtraction gives the same 1-instanton partition function as when using $USp(2)$.

The full index including the 1-instanton contribution, expressed in terms of $SU(7)_F$ characters, is then given by
\bea
\label{R04index2}
I^{R_{0,4}} & = & 1 + x^2\left(2 + \chi_{\bold{48}} + q' b^{\frac{5}{2}} \chi_{\bar{\bold{7}}} + \frac{1}{q' b^{\frac{5}{2}}} \chi_{\bold{7}} 
+ \frac{q'}{b^{\frac{7}{2}}} + \frac{b^{\frac{7}{2}}}{q'} \right) \\ 
& + & x^3 \Bigg((y+\frac{1}{y})\left(3 + \chi_{\bold{48}} + q' b^{\frac{5}{2}} \chi_{\bar{\bold{7}}} + \frac{1}{q' b^{\frac{5}{2}}} \chi_{\bold{7}} 
+ \frac{q'}{b^{\frac{7}{2}}} + \frac{b^{\frac{7}{2}}}{q'} \right) \nonumber \\ 
&& \mbox{} +  \left(b^3 + \frac{q'}{\sqrt{b}}\right)\chi_{\bold{35}} + \left(\frac{1}{b^3} + \frac{\sqrt{b}}{q'}\right)\chi_{\bar{\bold{35}}} \Bigg) + {\cal O}(x^4) \,.
\eea 
This exhibits the enhancement of $SU(7)_F\times U(1)_B\times U(1)_I \rightarrow SU(8)\times SU(2)$,
where the $SU(2)$ is spanned by $\sqrt{q'}\, b^{-\frac{7}{4}}$, and the $SU(8)$ by 
$\chi^{SU(8)}_{\bf 8} =\frac{1}{q'^{\frac{1}{8}}b^{\frac{5}{16}}} (\chi^{SU(7)}_{\bold{7}} + q' b^{\frac{5}{2}})$.

The superconformal indices of the dual gauge theories agree, and can be expressed in terms of 
$SU(8)\times SU(2)$ characters as
\bea
I^{R_{0,4}} & = & 1 + x^2(\chi_{({\bf 1},\bold{3})} + \chi_{(\bold{63},{\bf 1})}) + x^3 \left((y+\frac{1}{y})(1 + \chi_{({\bf 1},\bold{3})} 
+ \chi_{(\bold{63},{\bf 1})}) + \chi_{(\bold{70},{\bf 2})} \right) 
+ {\cal O}(x^4) . \nonumber \\
\eea

\subsection{The $\chi_N^k$ theories}
\label{sec:chi}

\begin{figure}[h]
\center
\includegraphics[width=0.2\textwidth]{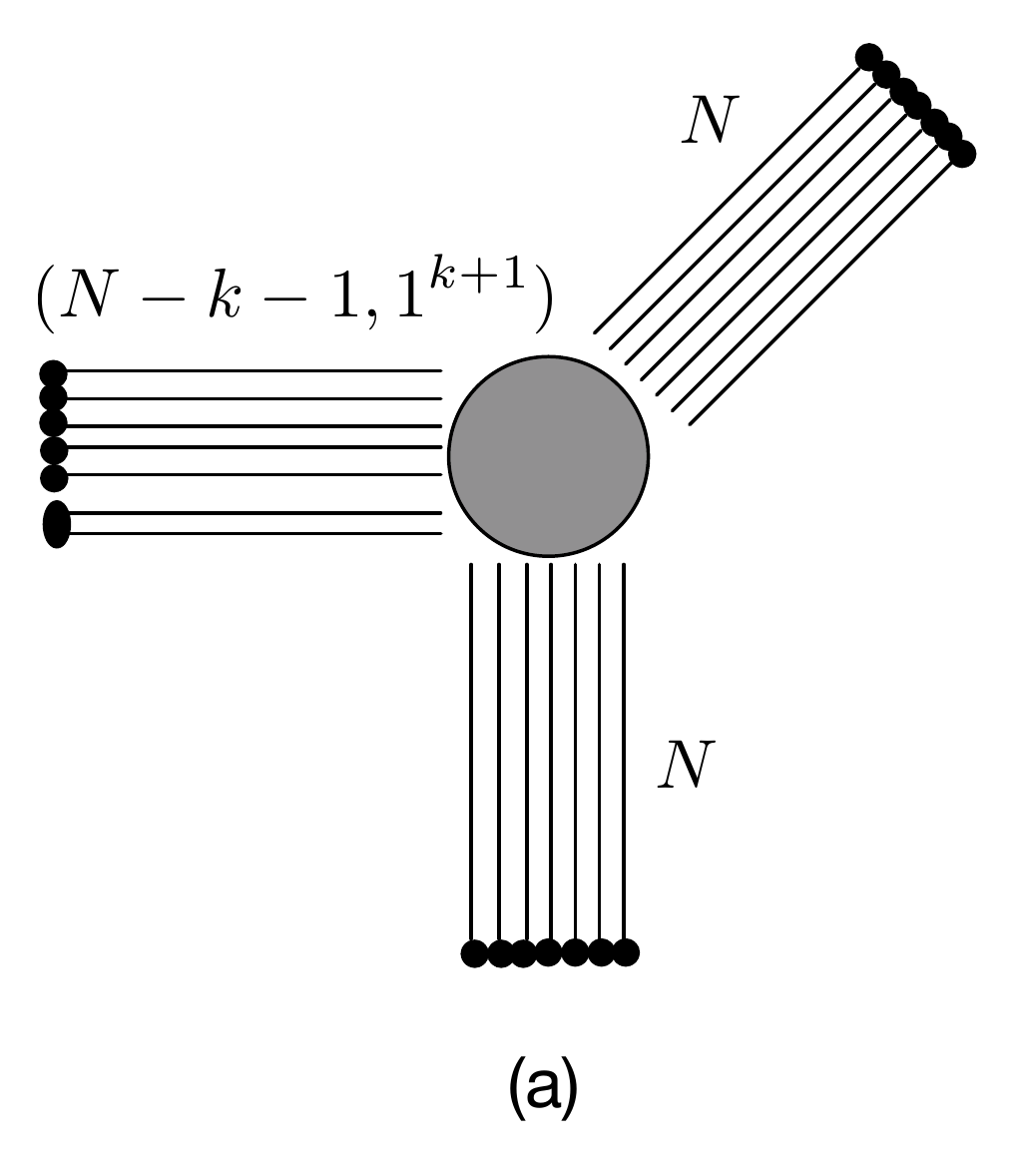} 
\hspace{0.5cm}
\includegraphics[width=0.38\textwidth]{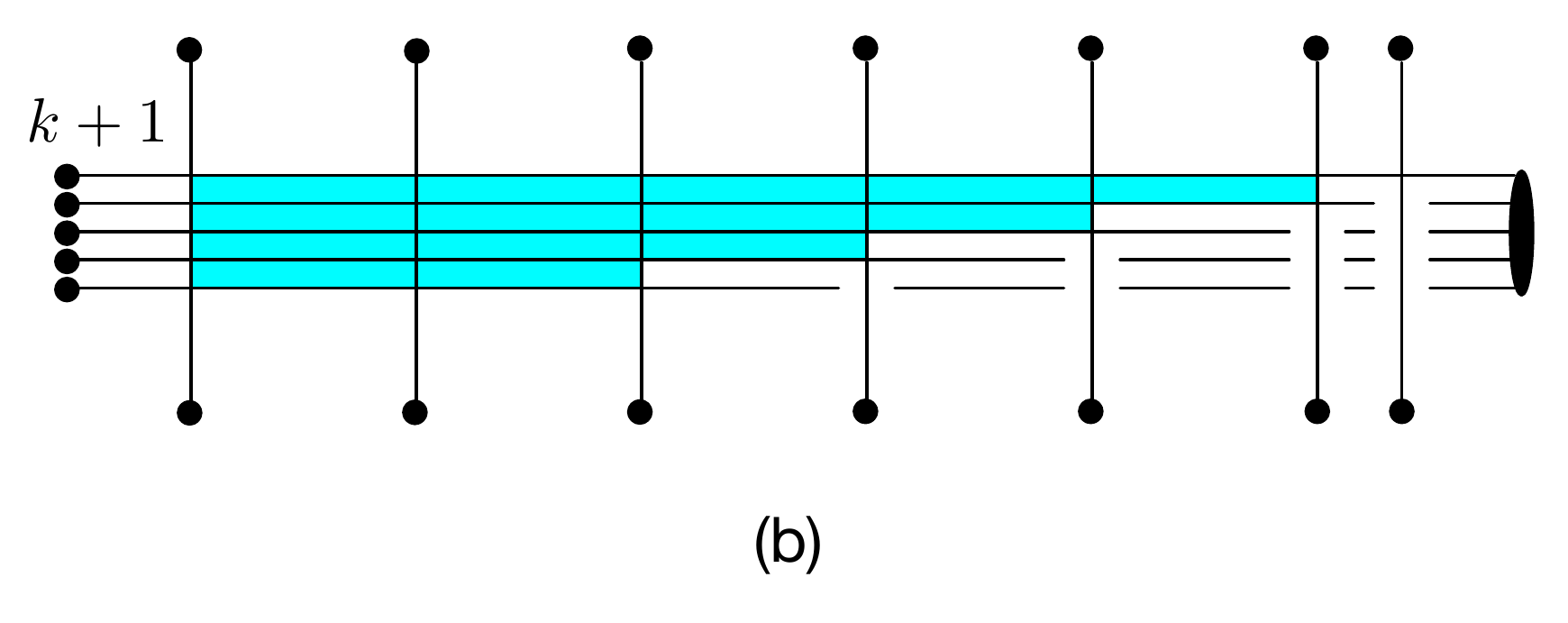} 
\hspace{0.5cm}
\includegraphics[width=0.3\textwidth]{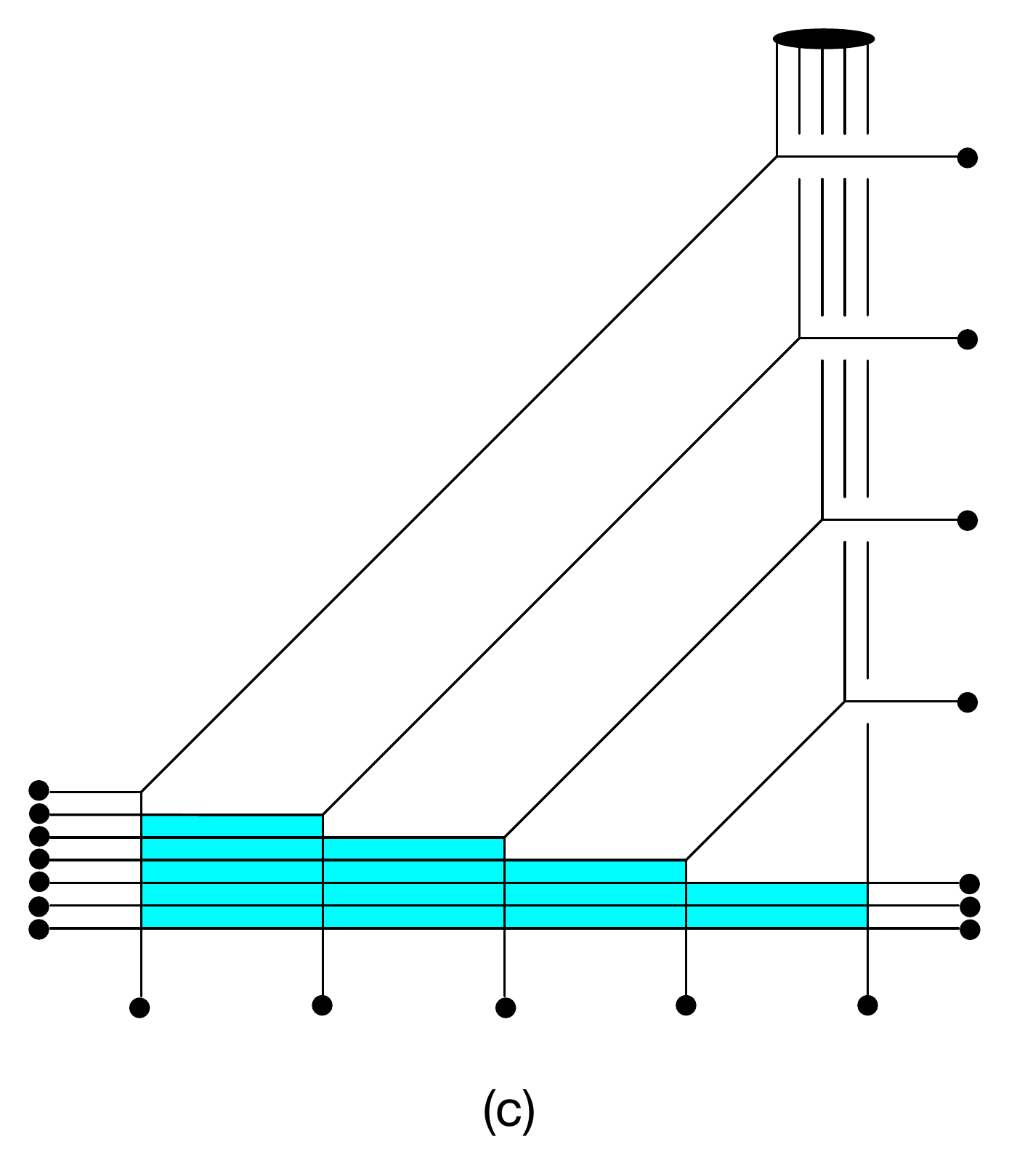} 
\caption{5-brane webs for $\chi_N^k$ ($N=7, k=4$)}
\label{chiwebs}
\end{figure}

A more general class of theories that will be useful below is gotten by replacing one of the maximal
punctures of the $T_N$ theory by a puncture labelled by $(N-k-1,1^{k+1})$. 
We call these theories $\chi_N^k$.
For $k=N-2$ this is just $T_N$.
For $k=1$ this is the $R_{0,N}$ theory.
The $k=2$ and $k=3$ cases were considered in \cite{Distler}, where they were called $U_N$ and $W_N$, respectively.
The $3<k<N-2$ cases have not been considered previously. 
The global symmetry that follows from the puncture structure is apparently $SU(N)^2\times SU(k+1)\times U(1)$.
We can also read-off the structure of the Coulomb branch using the formulas in \cite{Distler}.
We find that it has a graded dimension $(0,1,2,\ldots,k-1,k,\ldots,k)$ associated to operators
of mass dimensions $(2,3,4\ldots,k+1,k+2,\ldots,N)$.
The total dimension of the Coulomb branch is therefore $d_C = k (N-1)-k(k+1)/2$.
The dimension of the Higgs branch is easiest to determine from the 5-brane junction describing 
the 5d lift of the theory,
Fig.~\ref{chiwebs}a. This gives $d_H = N^2 + k(k+3)/2$.

\begin{figure}[h!]
\center
\includegraphics[width=0.5\textwidth]{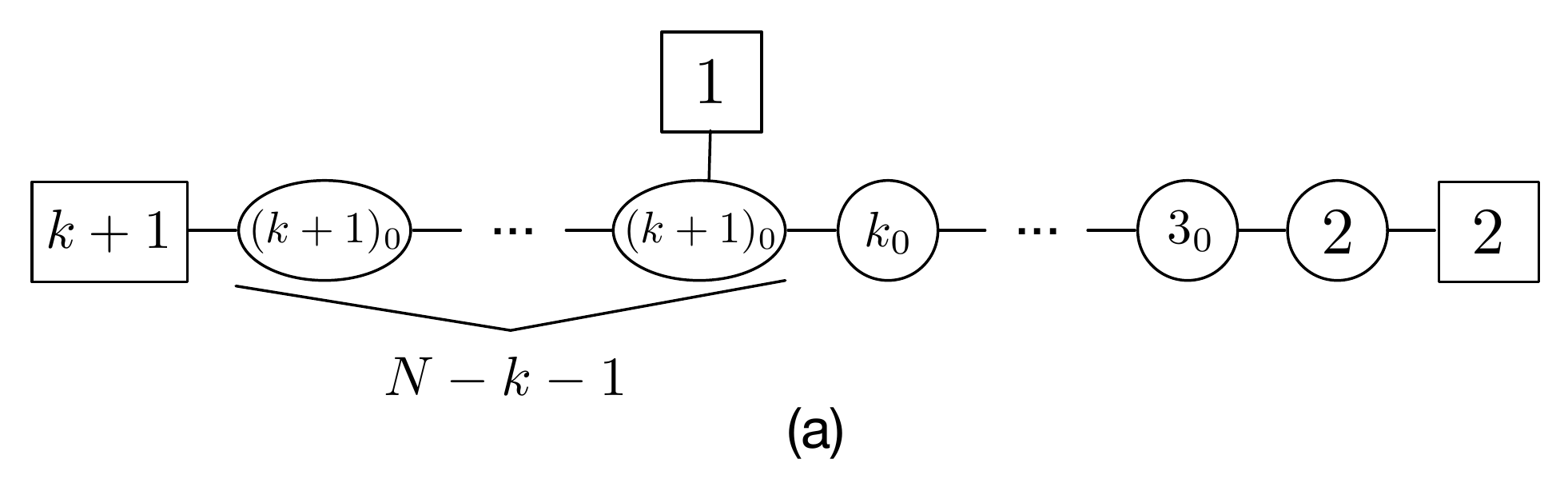} 
\hspace{1cm}
\includegraphics[width=0.4\textwidth]{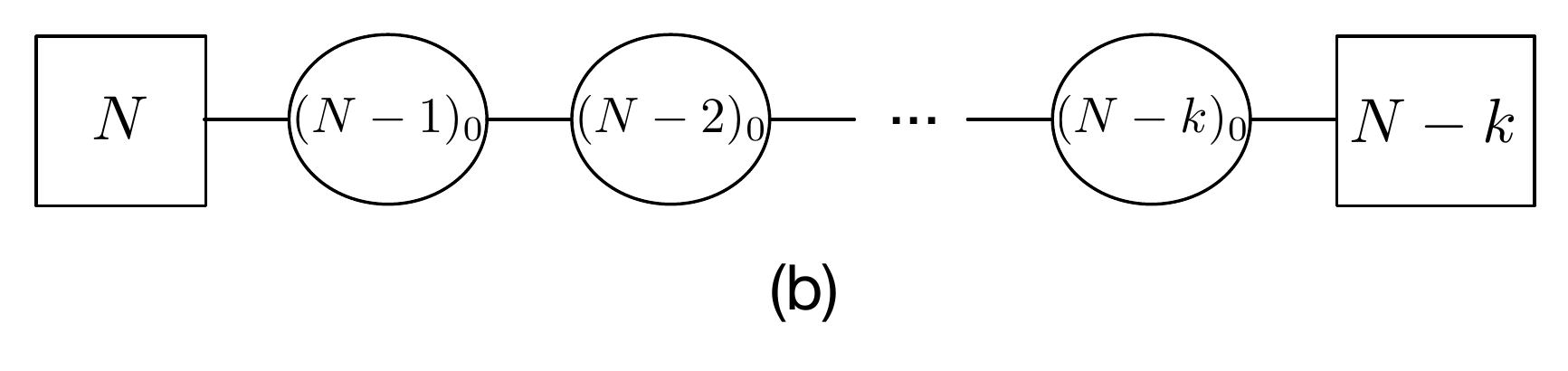} 
\caption{Gauge theories for $\chi_N^k$}
\label{chiquivers}
\end{figure}

Gauge theories for the 5d $\chi_N^k$ theory are found as before by considering the deformation 
of the web, Fig.~\ref{chiwebs}b, and its S-dual, Fig.~\ref{chiwebs}c.
This is shown for the first ``new" theory, $\chi_7^4$, for which the gauge theories are
$5+SU(5)_0\times [SU(5)_0+1] \times SU(4)_0 \times SU(3)_0 \times SU(2)+2$ and
$7+SU(6)_0\times SU(5)_0\times SU(4)_0\times SU(3)_0+3$, respectively.
The quiver diagrams for the two gauge theories in the general case are shown in Fig.~\ref{chiquivers}.

The computation of the superconformal index becomes technically challenging as $N$ and $k$ are increased,
so we will not pursue it presently.

\subsection{More theories}

\subsubsection{$R_{1,N}$}
\label{section:R1N}

The $R_{1,N}$ series ($N\geq 4$) is defined by replacing two of the maximal punctures of $T_N$ by
$(n,n-1,1)$ and $((n-1)^2,1^2)$ for $N=2n$, and by
$(n,n,1)$ and $(n,n-1,1^2)$ for $N=2n+1$.
The first new theory is $R_{1,5}$, since $R_{1,4}$ is equivalent to $R_{0,4}$.
These theories have $d_C = N-2$, $d_H = \frac{1}{2}(N^2 + 3N +8)$,
and a global symmetry $SU(N+2)\times SU(2)\times U(1)^2$ (except for $N=5$, where the global
symmetry is $SU(7)\times SU(3)\times U(1)$) \cite{Distler}.

\begin{figure}[h]
\center
\includegraphics[width=0.3\textwidth]{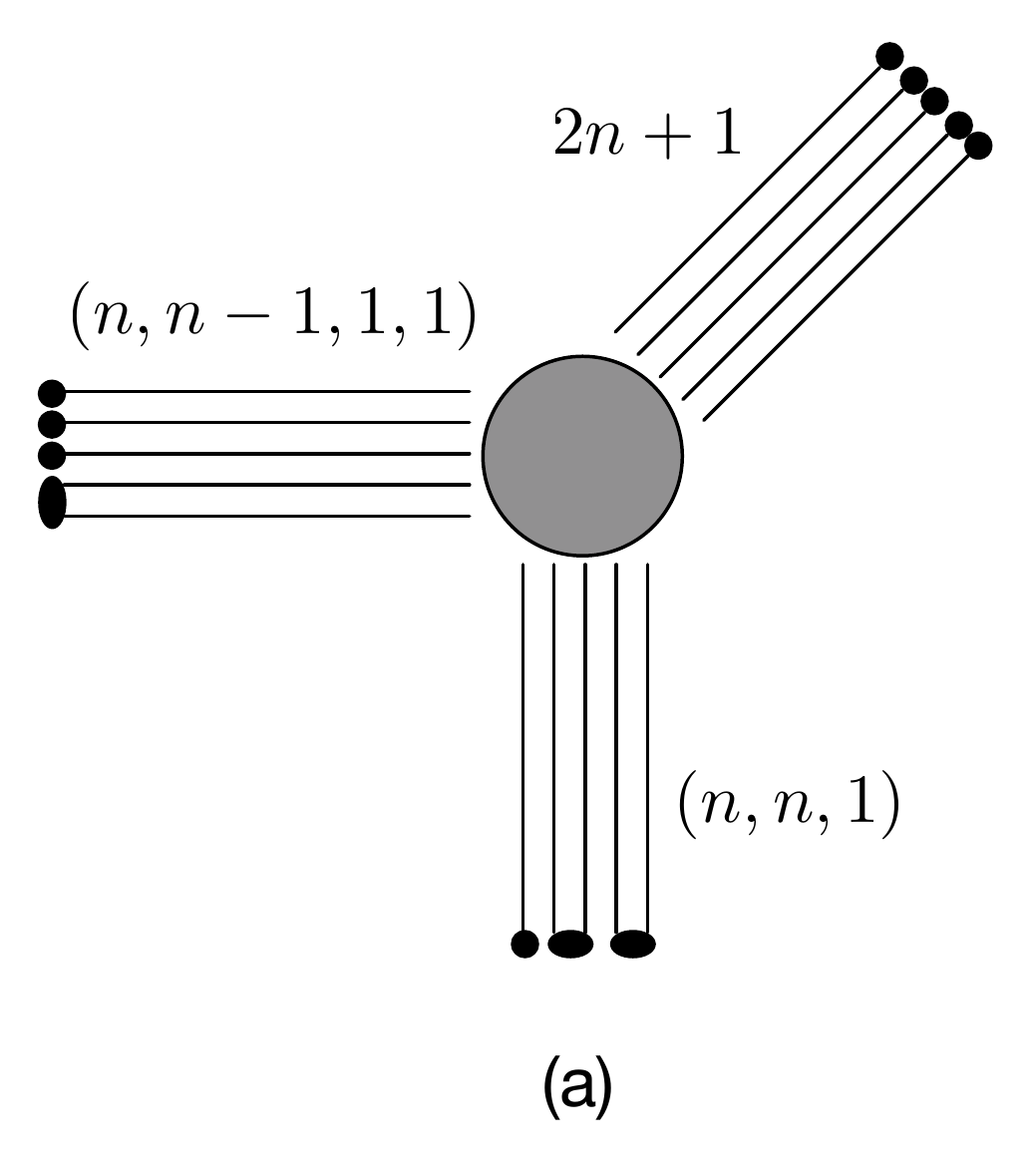} 
\hspace{0.2cm}
\includegraphics[width=0.27\textwidth]{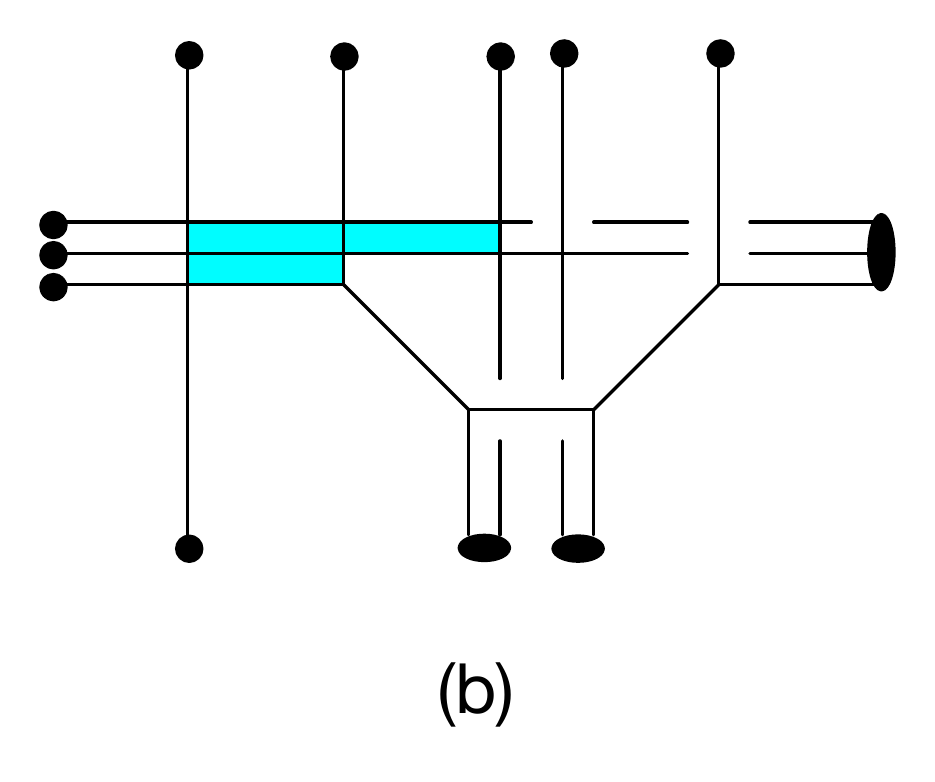} 
\hspace{0.5cm}
\includegraphics[width=0.2\textwidth]{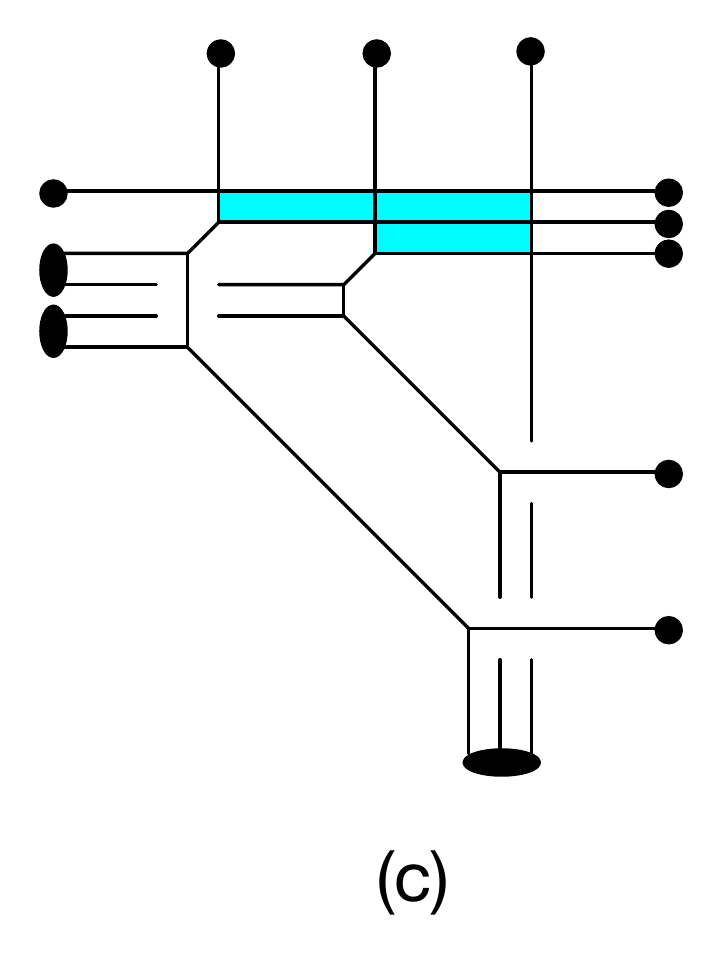} 
\caption{$R_{1,2n+1}$ webs (shown for $n=2$): }
\label{R1(2n+1)web}
\end{figure}

\begin{figure}[h]
\center
\includegraphics[width=0.4\textwidth]{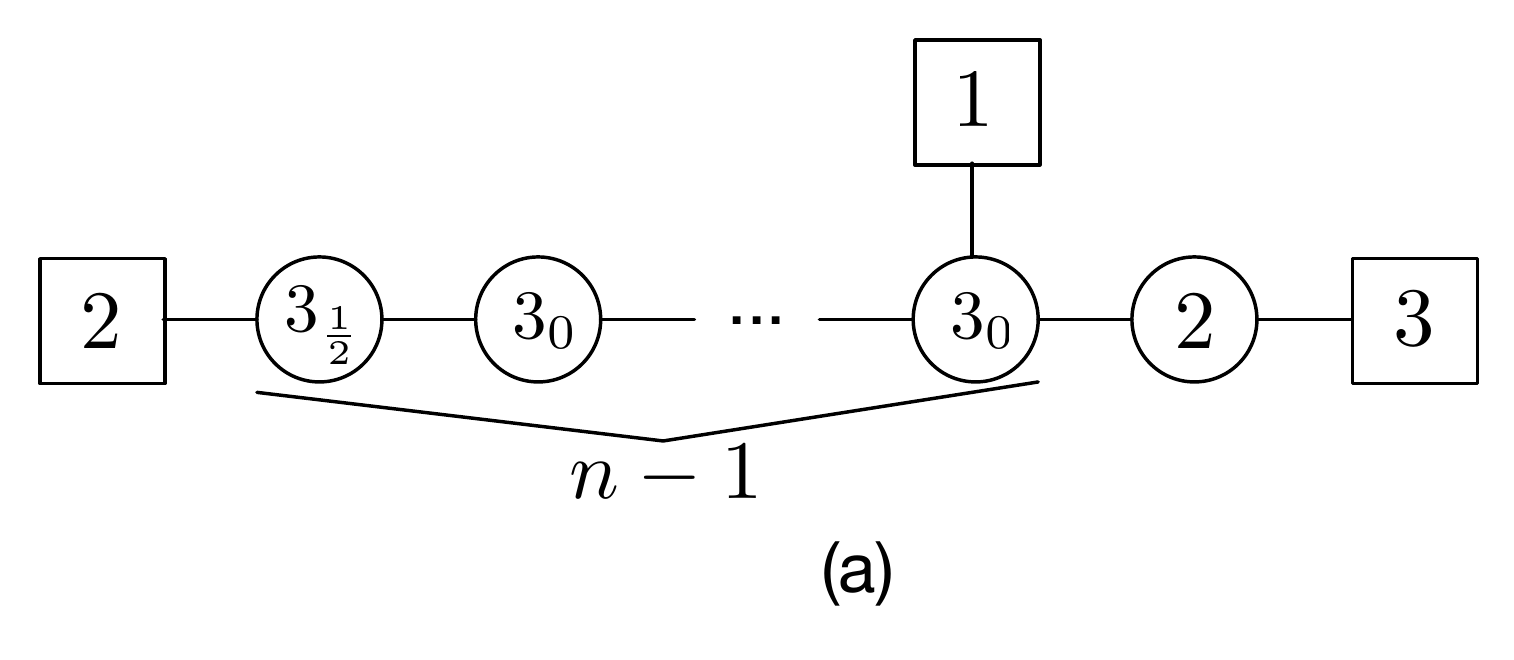} 
\hspace{0.5cm}
\includegraphics[width=0.4\textwidth]{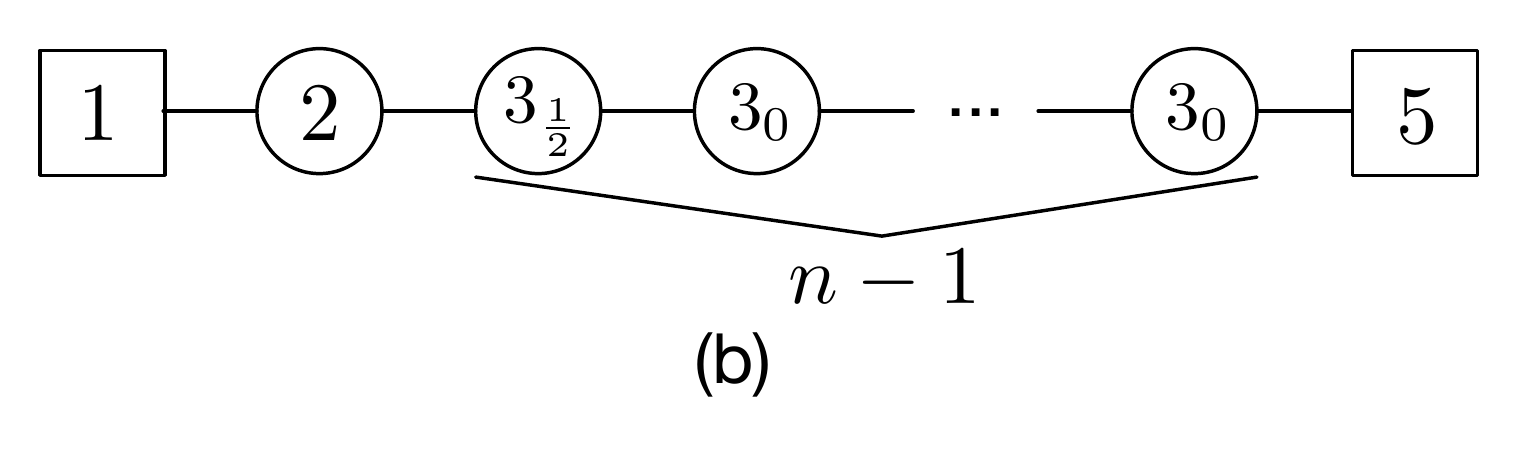} 
\caption{Quiver gauge theories for $R_{1,2n+1}$.} 
\label{R1(2n+1)quivers}
\end{figure}

The 5-brane junction for $N=2n+1$, represented by $R_{1,5}$, is shown in Fig.~\ref{R1(2n+1)web}a.
The mass-deformed web resulting from moving the D7-brane with the two D5-branes to the right, shown in Fig.~\ref{R1(2n+1)web}b, 
describes the linear quiver theory $3+SU(3)_{\frac{1}{2}}\times SU(2)+3$.
The S-dual web (Fig.~\ref{R1(2n+1)web}c) on the other hand gives $1+SU(2)\times SU(3)_{\frac{1}{2}}+5$.
The flavor structure is made more manifest using the manipulations described in Appendix~\ref{sec:flavors}.
The quiver diagrams for the general cases are shown in Fig.~\ref{R1(2n+1)quivers}.

\begin{figure}[h!]
\center
\includegraphics[width=0.3\textwidth]{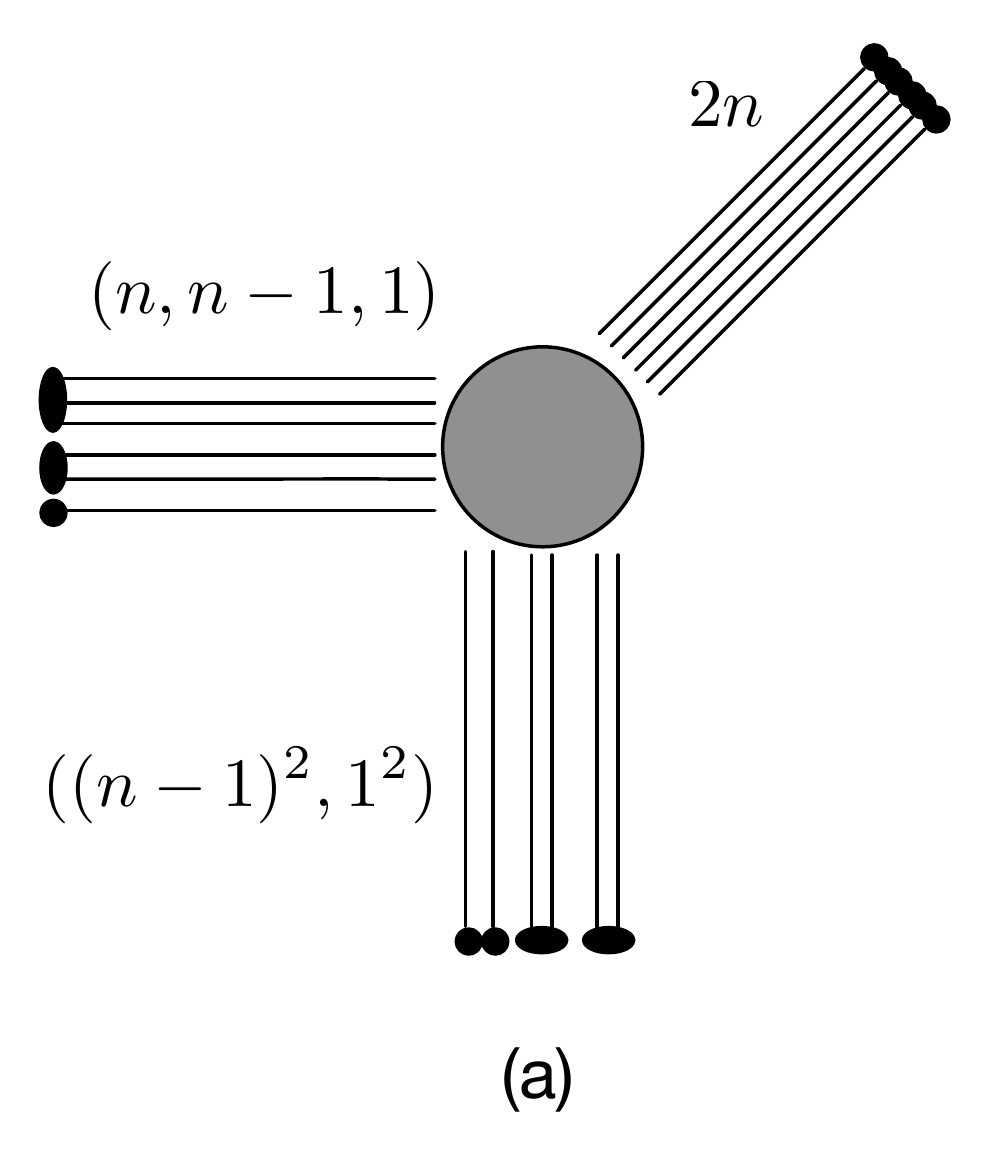} 
\hspace{0.2cm}
\includegraphics[width=0.37\textwidth]{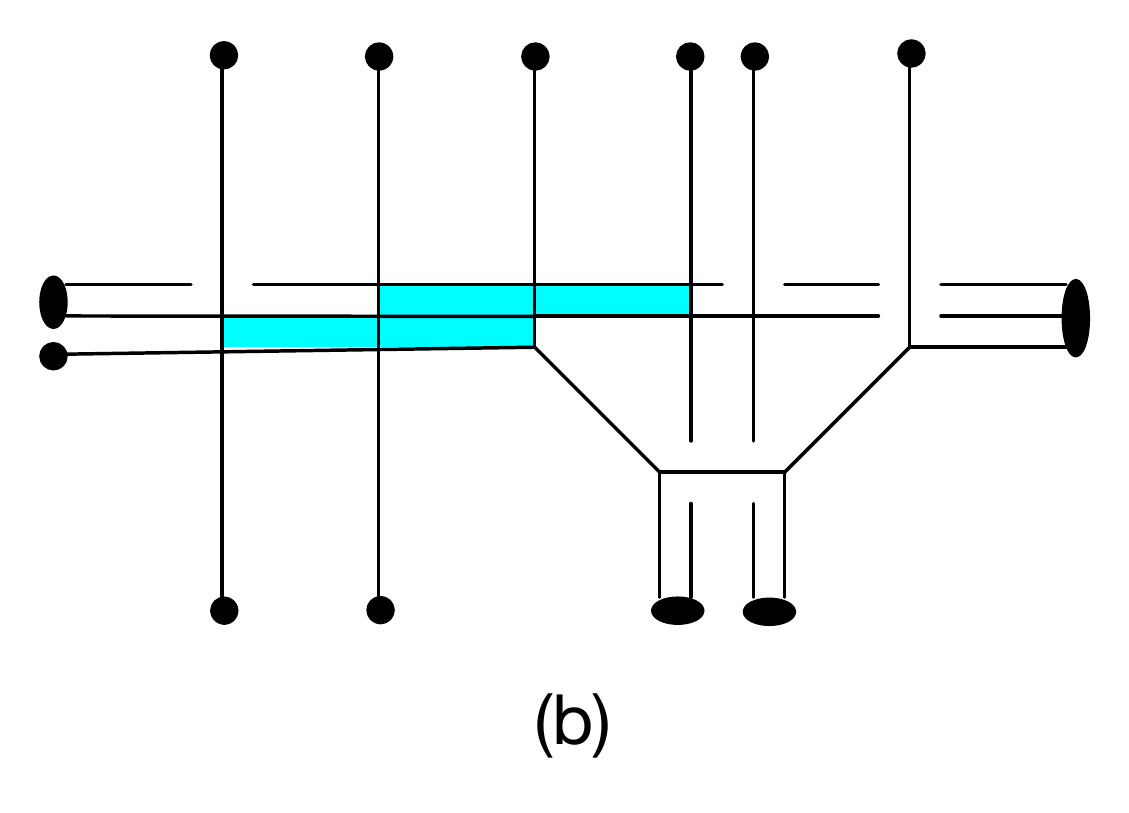} 
\hspace{0.5cm}
\includegraphics[width=0.22\textwidth]{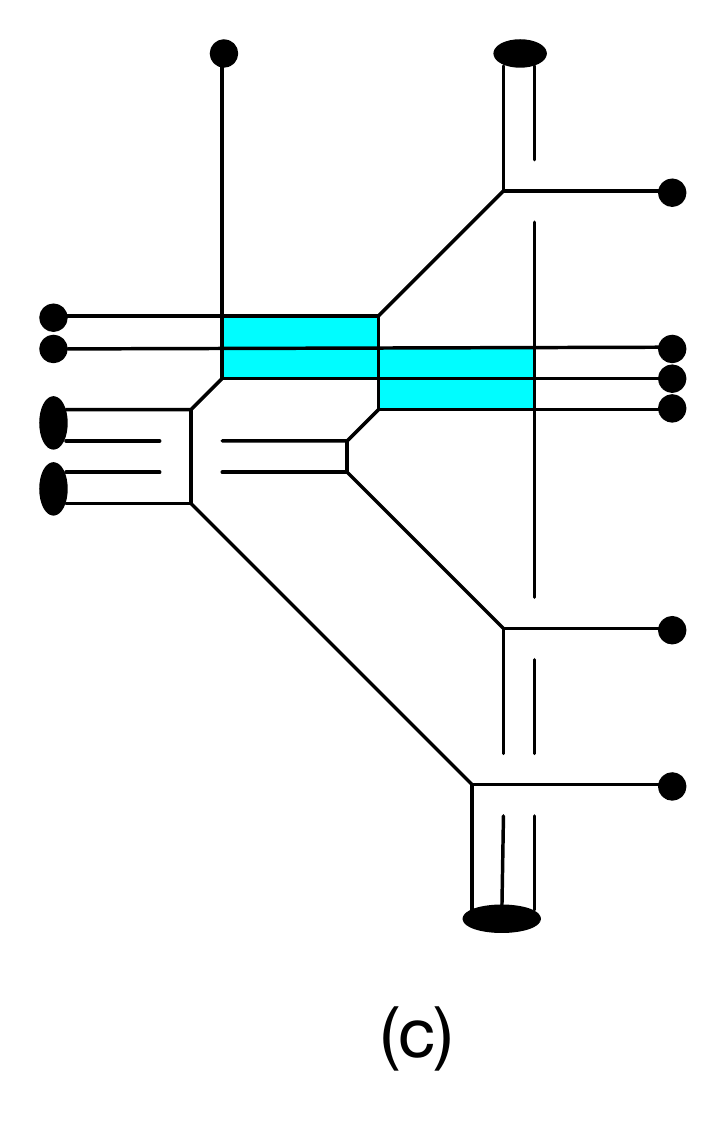} 
\caption{$R_{1,2n}$ webs (shown for $n=3$): }
\label{R1(2n)web}
\end{figure}

\begin{figure}[h!]
\center
\includegraphics[width=0.35\textwidth]{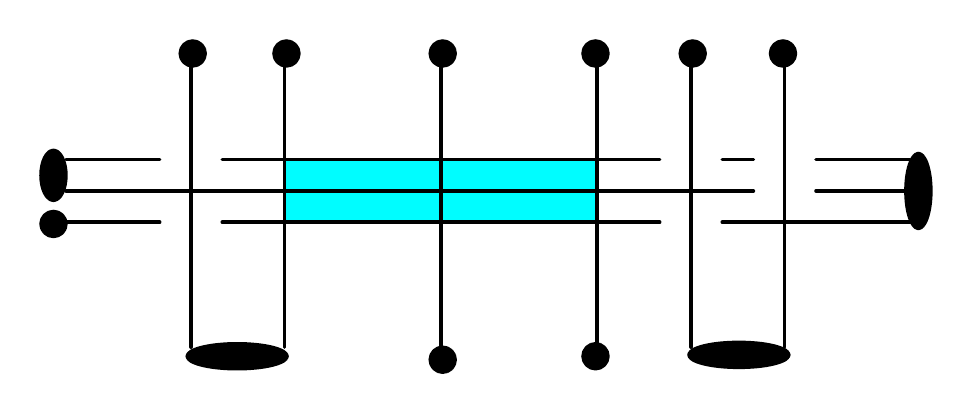} 
\caption{Another $R_{1,2n}$ web ($n=3$)}
\label{R1(2n)web-alternate}
\end{figure}

\begin{figure}[h!]
\center
\includegraphics[width=0.23\textwidth]{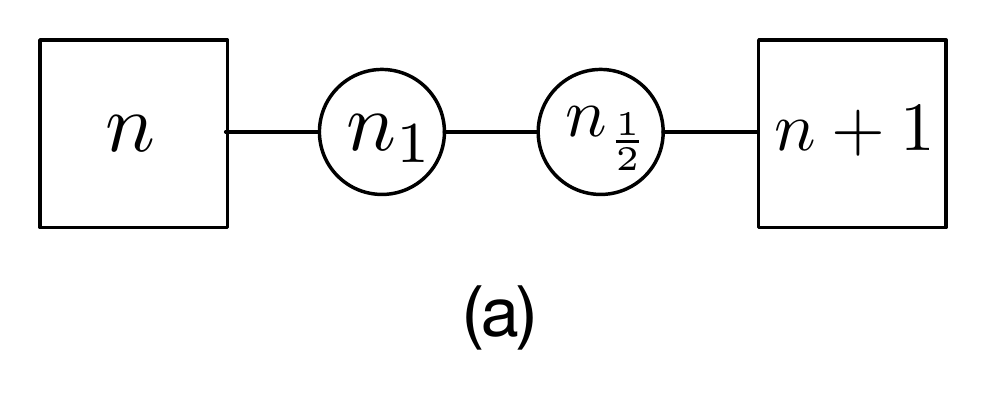} 
\hspace{0.2cm}
\includegraphics[width=0.4\textwidth]{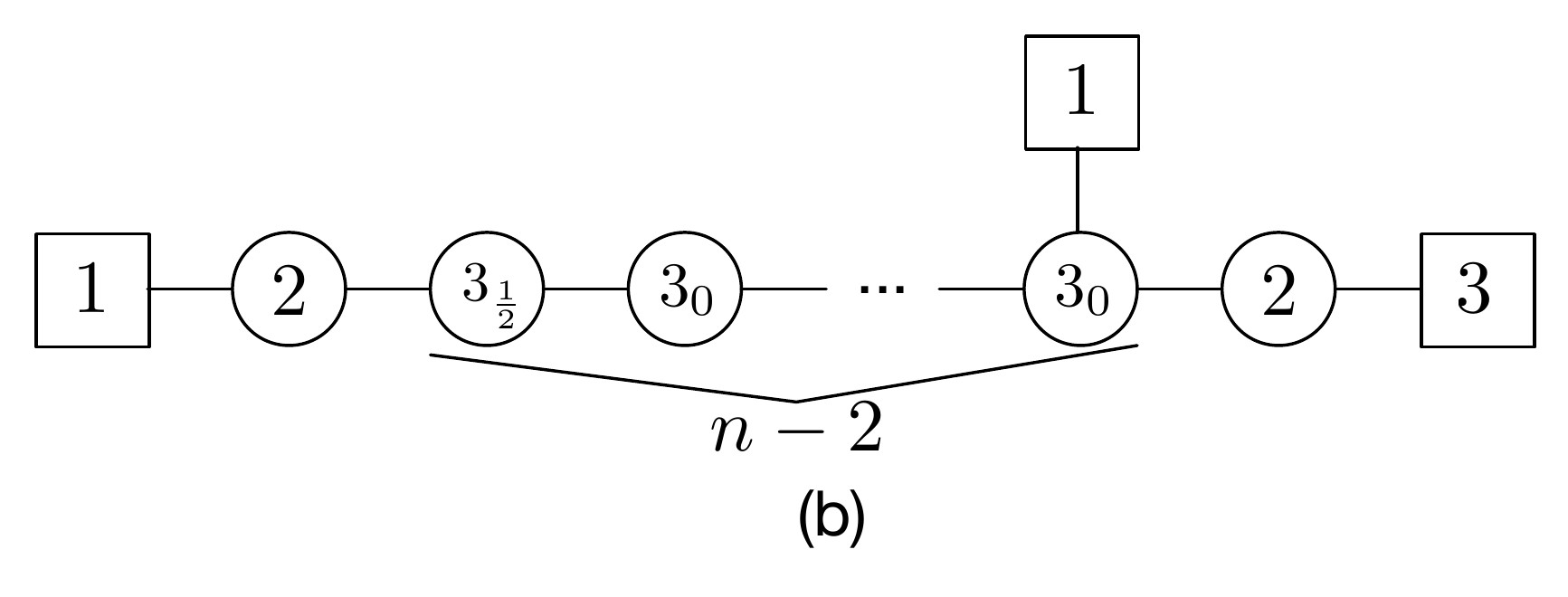} 
\hspace{0.2cm}
\includegraphics[width=0.3\textwidth]{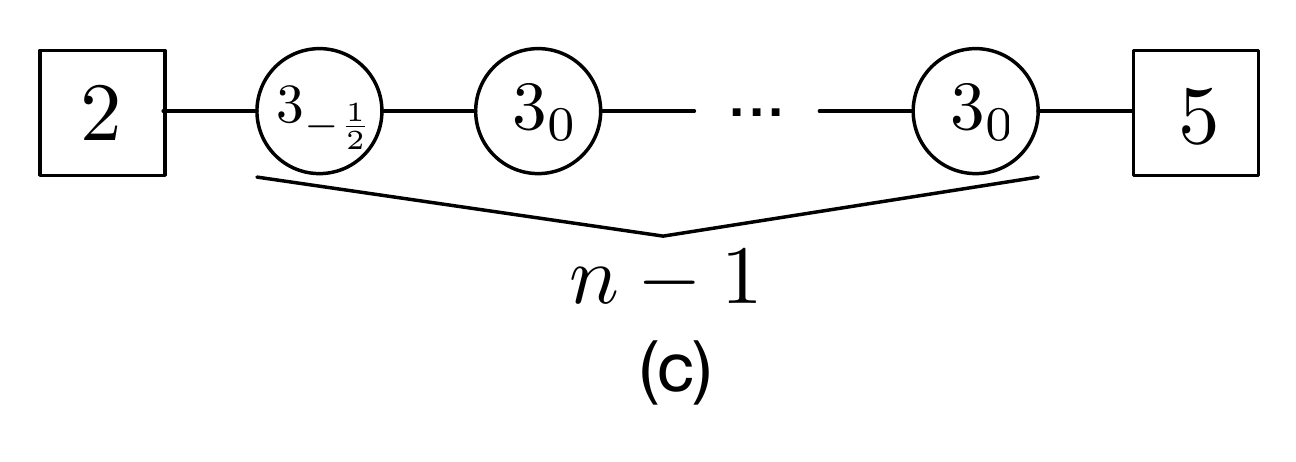} 
\caption{Quiver gauge theories for $R_{1,2n}$.} 
\label{R1(2n)quivers}
\end{figure}

The analogous 5-brane webs for $N=2n$ (with $n=3$) are shown in Fig.~\ref{R1(2n)web}.
In particular the $R_{1,6}$ SCFT admits the dual quiver gauge theory descriptions: 
$1+SU(2)\times[SU(3)_{\frac{1}{2}}+1]\times SU(2)+3$ and $2+SU(3)_{-\frac{1}{2}}\times SU(3)_0+5$.
There is actually a third gauge theory description that can be obtained by a rearrangement 
of the $(0,1)$ 7-branes at the bottom, Fig.~\ref{R1(2n)web-alternate}.
Going through some ``7-brane gymnastics" shows that this describes $3+SU(3)_1\times SU(3)_{\frac{1}{2}}+4$.
The S-dual web in this case yields the same gauge theory.
The quiver diagrams for the general cases are shown in Fig.~\ref{R1(2n)quivers}.

\begin{figure}[h]
\center
\includegraphics[width=0.2\textwidth]{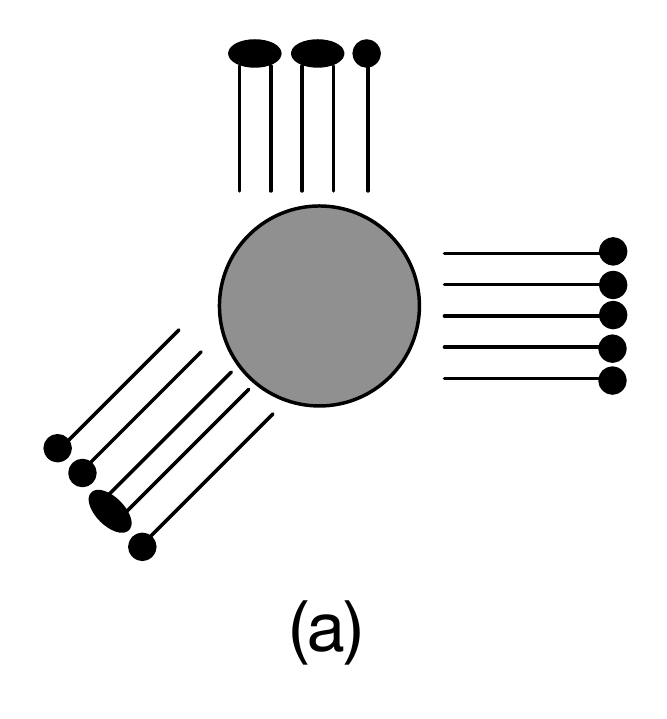} 
\hspace{1cm}
\includegraphics[width=0.23\textwidth]{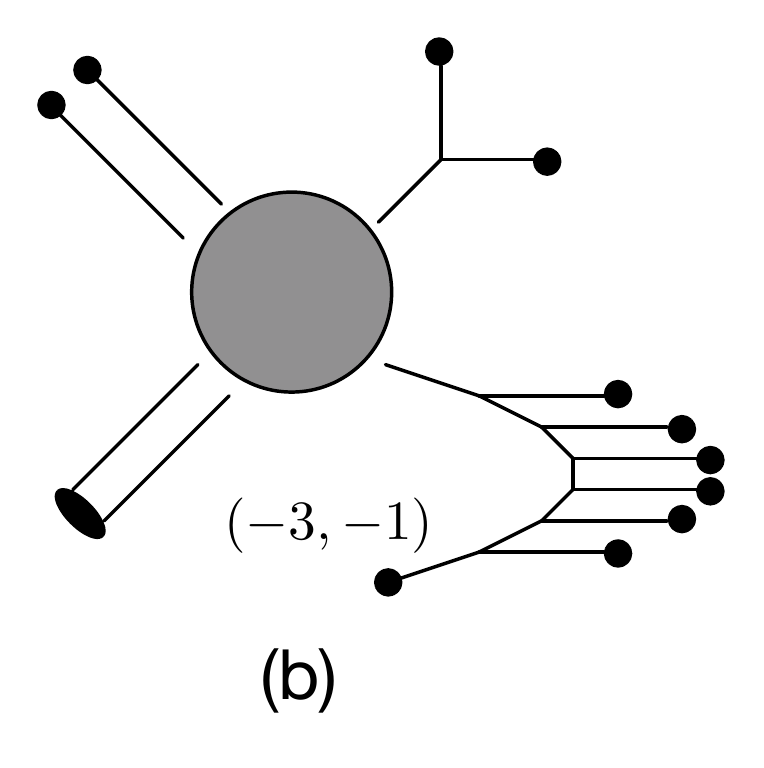} 
\hspace{1cm}
\includegraphics[width=0.33\textwidth]{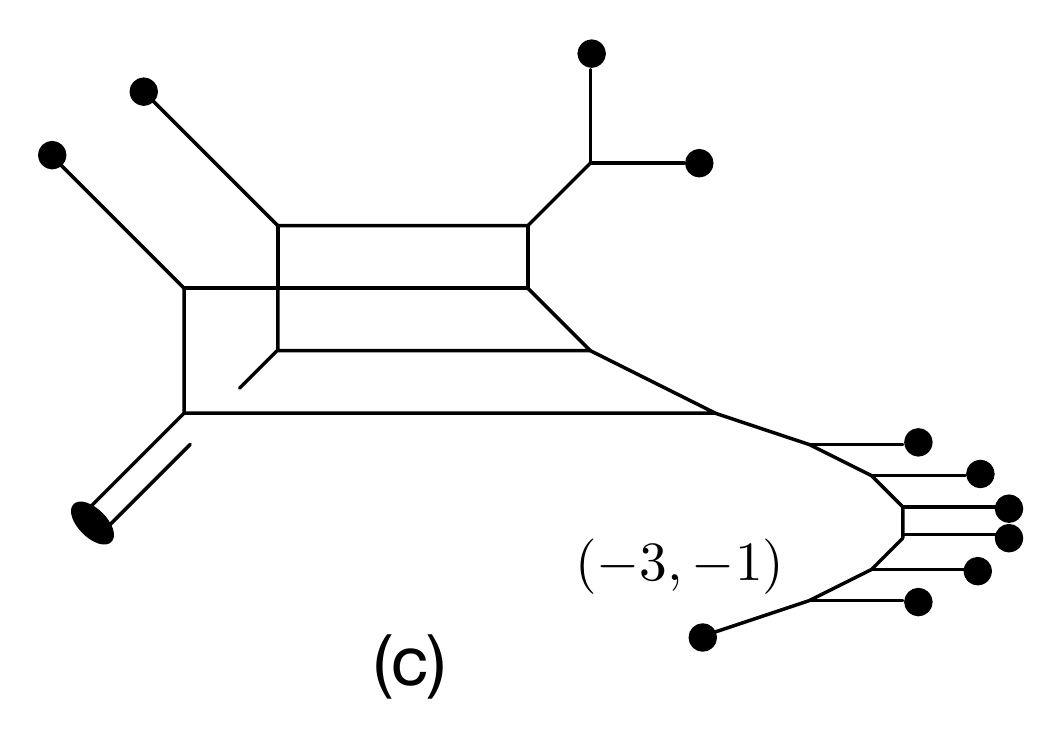} 
\caption{Another gauge theory for $R_{1,5}$: $SU(4)_{-\frac{1}{2}} + \asymm + 7\, \funda$.}
\label{R1(2n+1)junction2}
\end{figure}

\begin{figure}[h]
\center
\includegraphics[width=0.4\textwidth]{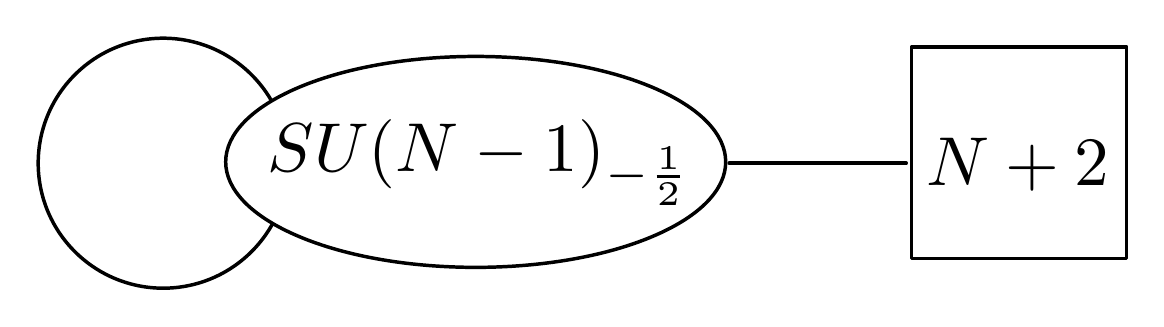} 
\caption{A single gauge group gauge theory for $R_{1,N}$.}
\label{R1Nquiver}
\end{figure}

One may wonder whether there exists a gauge theory for $R_{1,N}$ with a just single rank $N-2$ gauge group,
as for $R_{0,N}$. There does, but it takes a bit more work to see.
Consider $R_{1,5}$.
Using $SL(2,\mathbb{Z})$, the original 5-brane junction for $R_{1,5}$ can be transformed to that shown in 
Fig.~\ref{R1(2n+1)junction2}a.
Through a series of HW transitions this is transformed to the web of Fig.~\ref{R1(2n+1)junction2}b (see Appendix~\ref{sec:antisymmetric}),
which is in turn deformed into the web of Fig.~\ref{R1(2n+1)junction2}c.
The latter describes a low energy gauge theory with gauge group $SU(4)$, bare CS level $\kappa_0 = -\frac{1}{2}$,
seven fundamental hypermultiplets, and one hypermultiplet in the rank 2 antisymmetric representation (see Appendix~\ref{sec:antisymmetric} for details).
More generally, we find that $R_{1,N}$ possesses a gauge theory description as 
$SU(N-1)_{-\frac{1}{2}} + \asymm + (N+2)\, \funda$ (Fig.~\ref{R1Nquiver}).
As in the $R_{0,N}$ theory, this description allows one to guess the full global symmetry of the $R_{1,N}$ theory, by comparing
its classical global symmetry to that of the original 5-brane junction.
The former is $SU(N+2)_F\times U(1)_B \times U(1)_A \times U(1)_I$, and the latter is 
$SU(N)\times SU(2)^2\times U(1)^3$.
The full global symmetry should therefore be at least $SU(N+2)\times SU(2)\times U(1)^2$, which is exactly what it is
(except for $N=5$).

We can again use the gauge theory descriptions to compute the superconformal index, and verify the global symmetry.
We will do this for $R_{1,5}$, which should exhibit a further enhancement to $SU(7)\times SU(3)\times U(1)$.
Let us start with the $SU(4)_{\frac{1}{2}}+ \asymm + 7$ description.
The classical global symmetry of this gauge theory is $SU(7)_F\times SU(2)_A\times U(1)_B\times U(1)_I$
($U(1)_A$ is perturbatively enhanced to $SU(2)_A$ since the $\asymm = {\bf 6}$ of $SU(4)$ is real).
The perturbative contribution to the index is given by
\bea
I_{pert}^{SU(4) + \asymm + 7} & = & 1 + x^2 (2 + \chi_{({\bf 1},\bold{3})} + \chi_{(\bold{48},{\bf 1})})  \\
&+& x^3 \left((y + \frac{1}{y})(3 + \chi_{({\bf 1}, \bold{3})} + \chi_{(\bold{48},{\bf 1})})  + b'^2 \chi_{(\bold{21},{\bf 2})} 
+ \frac{1}{b'^2} \chi_{(\bar{\bold{21}},{\bf 2})} \right) + {\cal O}(x^4) \,. \nonumber
\eea 

To order $x^3$ there is also a contribution from one-instanton states.
As in other cases, the instanton partition function computed via $U(4)$ does not respect
the $x\rightarrow 1/x$ symmetry or the full classical global symmetry (the $SU(2)_A$ part in this case),
suggesting that there are ``$U(1)$ remnant" states that must be removed.
The nature of these states is not obvious from the web in Fig.~\ref{R1(2n+1)junction2},
but we can figure out the necessary correction factor by requiring $x\rightarrow 1/x$ and $SU(2)_{A}$ invariance.\footnote{This can be
derived from an alternative 5-brane web with O7-planes \cite{BZ}.}
The appropriate correction is given by
\be
\mathcal{Z}_{c} = PEֿ\left[\frac{x^2 q' z b'^{\frac{7}{2}}}{(1-x y)(1-\frac{x}{y})}\right] \mathcal{Z} \,,
\ee 
and the resulting 1-instanton contribution is
\bea
I_{1}^{SU(4)+ \asymm + 7} & = & x^2 \left(q' b'^{\frac{7}{2}} + \frac{1}{q' b'^{\frac{7}{2}}} \right)\chi_{({\bf 1},\bold{2})} 
+ x^3 \Bigg( (y+\frac{1}{y}) \left(q' b'^{\frac{7}{2}} + \frac{1}{q' b'^{\frac{7}{2}}}\right)\chi_{({\bf 1},\bold{2})}  \nonumber \\ 
 & + &  \frac{q'}{b'^{\frac{5}{2}}}\chi_{(\bold{7},\bold{1})} + \frac{b'^{\frac{5}{2}}}{q'}\chi_{(\bar{\bold{7}},\bold{1})} 
 + \frac{1}{q' b'^{\frac{3}{2}}}\chi_{(\bold{21},\bold{1})} + q' b'^{\frac{3}{2}}\chi_{(\bar{\bold{21}},{\bf 1})} \Bigg) + {\cal O}(x^4) .
\eea 
The $x^2$ terms provide the additional currents to enhance $SU(2)_A \times U(1)_B \times U(1)_I \rightarrow SU(3)\times U(1)$,
where the $SU(3)$ is spanned by $\chi^{SU(3)}_{\bf 3} = q'^{-\frac{1}{3}} b'^{-\frac{7}{6}} \chi^{SU(2)}_{\bold{2}} + q'^{\frac{2}{3}} b'^{\frac{7}{3}}$. 

As a further check, let us also compute the index in one of the other gauge theory descriptions, specifically
in the $3+SU(3)\times SU(2)+3$ theory (Fig.~\ref{R1(2n+1)quivers}a).
The classical global symmetry is $SU(3)_F \times SO(6)_F \times U(1)_B \times U(1)_I^2$, and 
the perturbative contribution is now
\bea
I_{pert}^{3+SU(3)\times SU(2)+3}  =  1 + x^2 \left(4 + \chi_{(\bold{8},\bold{1})} + \chi_{(\bold{1},\bold{15})}\right) 
+ x^3 \Bigg((y+\frac{1}{y})\left(5 + \chi_{(\bold{8},\bold{1})} + \chi_{(\bold{1},\bold{15})}\right)  \nonumber \\  
  \mbox{} +  \frac{b}{z} \chi_{(\bold{3},\bold{6})} + \frac{z}{b} \chi_{(\bar{\bold{3}},\bold{6})} + z^2 b\chi_{(\bold{3},\bold{1})} 
+ \frac{1}{b z^2}\chi_{(\bar{\bold{3}},\bold{1})} + b^3 + \frac{1}{b^3}\Bigg) +  {\cal O}(x^4) .
\eea  

For the instanton calculation we again treat $SU(2)$ as $USp(2)$.
An $x\rightarrow 1/x$ invariant partition function is obtained by 
\be
\mathcal{Z}_{c} = PE\left[\frac{x^2(q_1 z b^{\frac{3}{2}} + q^2_2 + q_1 q^2_2  z b^{\frac{3}{2}})
}{(1-x y)(1-\frac{x}{y})}\right] \mathcal{Z} \,,
\ee 
where $q_1, q_2$ are the $SU(3)$ and $SU(2)$ instanton fugacities, respectively.
This apparently includes ``$U(1)$ remnant" states associated both with the $SU(3)$ part, visible in the web 
of Fig.~\ref{R1(2n+1)web}b in terms of a D1-brane between the external NS5-branes, as well as with the $SU(2)$ part.
The latter are not obvious in the given 5-brane web, but appear to be a necessary ingredient 
in computing the instanton partition function for $USp(2)+6$ (see \cite{Hwang:2014uwa,Zaf}).\footnote{This can also be derived
from an alternative 5-brane web with O7-planes \cite{BZ}.}
With this, we find the instanton contributions (for conciseness we show to ${\cal O}(x^2)$, but we computed to ${\cal O}(x^3)$)

\bea
I_{(1,0)+(0,1)+(1,1)+(1,2)}^{3+SU(3)\times SU(2)+3} =
x^2 \Bigg( (q_2 + \frac{1}{q_2})\left(z^{\frac{3}{2}} \chi_{(\bold{1},\bold{4})} + \frac{1}{z^{\frac{3}{2}}}\chi_{(\bold{1},\bar{\bold{4}})}\right) 
+ q_1 z b^{\frac{3}{2}} +\frac{1}{q_1 z b^{\frac{3}{2}}} \nonumber\\
\mbox{} + \frac{q_1 q_2 b^{\frac{3}{2}}}{\sqrt{z}}\chi_{(\bold{1},\bar{\bold{4}})}
 +   \frac{\sqrt{z}}{q_1 q_2 b^{\frac{3}{2}}} \chi_{(\bold{1},\bold{4})} + q^2_2 + \frac{1}{q^2_2} + q_1 q^2_2  z b^{\frac{3}{2}} 
 + \frac{1}{q_1 q^2_2  z b^{\frac{3}{2}}} \Bigg) + {\cal O}(x^3) .
\eea     
The instantons provide additional conserved currents enhancing 
$SO(6)_F\times U(1)_B \times U(1)_I^2 \rightarrow SU(7)$, with 
$\chi^{SU(7)}_{\bf 7} = \sqrt{z}\, (q_1 q_2)^{-\frac{1}{7}} b^{-\frac{3}{14}}
(\chi^{SO(6)}_{\bold{4}} + z^{-\frac{3}{2}}\, (q_2 + \frac{1}{q_2}) + q_1 q_2 b^{\frac{3}{2}} z^{-\frac{1}{2}} )$.

Thus the full quantum symmetry in both descriptions is $SU(7)\times SU(3)\times U(1)$.
Furthermore the indices agree to ${\cal O}(x^3)$, and both can be expressed using characters of the full symmetry as
\bea
I^{R_{1,5}} & = & 1 + x^2 \left(1 + \chi^0_{({\bf 1},\bold{8})} + \chi^0_{(\bold{48},{\bf 1})}\right) \\ \nonumber 
& + & x^3 \left( (y+\frac{1}{y})(2 + \chi_{(\bold{1},\bold{8})} + \chi_{(\bold{48},\bold{1})}) + \chi^{-3}_{(\bold{7},\bold{1})} 
+ \chi^3_{(\bar{\bold{7}},{\bf 1})}  + \chi^1_{(\bold{21},\bar{\bold{3}})} + \chi^{-1}_{(\bar{\bold{21}},\bold{3})} \right) + {\cal O}(x^4).
\eea 
The superscript refers to the $U(1)$ charge, given by
$\frac{5}{6}B' - \frac{1}{3}I$ in the $SU_{\frac{1}{2}}(4)+ \asymm +7$ theory, and by
$\frac{2}{7}(I_1 + I_2 - 2B)$ in the $3+SU(3)\times SU(2)+3$ theory.

\subsubsection{$R_{2,2n+1}$}

\begin{figure}[h]
\center
\includegraphics[width=0.33\textwidth]{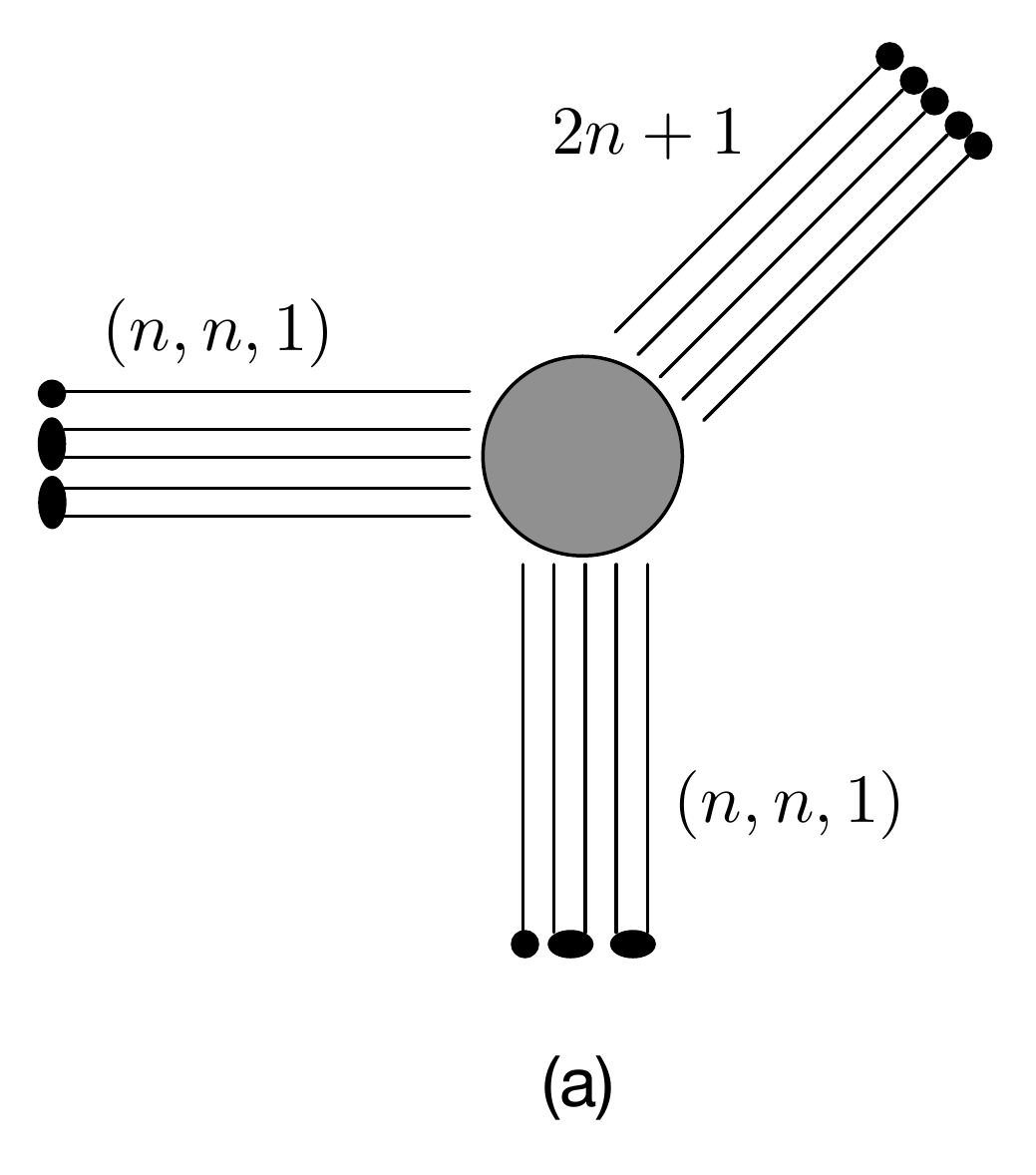} 
\hspace{0.5cm}
\includegraphics[width=0.3\textwidth]{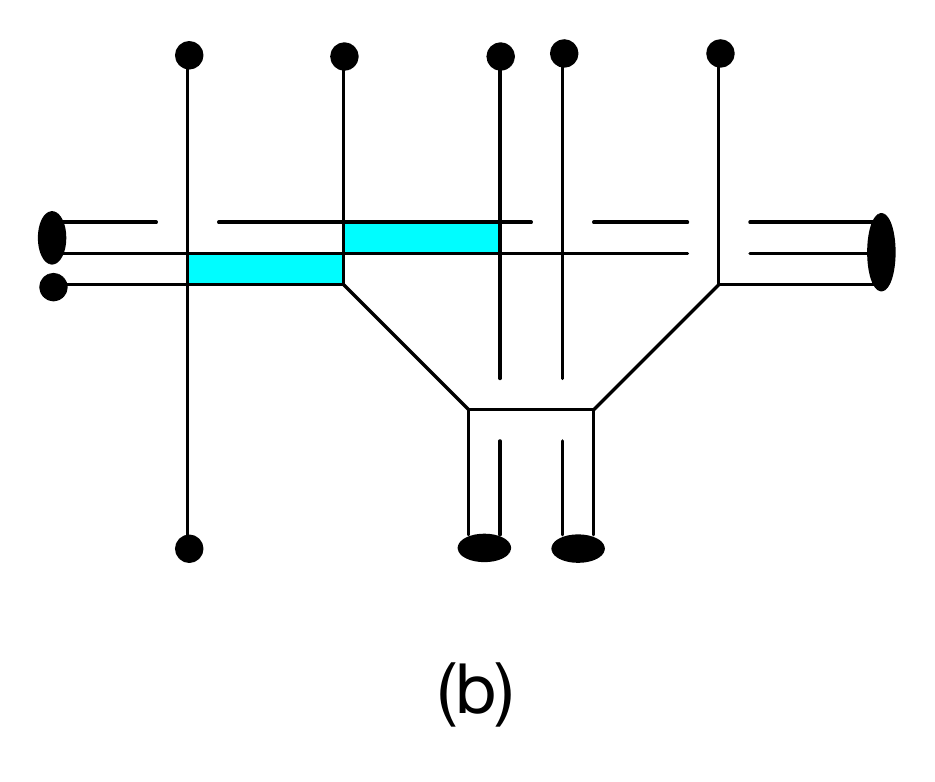} 
\hspace{0.5cm}
\includegraphics[width=0.23\textwidth]{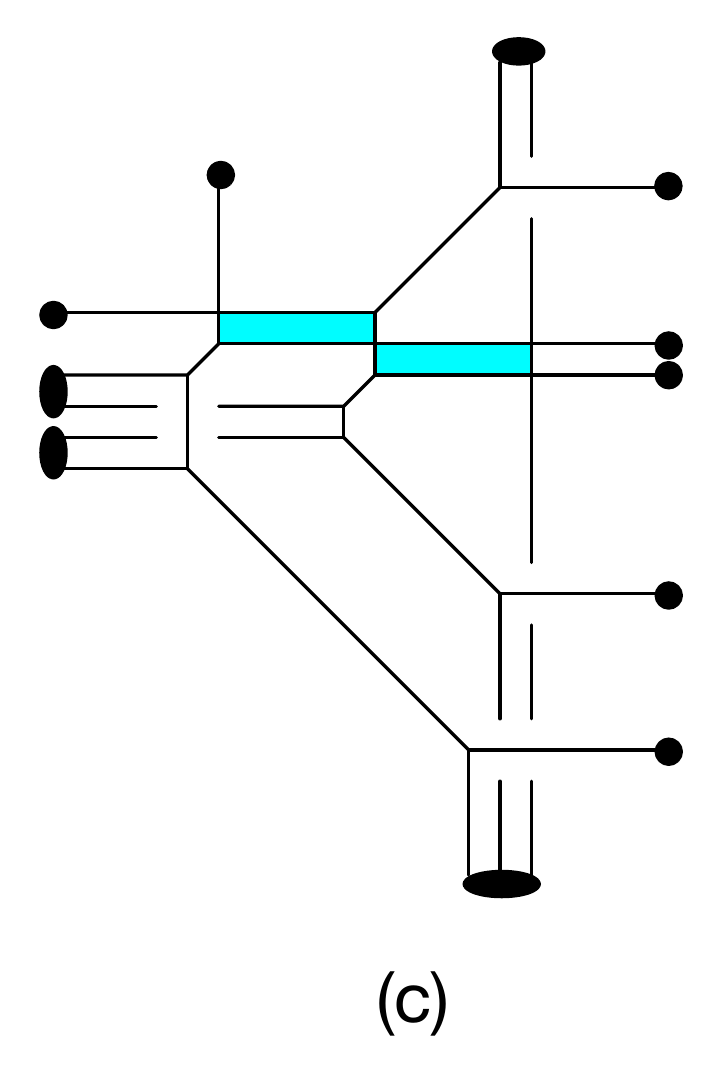} 
\caption{$R_{2,2n+1}$ webs ($n=2$)}
\label{R2Njunction}
\end{figure}

\begin{figure}[h]
\center
\includegraphics[width=0.35\textwidth]{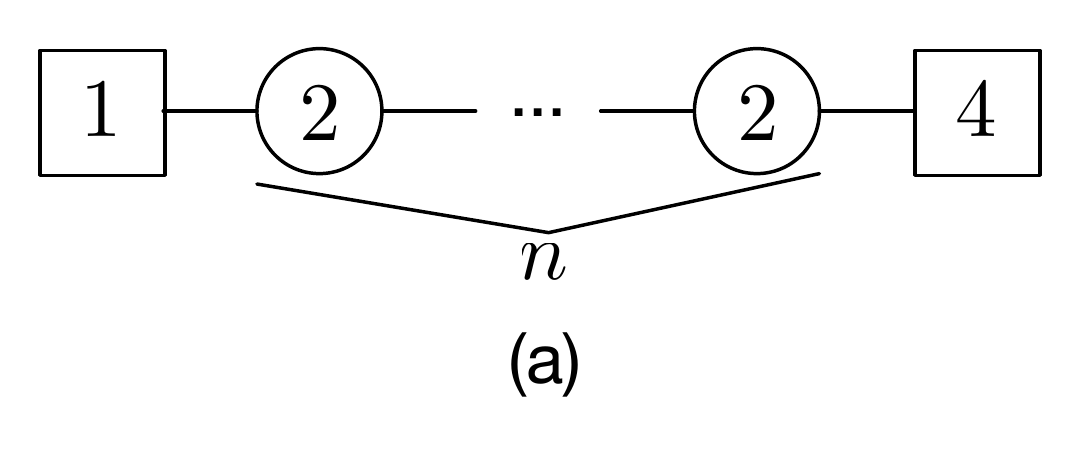} 
\hspace{1cm}
\includegraphics[width=0.25\textwidth]{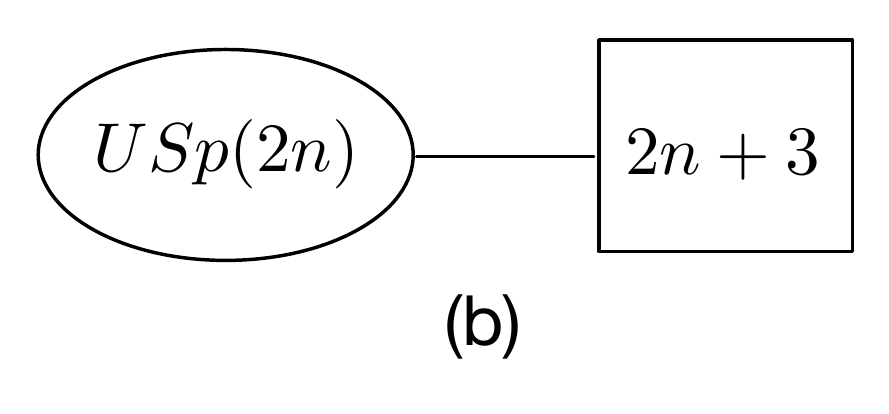}
\caption{Gauge theories for $R_{2,2n+1}$.} 
\label{R2(2n+1)quivers}
\end{figure}

The $R_{2,2n+1}$ theory is defined by replacing two of the maximal punctures by $(n,n,1)$.
This theory has an $n$-dimensional Coulomb branch, a $(2n^2 + 5n + 4)$-dimensional Higgs branch,
and a global symmetry $SO(4n+6)\times U(1)$.
The corresponding 5-brane junction is shown for $R_{2,5}$ in Fig.~\ref{R2Njunction}a.
Note that this theory corresponds to a limit on the Higgs branch of the $R_{1,5}$ (and more generally $R_{1,2n+1}$) theory.
By appropriately modifying the mass-deformed $R_{1,5}$ webs we then get the $R_{2,5}$ webs 
shown in Fig.~\ref{R2Njunction}b,c.
In either case, the quiver gauge theory is $1+SU(2)\times SU(2)+4$,
and in the general case $1+SU(2)\times SU(2)^{n-3}_0 \times SU(2)+4$ (Fig.~\ref{R2(2n+1)quivers}a).

The single gauge group dual gauge theory can likewise be obtained from that of the $R_{1,5}$ theory.
In this case we notice that the Higgsing leading to $R_{2,5}$ corresponds simply to a VEV for the $SU(4)$ antisymmetric field.
The gauge theory for $R_{2,5}$ is therefore $USp(4) + 7$.
More generally for $R_{2,2n+1}$ this is $USp(2n)+2n+3$ (Fig.~\ref{R2(2n+1)quivers}b).
The global symmetry in this case is $SO(4n+6)_F\times U(1)_I$, which is the full symmetry of the fixed point theory.

In the quiver theory description the global symmetry should be enhanced by instantons.
Let us demonstrate this explicitly with the superconformal index of the gauge theory.
The perturbative contribution, expressed in terms of $SU(2)_{BF}\times SO(8)_F$ characters and the $U(1)_F$
fugacity $f$, is
\bea
I_{pert}^{1+SU(2)\times SU(2)+4} & = & 1 + x^2 \left(3+\chi_{(\bold{3},\bold{1})} + \chi_{(\bold{1},\bold{28})}\right)  \\ 
& + & x^3 \left( (y+\frac{1}{y}) \left(4+\chi_{(\bold{3},\bold{1})} + \chi_{(\bold{1},\bold{28})}\right) 
+ (f + \frac{1}{f})\chi_{(\bold{2},\bold{8}_v)} \right) + {\cal O}(x^4) \,. \nonumber 
\eea

The correction factor for the instanton partition function is quite similar to the one in the previous section
for the $3+SU(3)\times SU(2)+3$ theory, and is given by
\be
\mathcal{Z}_{c} = PE\left[\frac{x^2 (q_1 z \sqrt{f} + q^2_2 + q_1 q^2_2 z \sqrt{f})}{(1-x y)(1-\frac{x}{y})} \right] \mathcal{Z} \,.
\ee 
The resulting instanton contribution to the index is then (shown just to ${\cal O}(x^2)$)
\bea
I_{inst}^{1+SU(2)\times SU(2)+4} & = & x^2 \Bigg( (q_2+\frac{1}{q_2})\chi_{(\bold{2},\bold{8}_s)} 
+ (q_1 \sqrt{f} + \frac{1}{q_1 \sqrt{f}})\chi_{(\bold{2},\bold{1})} + q^2_2 + \frac{1}{q^2_2} \\
&+& \left(q_1 q_2\sqrt{f} + \frac{1}{q_1 q_2\sqrt{f}}\right)\chi_{(\bold{1},\bold{8}_s)} 
 + (q_1 q^2_2\sqrt{f} + \frac{1}{q_1 q^2_2\sqrt{f}})\chi_{(\bold{2},\bold{1})}\Bigg)  + {\cal O}(x^3) . \nonumber
\eea
The $x^2$ terms provide the necessary conserved currents to enhance $SU(2)_{BF}\times SO(8)\times U(1)_B\times U(1)_I^2 \rightarrow SO(14)\times U(1)$,
where $\chi^{SO(14)}_{\bf 14} = \chi_{(\bold{1},\bold{8}_s)} + (q_2+ q_2^{-1})\chi_{(\bold{2},\bold{1})} + q_1 q_2\sqrt{f} + (q_1 q_2\sqrt{f})^{-1}$. 
The full index can then be expressed in terms of $SO(14)$ characters as:
\bea
I^{R_{2,5}} & = & 1 + x^2 (1+ \chi_{\bold{91}}) + x^3 \left( (y+\frac{1}{y}) (2+ \chi_{\bold{91}}) + q'\chi_{\bold{64}} + \frac{1}{q'} \chi_{\bar{\bold{64}}} \right) + O(x^4) \label{eq:inR2}
\eea
where $q'=f^{\frac{3}{4}}/\sqrt{q_1 q_2}$.
This also agrees with the index of the $USp(4)+7$ theory, where we identify $q'$ as the $U(1)_I$ fugacity.

\subsubsection{$S_N$ (and $E_7$)}

The $S_N$ theory is defined by replacing two of the maximal punctures by 
$(n,n)$ and $(n-1,n-2,1^3)$ for $N=2n$, and by $(n+1,n)$ and $((n-1)^2,1^3)$ for $N=2n+1$.
This theory has an $(N-3)$-dimensional Coulomb branch, an $(N^2 + 3N +12)/2$-dimensional Higgs branch,
and a global symmetry $SU(N+2)\times SU(3)\times U(1)$.
The $N=4$ case is special and the global symmetry is enhanced to $E_7$.\footnote{Strictly speaking,
the $S_N$ theories are defined for $N\geq 5$.
For $N=4$ one of the punctures degenerates to $(1,0,1^3)$.
For all intensive purposes this gives the $E_7$ theory, but the formula for the dimension
of the Higgs branch is wrong. The $E_7$ theory has a 17-dimensional Higgs branch.}
The cases $N=5$ and $N=6$ also have a further enhancement of symmetry to $SU(10)$ and $SU(4)\times SU(8)$ respectively. 

\begin{figure}[h]
\center
\includegraphics[width=0.35\textwidth]{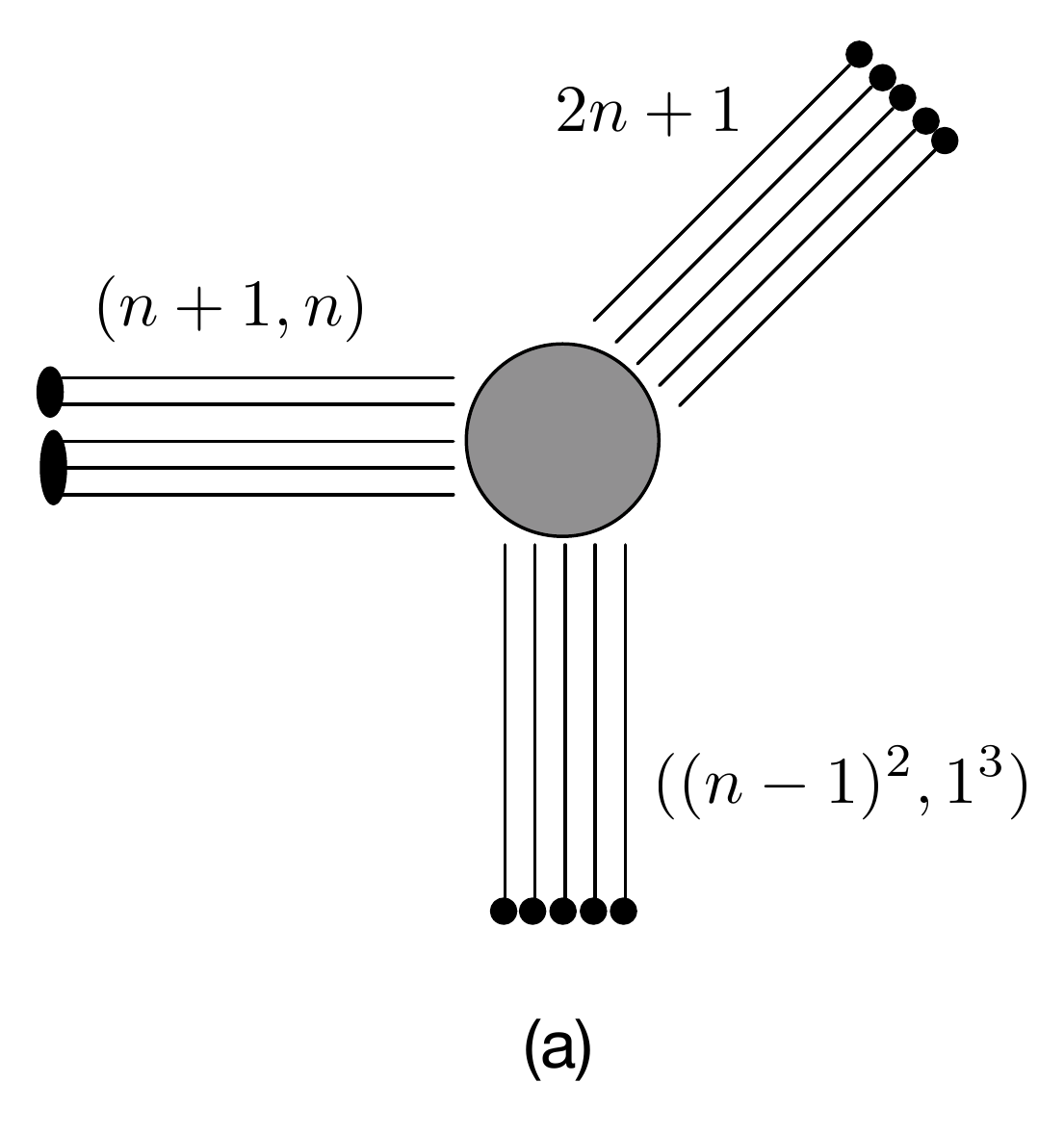} 
\hspace{0.5cm}
\includegraphics[width=0.3\textwidth]{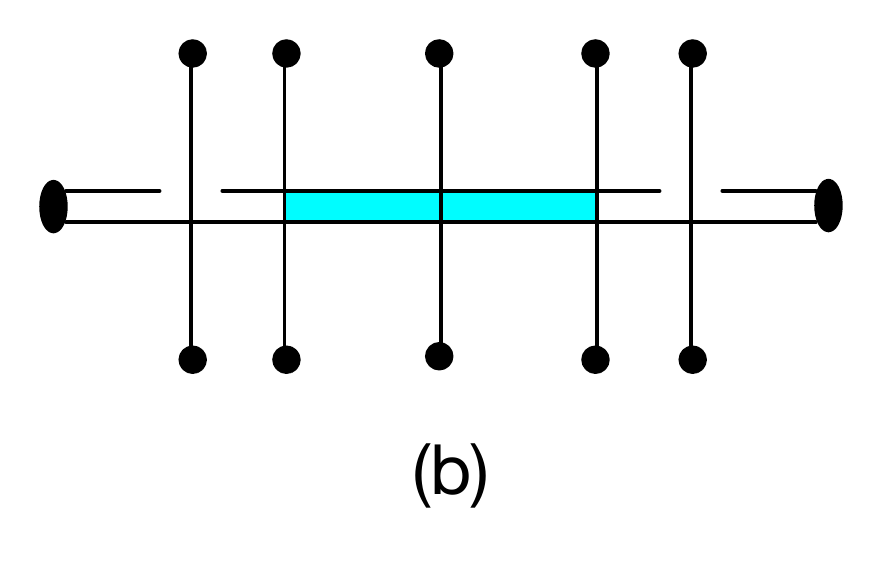} 
\hspace{0.5cm}
\includegraphics[width=0.2\textwidth]{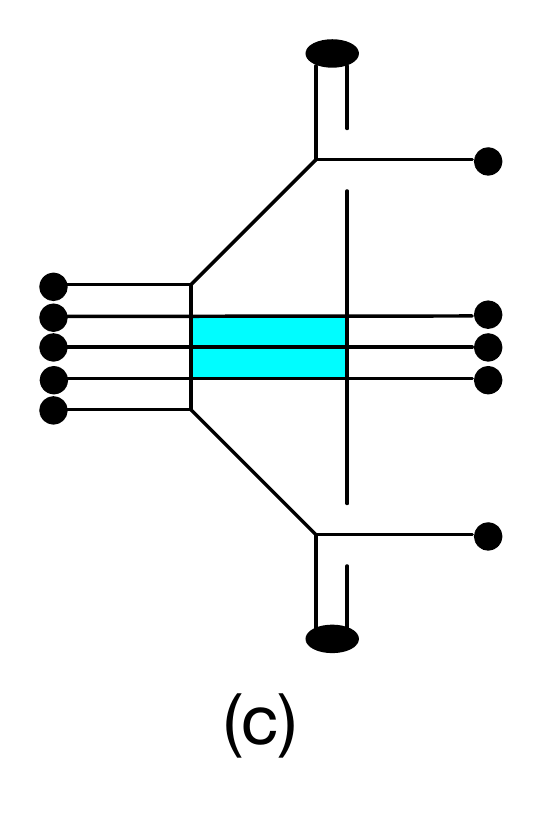}
\caption{$S_5$ webs}
\label{S5web}
\end{figure}

\begin{figure}[h]
\center
\includegraphics[width=0.25\textwidth]{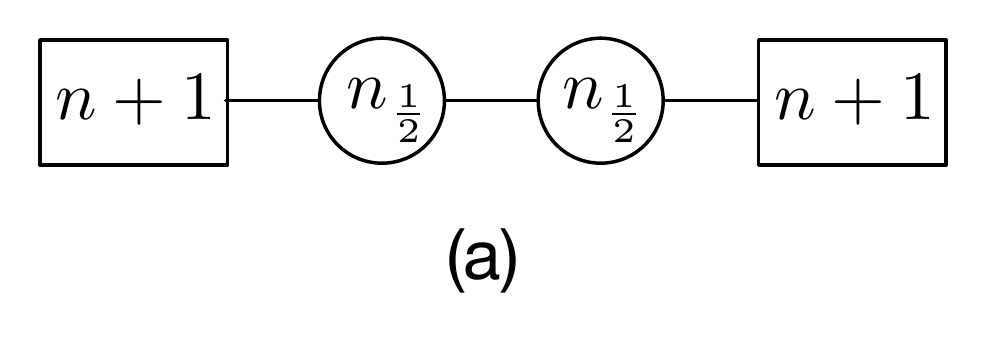} 
\hspace{0.5cm}
\includegraphics[width=0.25\textwidth]{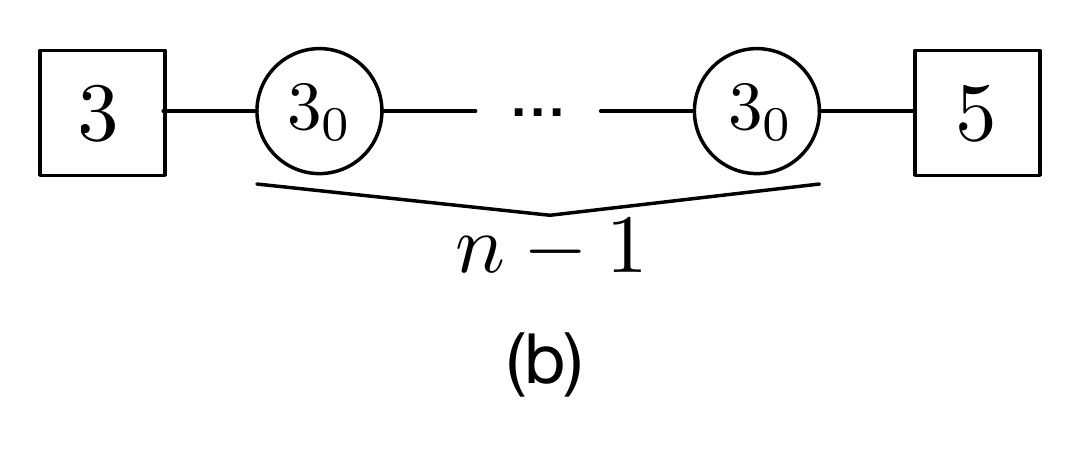}
\hspace{0.5cm}
\includegraphics[width=0.35\textwidth]{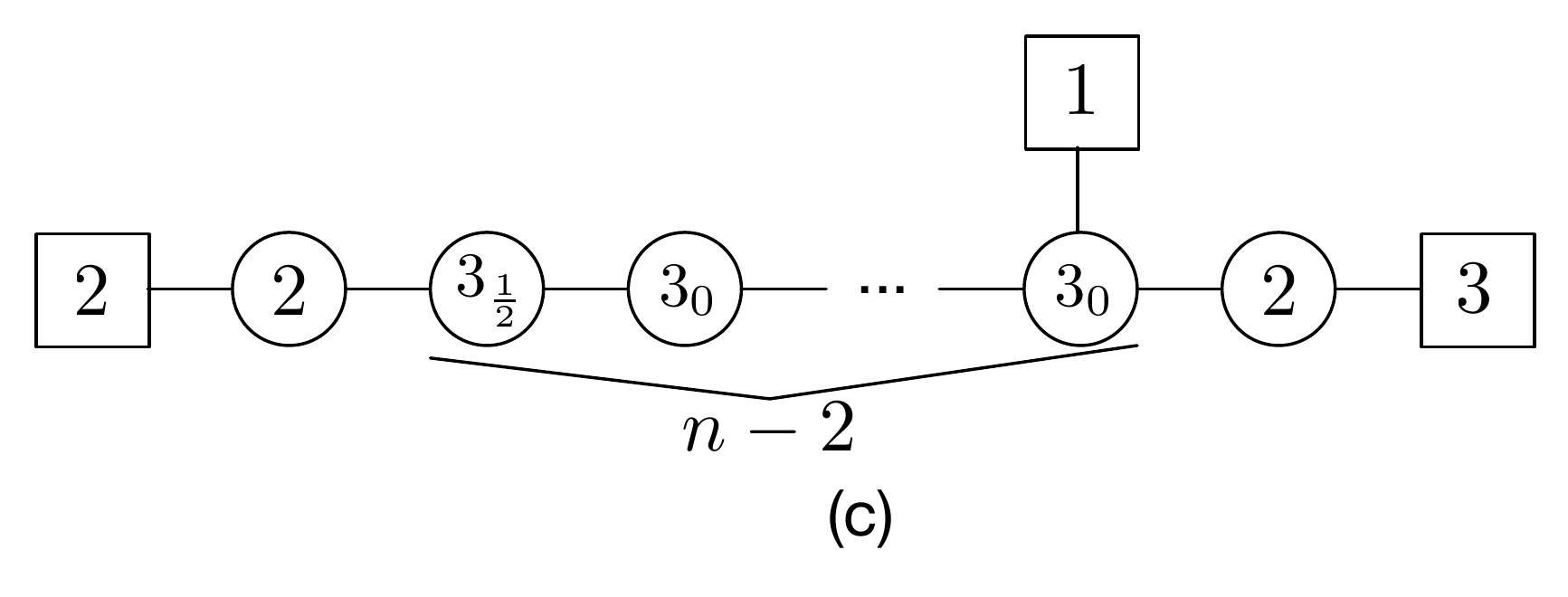}
\caption{Gauge theories for $S_{2n+1}$.} 
\label{S(2n+1)quivers}
\end{figure}

The 5-brane junction for $S_{2n+1}$ (represented by $S_5$) is shown in Fig.~\ref{S5web}a.
Moving the D7-brane with the three D5-branes attached to the right and mass-deforming leads to the web shown
in Fig.~\ref{S5web}b, which corresponds to $3+SU(2)\times SU(2)+3$.
The S-dual web, shown in Fig.~\ref{S5web}c, gives $SU(3)_0+8$.
For $N=2n+1$, the former generalizes to $n+1+SU(n)_{\frac{1}{2}}\times SU(n)_{\frac{1}{2}}+n+1$, and the latter to $3+SU(3)_0^{n-1}+5$.
For $n\geq 3$ we find (at least) one more gauge theory description by rearranging the $(0,1)$ 7-branes
at the bottom, which is $2+SU(2)\times[SU(3)_0^{n-2}+1]\times SU(2)+3$. 
The three quiver diagrams for the general case $S_{2n+1}$ are shown in Fig.~\ref{S(2n+1)quivers}.

\begin{figure}[h!]
\center
\includegraphics[width=0.3\textwidth]{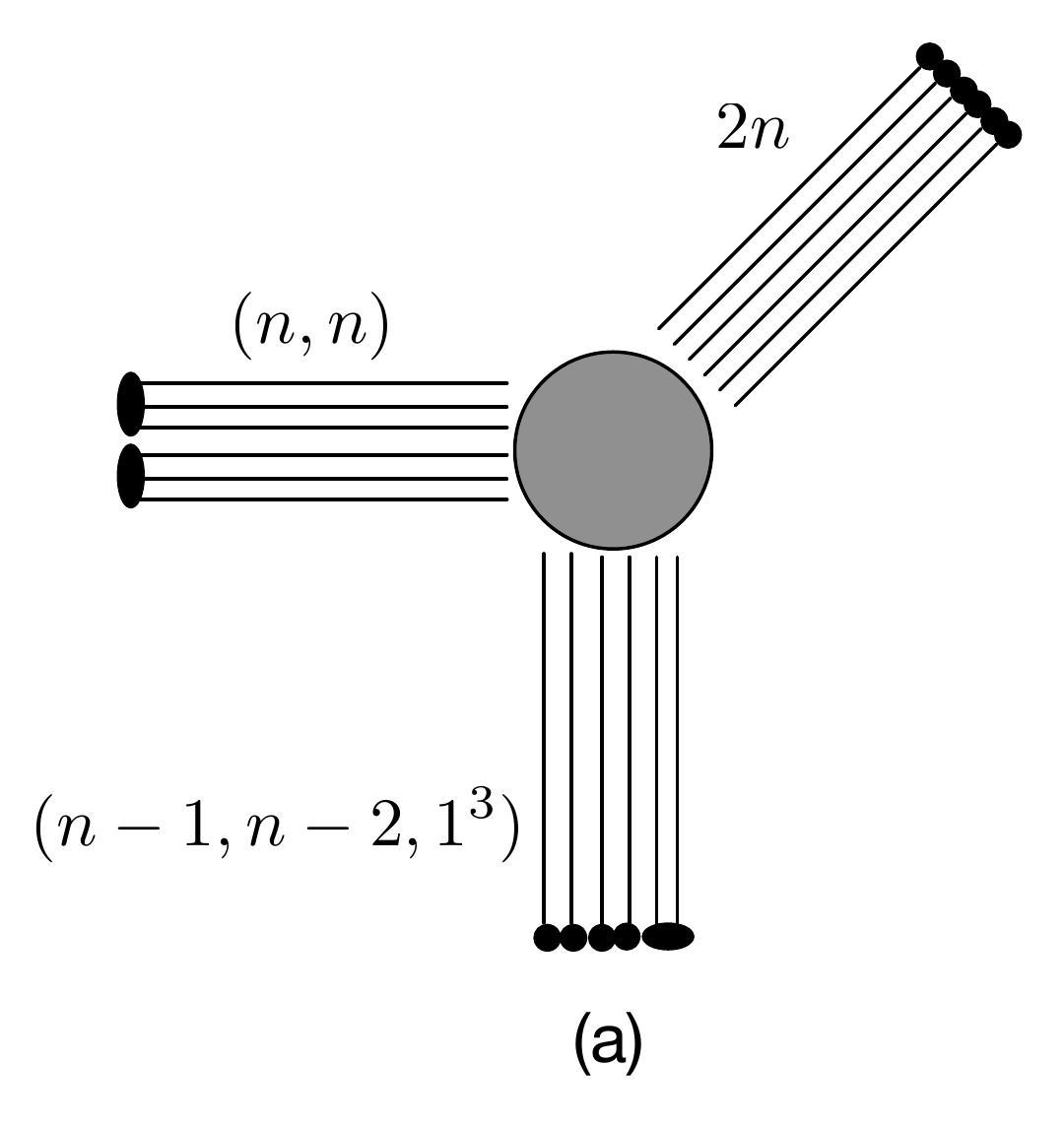} 
\hspace{0.5cm}
\includegraphics[width=0.27\textwidth]{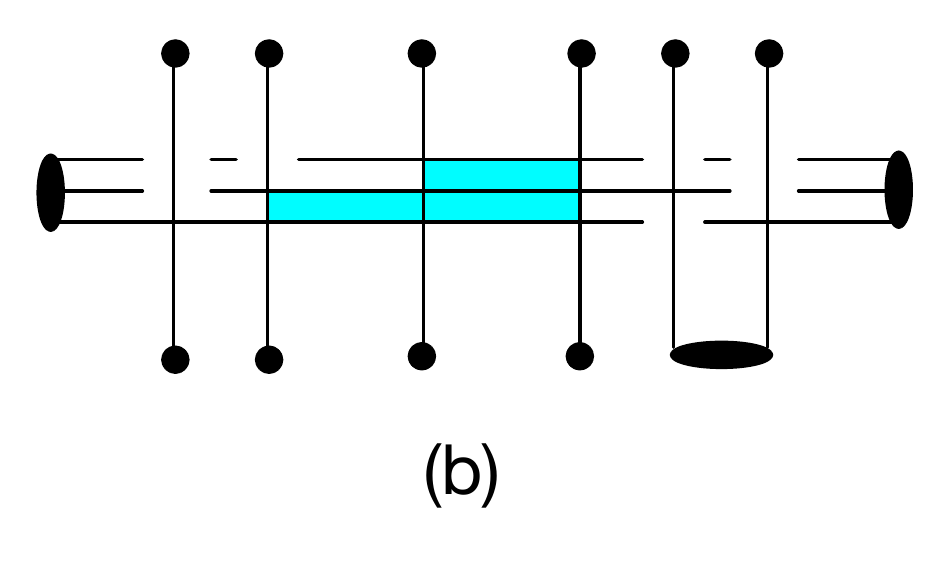} 
\hspace{0.5cm}
\includegraphics[width=0.2\textwidth]{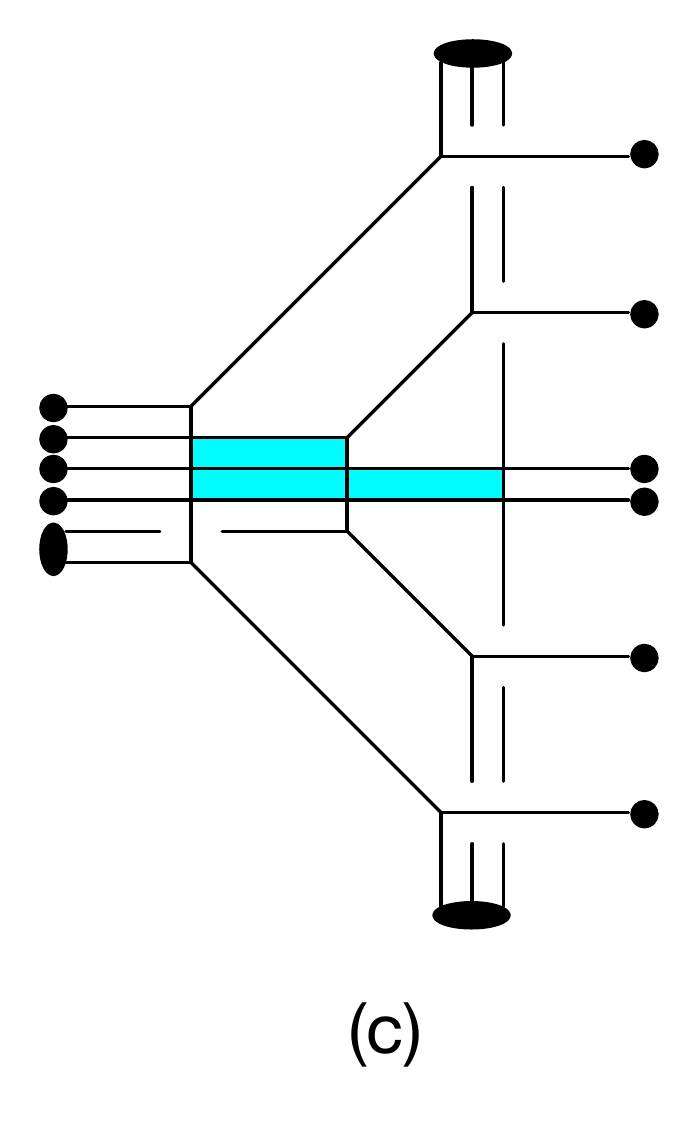}
\caption{$S_6$ webs}
\label{S6web}
\end{figure}

\begin{figure}[h!]
\center
\includegraphics[width=0.4\textwidth]{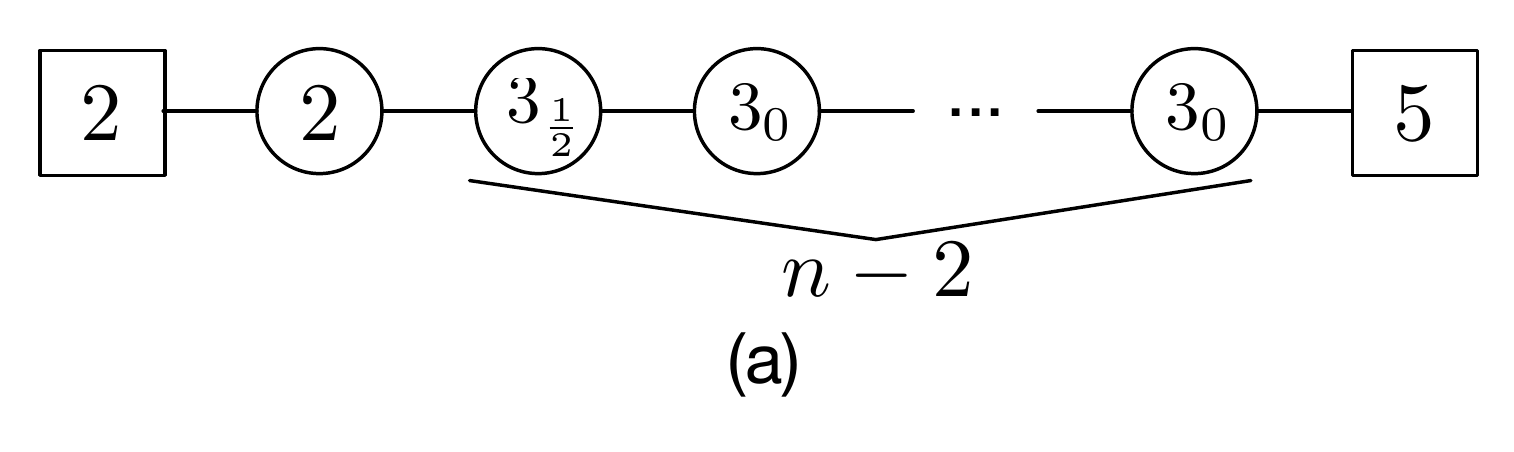} 
\hspace{0.5cm}
\includegraphics[width=0.4\textwidth]{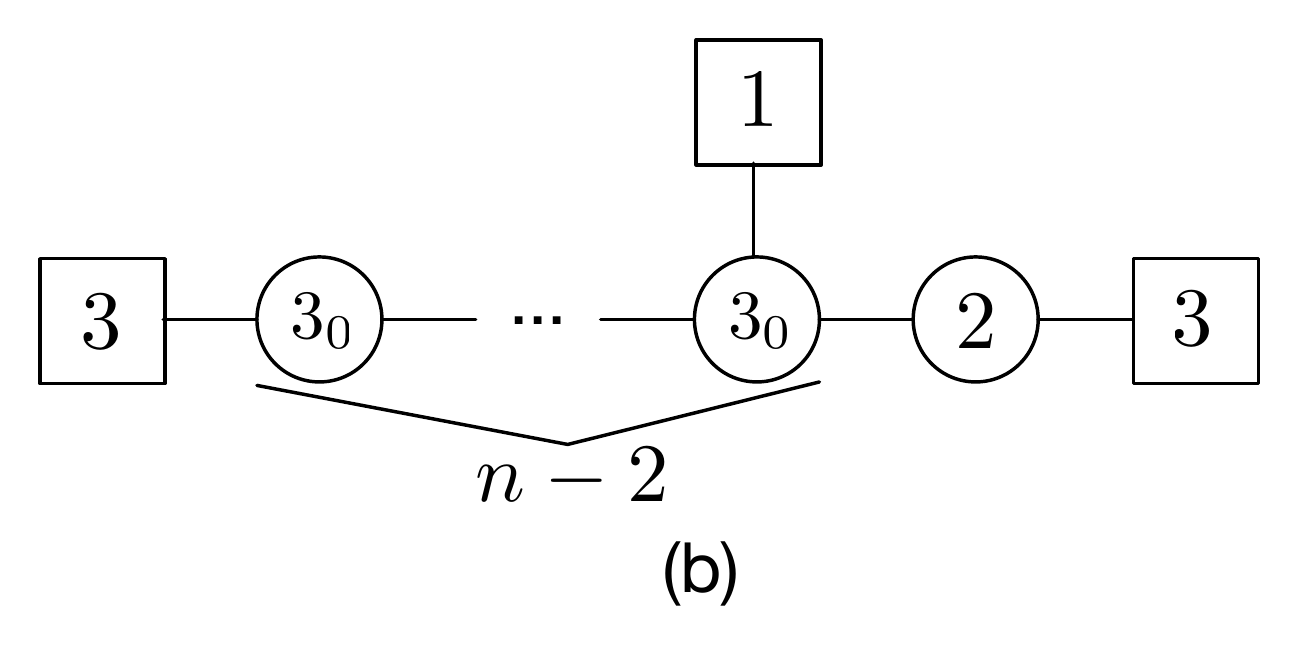} 
\caption{Gauge theories for $S_{2n}$}
\label{S(2n)quivers}
\end{figure}

The $S_{2n}$ series (for $n\geq 3$)  is represented by $S_6$ in Fig.~\ref{S6web}.
The web in Fig.~\ref{S6web}b corresponds to $2+SU(2)\times SU(3)_{\frac{1}{2}}+5$,
and its S-dual in Fig.~\ref{S6web}c to $4+SU(3)_0\times SU(2)+3$.
The generalizations to $S_{2n}$ are shown in Fig.~\ref{S(2n)quivers}.

\begin{figure}[h!]
\center
 \includegraphics[width=0.3\textwidth]{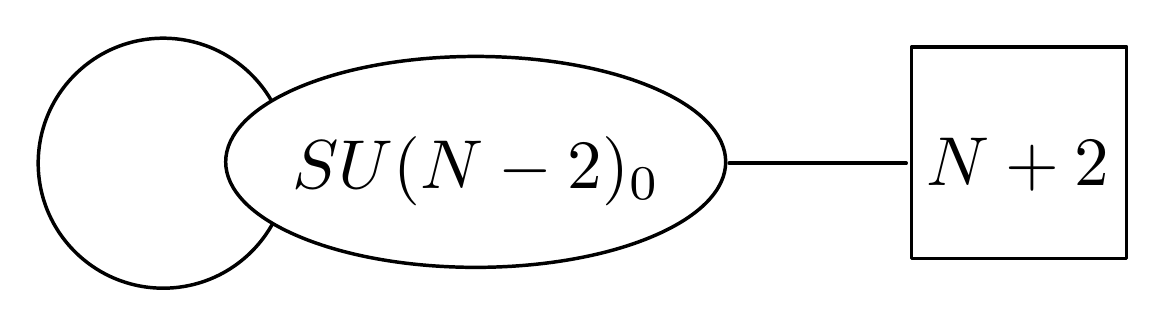}
\caption{Single gauge group gauge theory for $S_N$: $SU(N-2)_0 + \asymm + (N+2)\,\funda$.}
\label{SNquiver}
\end{figure}

As in previous cases, we also find a single-gauge-group theory for $S_N$, Fig.~\ref{SNquiver}.
This is valid for $N\geq 4$, 
and reproduces the known $SU(2)+6$ gauge theory description of the $E_7$ theory for $N=4$.

\subsubsection{$E_8$}

\begin{figure}[h!]
\center
\includegraphics[width=0.25\textwidth]{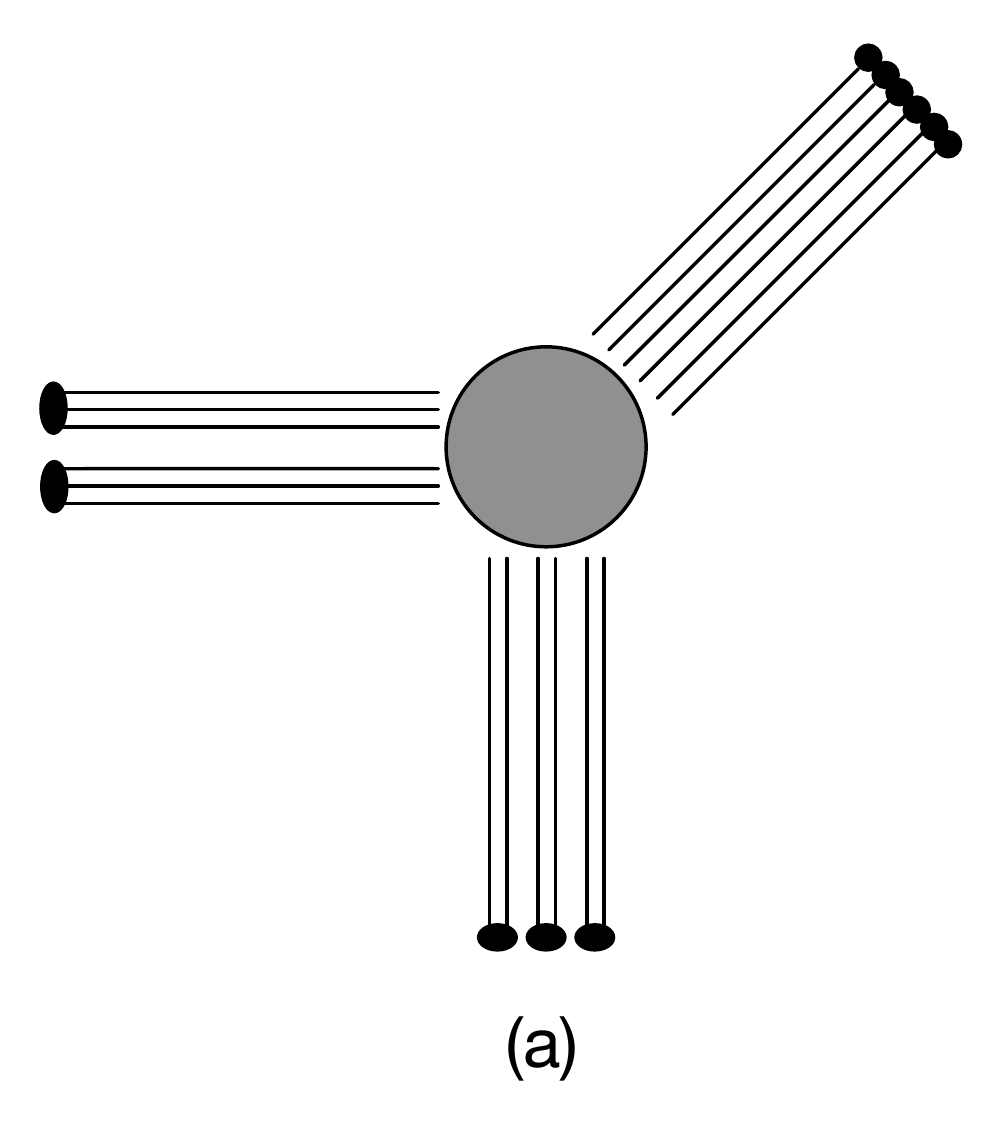} 
\hspace{1cm}
\includegraphics[width=0.25\textwidth]{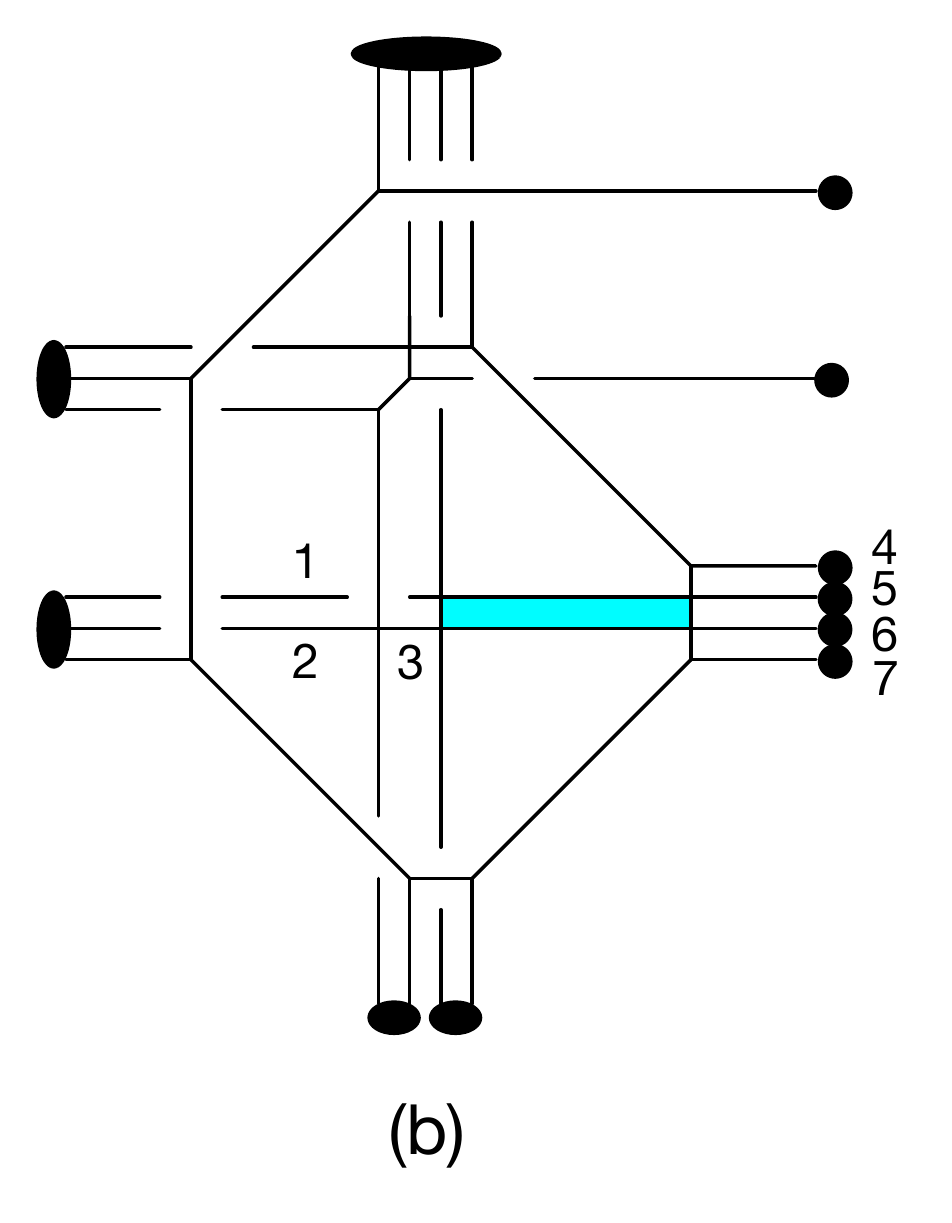} 
\caption{$E_8$ theory: $SU(2) +7$}
\label{E8web}
\end{figure}

Since we already discussed the $E_6$ and $E_7$ theories as special cases of $T_N$ and $S_N$,
we close this section by discussing the 5d lift of the $E_8$ theory \cite{BBT}.
The $E_8$ theory is obtained from the $T_6$ theory by replacing one of the maximal punctures with
a $(3,3)$ puncture, and a second one by a $(2,2,2)$ puncture, resulting in the 5-brane junction shown in Fig,~\ref{E8web}a.
An HW transition followed by an appropriate (and somewhat tricky) mass deformation (Fig.~\ref{E8web}b),
then reveals the known $SU(2) +7$ gauge theory description.


\section{5d duality for 4d duality}

In the previous section we identified 5d ${\cal N}=1$ quiver gauge theories that flow in the UV
to 5d SCFT's that reduce to known isolated ${\cal N}=2$ SCFT's in 4d.
In most cases there are several dual gauge theories associated to the same SCFT.
In this section we will use this realization to relate cases of S-duality in 4d ${\cal N}=2$ theories to this kind of duality in 5d.
The 4d dualities usually involve weakly gauging an isolated SCFT of the type that we discussed 
in the previous section.
The 5d gauge theoretic descriptions of these SCFT's allow us to describe also their weakly gauged versions 
in terms of weakly coupled quiver gauge theories.

\subsection{$SU(3)+6$ and Argyres-Seiberg duality}

The simplest example of a non-trivial duality in 4d ${\cal N}=2$ theories involves the theory with gauge group
$SU(3)$ and $N_F=6$. This is a SCFT with one marginal parameter, the $SU(3)$ gauge coupling.
It was conjectured in \cite{AS} that the strong coupling limit of this theory is dual to a weakly interacting
SCFT, defined by weakly gauging an $SU(2)$ subgroup of the $E_6$ global symmetry of the $E_6$ theory,
and adding one flavor hypermultiplet, denoted in short by $1+SU(2)\subset E_6$. This theory also has a single
marginal parameter, the $SU(2)$ gauge coupling. The gauging breaks the global symmetry to $SU(6)$,
and the flavor provides an additional $U(1)$, in agreement with the global symmetry of the $SU(3)+6$ theory.

\begin{figure}[h]
\center
\includegraphics[width=0.17\textwidth]{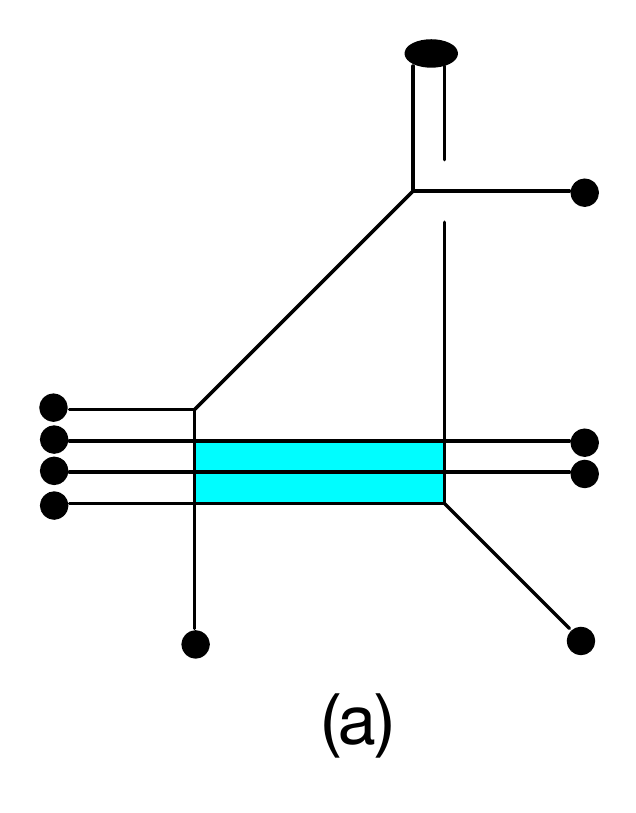} 
\hspace{0.5cm}
\includegraphics[width=0.15\textwidth]{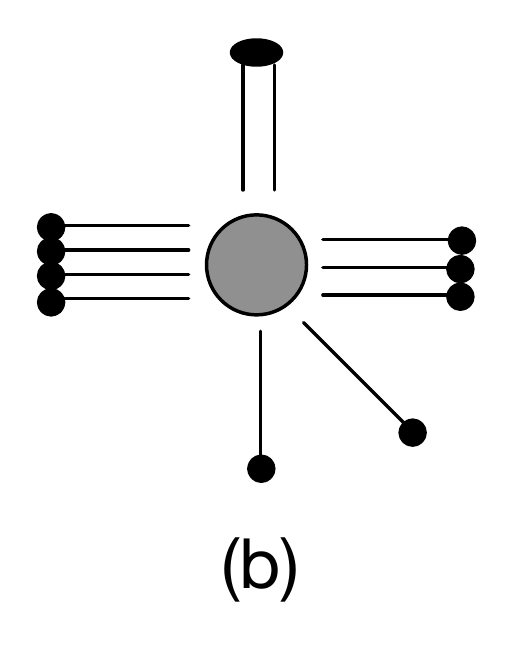} 
\hspace{0.5cm}
\includegraphics[width=0.23\textwidth]{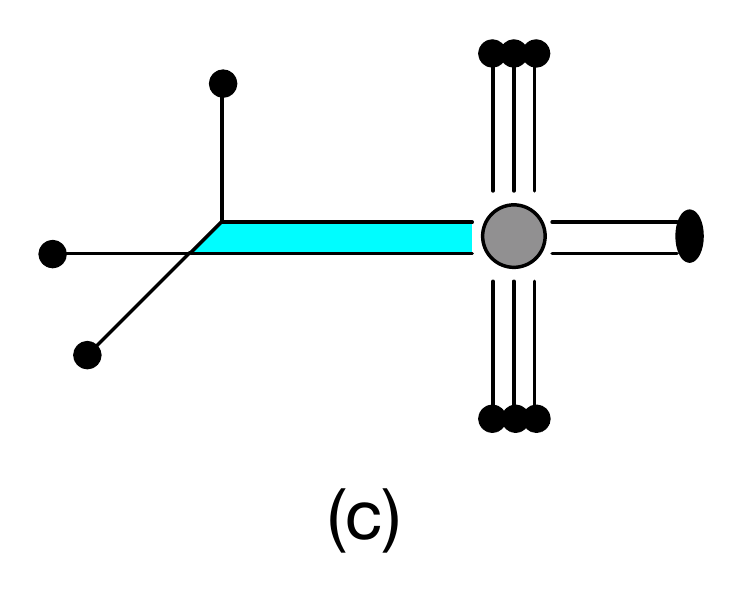} 
\hspace{0.5cm}
\includegraphics[width=0.3\textwidth]{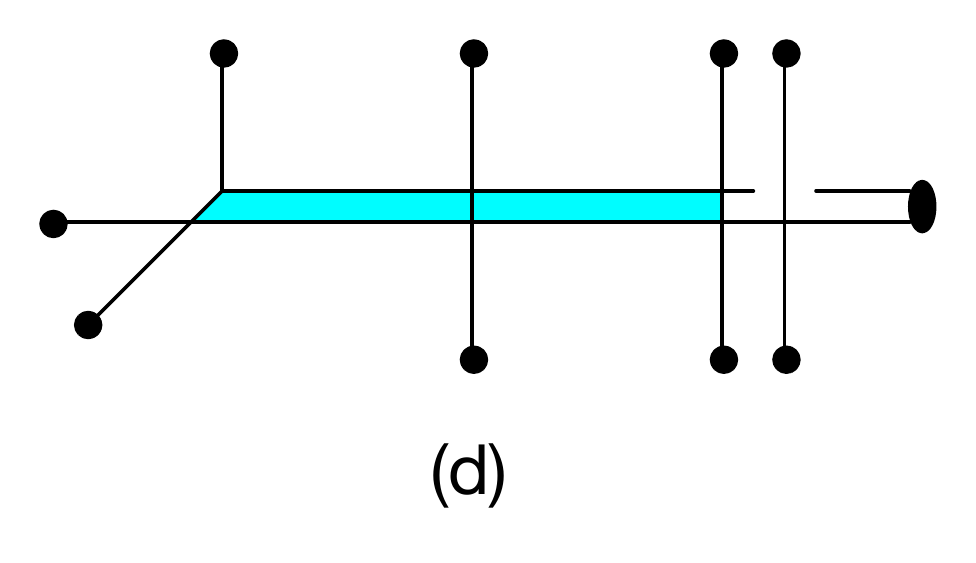}
\caption{5d lift of Argyres-Seiberg duality: (a) $SU(3)_1+6$ (b) fixed point theory (c) $1+SU(2)\subset E_6$ 
(d) $1+SU(2)^{(1)}\times SU(2)^{(2)}+3$.}
\label{SU(3)+6webs}
\end{figure}

We can relate this duality to a 5d duality between gauge theories as follows.
Begin by lifting the 4d superconformal $SU(3)+6$ gauge theory to 5d by constructing the appropriate 5-brane web.
Here one has to make a choice of the CS level in the 5d theory.
Not all choices are allowed. We are limited by the requirement for a UV fixed point to exist.
Take the theory described by the 5-brane web in Fig.~\ref{SU(3)+6webs}a.
This is an $SU(3)$ gauge theory with $N_F=6$. The bare CS level can be determined as before, 
by making all the matter fields massive and reading-off the renormalized CS level of the pure $SU(3)$ theory.
This gives $\kappa_0 = 1$. 

This gauge theory is the result of mass-deforming the 5d SCFT described by the 5-brane junction in 
Fig.~\ref{SU(3)+6webs}b, by moving apart the two 7-branes at the bottom.
Reversing this motion corresponds to changing the sign of the mass, and therefore to
a ``continuation past infinite coupling". This yields, after S-duality, the web in Fig.~\ref{SU(3)+6webs}c.
Note that this contains a sub-web corresponding to an interacting fixed point, 
which is in fact the $E_6$ (or $T_3$) theory.
This web is the 5d realization of weakly gauging an $SU(2)$ subgroup of the $E_6$ theory.
The two parallel external D5-branes have been fused into another sub-web corresponding to $SU(2)$ with one flavor.

The $E_6$ module can be further deformed, as before, to an $SU(2)+5$ gauge theory web,
where now two of the fundamentals become bi-fundamentals of the total $SU(2)\times SU(2)$ gauge symmetry
(Fig.~\ref{SU(3)+6webs}d).
The dual 5d gauge theory is therefore the quiver theory with 
$3+SU(2)^{(1)}\times SU(2)^{(2)}+1$.
The 4d duality is then obtained from the 5d duality by dimensional reduction in a specific scaling limit.
For the $SU(3)$ theory, we compactify on a circle of radius $R$, and take $R\rightarrow 0$ 
and $g_5\rightarrow 0$, holding the dimensionless combination $g_4^2 = g_5^2/R$ fixed. 
This gives the 4d $SU(3)+6$ theory with $g_4$ as its marginal YM coupling.
For the $SU(2)^{(1)}\times SU(2)^{(2)}$ theory, we take $R\rightarrow 0$ and $g_{5,2}\rightarrow 0$
holding $g_{4,2}^2=g_{5,2}^2/R$ fixed, but keep $g_{5,1}$ fixed to the 5d UV cutoff.
When we remove the 5d UV cutoff we end up with the dimensional reduction of the $E_6$
theory with a weakly gauged $SU(2)$ subgroup, with $g_{4,2}$
as its marginal coupling, plus one flavor.

Thus the 4d duality
\be
SU(3)+6 \leftrightarrow 1+SU(2)\subset E_6
\ee
is equivalent to the 5d duality\footnote{We can regard the 5d duality as a generalization 
of the basic duality between $SU(3)+2$ and $SU(2)\times SU(2)$ \cite{AH} with additional flavors.}
\be
SU(3)_1 + 6 \leftrightarrow 3+SU(2)^{(1)}\times SU(2)^{(2)}+1 \,.
\ee
One can therefore strengthen the case for the 4d duality by providing evidence for the 5d duality.
More significantly, one can use the 5d duality, where both sides have a Lagrangian description,
to derive the precise mapping of the BPS states by comparing the superconformal indices of the two theories.

To begin with, the two 5d gauge theories appear to have different global symmetries.
The $SU(3)+6$ theory has a global symmetry $SU(6)_F\times U(1)_B\times U(1)_I$,
whereas the dual theory has $SO(6)_F\times SU(2)_{BF}\times U(1)_F\times U(1)_{I_1}\times U(1)_{I_2}$.
Clearly there must be an instanton-led enhancement in the latter (and potentially also in the former).
We will show this explicitly by computing the superconformal index of the two theories.
This will also allow us to determine the precise map between the global symmetry charges.

But first let us give a more qualitative argument.
Had $SU(2)^{(2)}$ not been gauged,
we know that the classical $SO(10)_F\times U(1)_{I_1}$ symmetry would be enhanced via the $SU(2)^{(1)}$ 
instanton to $E_6$.
The gauging breaks $SO(10)_F$ to $SO(6)_F\times SU(2)_{BF}\times SU(2)^{(2)}$, and $E_6$ to $SU(6)_F$, so we expect the 
$(1,0)$ instanton to enhance $SO(6)_F\times SU(2)_{BF}\times U(1)_{I_1}$ to $SU(6)_F$.
The $SU(2)^{(1)}$ instanton is a spinor of $SO(10)_F$, and therefore decomposes as 
$({\bf 4},{\bf 2},{\bf 1})+(\bar{\bf 4},{\bf 1},{\bf 2})$ of $SO(6)_F\times SU(2)_{BF} \times SU(2)^{(2)}$.
The gauge invariant piece, together with the anti-instanton piece, provide the extra conserved currents
needed for $SU(6)_F$, since the adjoint representation of $SU(6)_F$ decomposes as
\be
{\bf 35} = ({\bf 15},{\bf 1})_0 + ({\bf 1},{\bf 3})_0 + ({\bf 1},{\bf 1})_0 + ({\bf 4},{\bf 2})_1 + (\bar{\bf 4},{\bf 2})_{-1} \,.
\ee

We can also see how the $U(1)$ factors map.
The basic $U(1)_B$-charged baryon operator of the $SU(3)$ theory is in the ${\bf 20}$ (rank 3 antisymmetric) of $SU(6)_F$,
and carries 3 units of charge under $U(1)_B$.
This decomposes as $({\bf 6},{\bf 2})_{0} + (\bar{\bf 4},{\bf 1})_{1} + ({\bf 4},{\bf 1})_{-1}$
of $SO(6)_F\times SU(2)_{BF}\times U(1)_{I_1}$.
In the $SU(2)^{(1)}\times SU(2)^{(2)}$ theory, 
the $({\bf 6},{\bf 2})_{0}$ state is formed by combining a flavor of $SU(2)^{(1)}$, the bi-fundamental, and
the flavor of $SU(2)^{(2)}$ in a gauge invariant way.
The $(\bar{\bf 4},{\bf 1})_{1}$ state is formed by combining a $(1,0)$ instanton
with the flavor of $SU(2)^{(2)}$, and the $({\bf 4},{\bf 1})_{-1}$ is formed in a similar way using a $(-1,0)$ instanton.
This implies that the $U(1)$'s of the two theories are related as $U(1)_B \leftrightarrow U(1)_F$
and $U(1)_I \leftrightarrow U(1)_{I_2}$. In fact, we see that the baryon charge on the $SU(3)$ side
is related to the $U(1)$ flavor charge on the $SU(2)^2$ side by $B=3F$.
We will see this also from the superconformal index below.

\subsubsection{Comparing superconformal indices}

The perturbative part of the index of the $SU(3)+6$ theory is given by 
\bea
I_{pert}^{SU(3)+6}  & = &  
 1 + x^2 \left( 2 + \chi_{\bold{35}} \right)   +   x^3 \Big( (y + \frac{1}{y}) \left( 3 + \chi_{\bold{35}} \right)  +  (b^3+\frac{1}{b^3})\chi_{\bold{20}} \Big) 
 + {\cal O}(x^4) .
\eea
For the $1+SU(2)^{(1)}\times SU(2)^{(2)}+3$ theory, the perturbative contribution is given by
\bea
I_{pert}^{3+SU(2)^2+1}  & = &  1 + x^2 \left( 3 + \chi_{(\bold{1},\bold{3})} + \chi_{(\bold{15},\bold{1})} \right) 
+ x^3 \Big( (y + \frac{1}{y}) \left(4 + \chi_{(\bold{1},\bold{3})} + \chi_{(\bold{15},\bold{1})} \right)   \nonumber \\   
& & \mbox{} + (f + \frac{1}{f})\chi_{(\bold{6},\bold{2})} \Big)  + {\cal O}(x^4) \,,
\label{eq:inft}
\eea
where $f$ denotes the $U(1)_F$ fugacity.
The contribution of the $(1,0)$ instanton (computed by treating $SU(2)^{(1)}$ as $USp(2)$) is
\bea
I_{(1,0)}^{3+SU(2)^2+1}  & = &  x^2\Big( q_1 \chi_{(\bold{4},\bold{2})} + \frac{1}{q_1} \chi_{(\bar{\bold{4}},{\bf 2})} \Big)   
 +  x^3 \Bigg( (y + \frac{1}{y}) \Big( q_1 \chi_{(\bold{4},\bold{2})} + \frac{1}{q_1} \chi_{(\bar{\bold{4}},{\bf 2})} \Big)  \label{eq:inft1}  \\ \nonumber  
& + &  (f + \frac{1}{f})\Big( q_1 \chi_{(\bar{\bold{4}},{\bf 1})} + \frac{1}{q_1} \chi_{(\bold{4},\bold{1})}  \Big) \Bigg) + {\cal O}(x^4) \, .
\eea
As anticipated, the $(1,0)$-instanton provides the states needed to enhance $SO(6)_F\times SU(2)_{BF}\times U(1)_{I_1}$
to $SU(6)_F$, where
$\chi^{SU(6)}_{\bf 6} = q_1^{-2/3} \chi_{(\bold{1},\bold{2})} + q^{1/3}_1 \chi_{(\bold{4},\bold{1})}$.
Furthermore, the indices of the two theories agree if we identify $f=b^3$.  
This confirms the relation between the $U(1)_B$ and $U(1)_F$ charges that we found above, $B=3F$.

This is the end of the story as far as the dimensionally-reduced theories (with the scaling limit mentioned above) are concerned.
The main lesson here is the above relation between the $U(1)$ charges.
However, there is more to be learned about the underlying 5d fixed point theory.
In particular, there is further enhancement of the global symmetry. 

\subsubsection{More on the 5d SCFT}

From the point of view of the $SU(2)^{(1)}\times SU(2)^{(2)}$ theory, 
the additional enhancement is due to instantons carrying $U(1)_{I_2}$ charge.
The contributions of these states is most easily computed by treating $SU(2)^{(2)}$ as $U(2)/U(1)$ (while $SU(2)^{(1)} = USp(2)$).
As usual, we encounter ``$U(1)$ remnants" that must be removed from the instanton partition function, this time using
\be
\mathcal{Z}_{c} = PE\left[\frac{x^2}{z\sqrt{f}\, (1-x y)(1-\frac{x}{y})}\right] \mathcal{Z} \,.
\ee
To order $x^3$ there are contributions from the $(0,1)$, $(1,1)$ and $(2,1)$ instantons:
\bea
I_{(0,1)+(1,1)+(2,1)}^{3+SU(2)^2+1}   &=&   x^2 \Bigg(\left(\frac{q_2}{\sqrt{f}}+\frac{\sqrt{f}}{q_2}\right)\chi_{(\bold{1},\bold{2})}
+ \left( \frac{q_1 q_2}{\sqrt{f}}\chi_{(\bold{4},\bold{1})} + \frac{\sqrt{f}}{q_1 q_2} \chi_{(\bar{\bold{4}},{\bf 1})} \right)\Bigg) \nonumber \\
&+& x^3 \Bigg( (y + \frac{1}{y})\left[\left(\frac{q_2}{\sqrt{f}}+\frac{\sqrt{f}}{q_2}\right)\chi_{(\bold{1},\bold{2})}  
+  \frac{q_1 q_2}{\sqrt{f}}\chi_{(\bold{4},\bold{1})} + \frac{\sqrt{f}}{q_1 q_2} \chi_{(\bar{\bold{4}},{\bf 1})}   \right] \nonumber \\
& + &  \left( q_2 \sqrt{f}+\frac{1}{q_2 \sqrt{f}}\right)\chi_{(\bold{6},\bold{1})} 
+   q_1 q_2\sqrt{f} \chi_{(\bar{\bold{4}},{\bf 2})} +  \frac{1}{q_1 q_2 \sqrt{f}} \chi_{(\bold{4},\bold{2})} \nonumber \\
& + &   \left(q_2 q_1^2 \sqrt{f} + \frac{1}{q_2 q_1^2 \sqrt{f}}\right)\Bigg)  + {\cal O}(x^4) \,.
\label{eq:inft2}
\eea
Combining this with the previous contributions shows that the global symmetry at the fixed point is further enhanced to 
$SU(7)\times U(1)$.
The full index can be expressed concisely in terms of $SU(7)\times U(1)$ characters:
\be
\label{fullSU(2)^2index}
I^{3+SU(2)^2+1} = 1 + x^2 (1+\chi^0_{\bf 48}) + x^3 \left( (y + \frac{1}{y})(2+\chi^0_{\bf 48}) + \chi^1_{\bf 35} + \chi^{-1}_{\overline{\bf 35}} \right) 
+ {\cal O}(x^4)\,,
\ee
where the $SU(7)$ is spanned by
$\chi^{SU(7)}_{\bf 7} = (q_2 q_1^{-4}/\sqrt{f})^{1/7}
(q_2^{-1} \sqrt{f} + \chi_{(\bold{1},\bold{2})} + q_1 \chi_{(\bold{4},\bold{1})}) $,
and the $U(1)$ by $(q_2^3 q_1^2 f^{11/2})^{-1/7}$.
The $U(1)$ has been normalized such that the state in the ${\bf 35}$ of $SU(7)$ carries one unit of charge.

From the point of view of the $SU(3)_1+6$ theory this enhancement must be due to the $SU(3)$ instanton.
The situation here is similar to the one encountered in the $SU(3)_{\frac{1}{2}}+7$ theory in section~\ref{sec:R0N},
as one can see from the similarity of the 5-brane webs.
In this case there are two remnant states: one corresponding to fundamental strings between the separated external D5-brane
and the flavor D5-branes, and the other to a 3-string junction.
The correction to the 1-instanton partition function is given by
\be
Z^c_1 = Z_1 + \frac{x^2 q b^2 \chi^{SU(6)}_{\bar{\bold{6}}}}{(1-x y)(1-\frac{x}{y})} - \frac{x^3 q b^3 \chi^{SU(3)}_{\bold{3}}}{(1-x y)(1-\frac{x}{y})} 
\ee
where $q$ is the $SU(3)$ instanton fugacity. 
This gives a 1-instanton contribution:\footnote{As a consistency check, we can also treat the theory as $SU(3)_1 + 5 + \asymm$, since for $SU(3)$ the antisymmetric is identical to the anti-fundamental. 
The calculation is similar to the one we did for $SU(4)_{\frac{1}{2}} + 7 + \asymm$ in section~\ref{section:R1N}.
The result agrees with what we find for $SU(3)+6$.}
\be
 \label{eq:chnn}
I_{1}^{SU(3)+6}   =  x^2 ( q b^2\chi_{\bar{\bold{6}}} +\frac{1}{q b^2} \chi_{\bold{6}})
+  x^3 \left( (y + \frac{1}{y})(q b^2\chi_{\bar{\bold{6}}} +\frac{1}{q b^2} \chi_{\bold{6}}) 
+ \frac{q}{b} \chi_{\bold{15}}  +  \frac{b }{q} \chi_{\bar{\bold{15}}} \right)  + {\cal O}(x^4) . 
\ee
Together with the perturbative contribution, this exhibits an
enhancement to $SU(7)\times U(1)$, where the $SU(7)$ and $U(1)$ are spanned respectively by
$\chi^{SU(7)}_{\bf 7} = (qb^2)^{-1/7} (qb^2 + \chi^{SU(6)}_{\bold{6}})$ and $(q/b^{5})^{3/7}$.
The full index can again be expressed in terms of $SU(7)\times U(1)$ characters as in (\ref{fullSU(2)^2index}).
Comparing with the $SU(2)\times SU(2)$ theory, we see that the duality relates the fugacities as 
$b^3 = f$ and $q = (q_2 q_1^{2/3} f^{1/6})^{-1}$.

\subsection{$SU(N) + 2N$}

The first natural generalization of the $SU(3)+6$ SCFT is to $SU(N)+2N$.
In \cite{Gaiotto:2009we} Gaiotto proposed that at strong coupling this theory is related to an isolated 4d SCFT with
a global symmetry $SU(2)\times SU(2N)$, later named $R_{0,N}$ \cite{Distler}, by weakly gauging the $SU(2)$ 
factor and adding one flavor. This gives a theory with one marginal parameter and a global symmetry $SU(2N)\times U(1)$.

\begin{figure}[h]
\center
\includegraphics[width=0.2\textwidth]{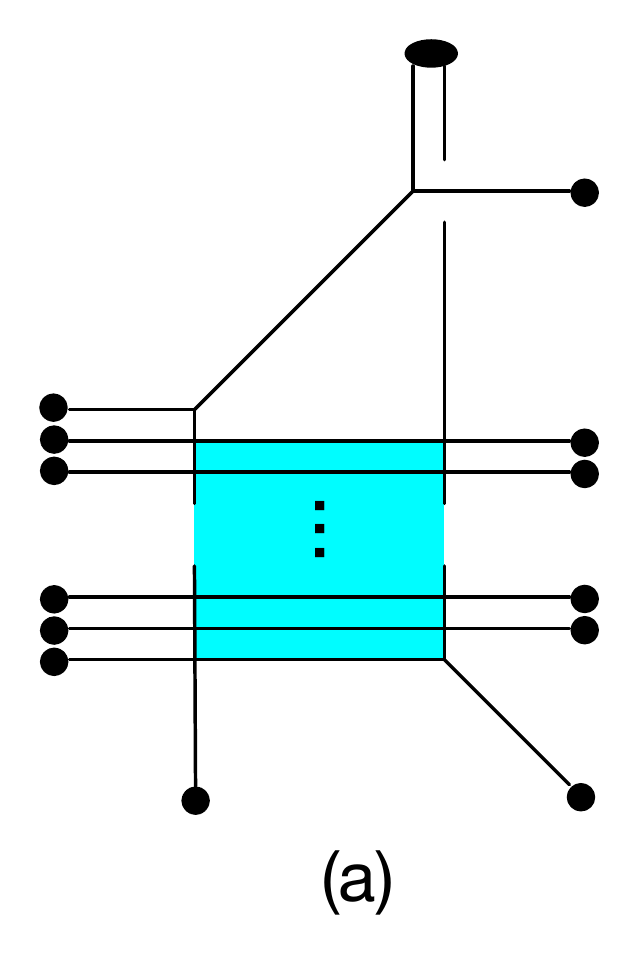} 
\hspace{1.5cm}
\includegraphics[width=0.45\textwidth]{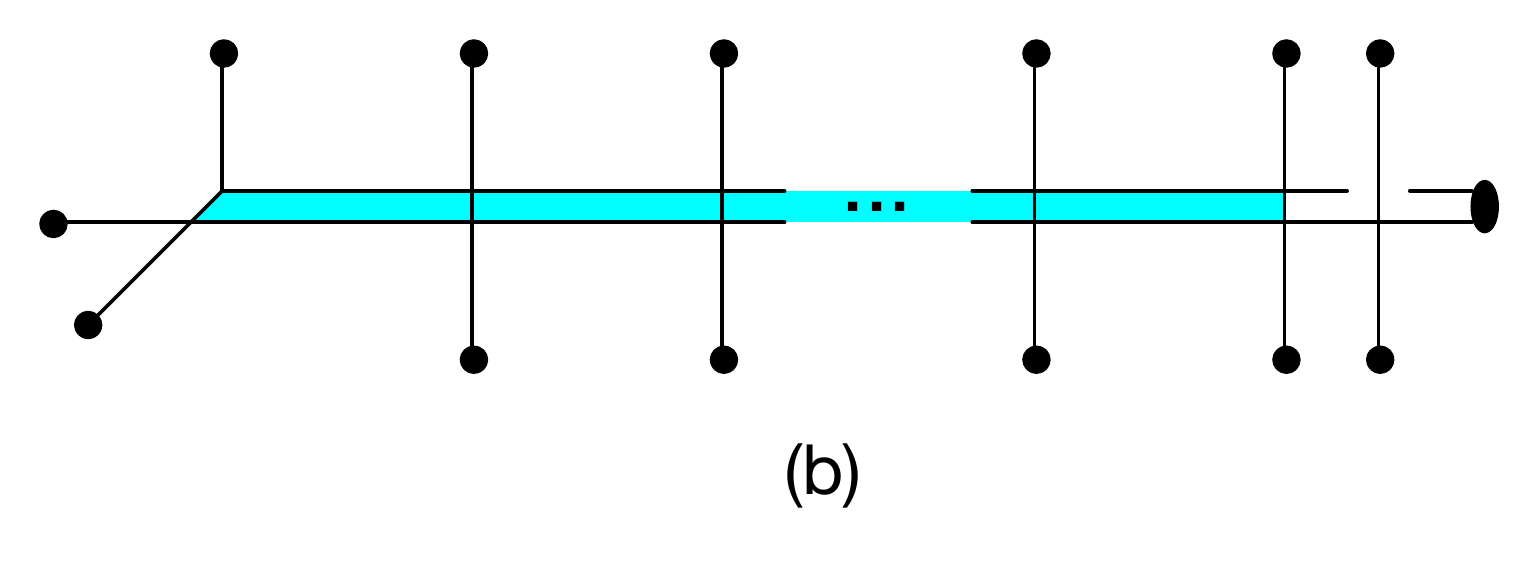}
\caption{5d lift of $SU(N)+2N \sim 1+SU(2)\subset R_{0,N}$.}
\label{SU(N)+2Nwebs}
\end{figure}

We can lift this duality to five dimensions as before.
Start with the 5-brane web in Fig.~\ref{SU(N)+2Nwebs}a. This describes an $SU(N)$ gauge theory with $2N$
flavors and a bare CS level $\kappa_0=1$.
This reduces to the ${\cal N}=2$ $SU(N)+2N$ SCFT in four dimensions.
The mass deformed S-dual web shown in Fig.~\ref{SU(N)+2Nwebs}b clearly describes a quiver gauge theory
with $3+SU(2)^{N-1}+1$. In the limit where the coupling of the last $SU(2)$ is scaled with $R$, this corresponds
to weakly gauging the $SU(2)$ subgroup of the global symmetry of the $R_{0,N}$ theory, as described
by the $3+SU(2)^{N-2}+2$ quiver gauge theory in section \ref{sec:R0N}.
The $\theta$ parameters of the unflavored $SU(2)$ factors follow from those of the $R_{0,N}$ theory.
The 4d duality
\be
SU(N)+2N \leftrightarrow 1+SU(2)\subset R_{0,N}
\ee
is therefore equivalent to the 5d duality
\be
SU(N)_{\pm1} + 2N \leftrightarrow 3+SU(2) \times SU(2)_0^{N-3} \times SU(2)+1 \,.
\ee

As before, the duality requires an enhancement of the global symmetry, at least in the $SU(2)^{N-1}$ theory.
The global symmetry of the 5d $SU(N)+2N$ theory is $SU(2N)_F\times U(1)_B\times U(1)_I$, and that of the 
quiver theory is $SU(4)_F\times SU(2)^{N-2}_{BF}\times U(1)_I^{N-1}\times U(1)_F$.
Evidently, the second theory should exhibit an enhancement of $SU(4)_F\times SU(2)^{N-2}_{BF}$ together
with $N-2$ combinations of the $U(1)$'s to $SU(2N)_F$.
Interpolating from the $N=3$ case suggests that this involves the topological symmetries
of all but the last (single-flavored) $SU(2)$ factor, namely $U(1)_{I_k}$ with $k=1,\ldots,N-2$.

The additional conserved currents transform in all possible bi-fundamental representations
of the non-abelian factors of the global symmetry, and carry charges under $U(1)_{I_k}$.
For example, the $(1,0^{N-2})$ instanton gives a current in the $({\bf 4},{\bf 2},{\bf 1}^{N-3})_{(1,0^{N-2})}$ representation
of $SU(4)_F\times SU(2)^{N-2}_{BF}\times U(1)_I^{N-1}$,
and the $(0^{k},1,0^{N-k-2})$ instanton, with $1\leq k \leq N-3$, gives a current in the 
$({\bf 1}^{k},{\bf 2},{\bf 2},{\bf 1}^{N-k-3})_{(0^{k},1,0^{N-k-2})}$ representation.\footnote{This follows from the
decomposition of the spinor representation.
The $(1,0^{N-3})$ instanton sees effectively 5 flavors, which generate a spinor of 
$SO(10)_F \supset SU(4)_F\times SU(2)_{BF_1}\times SU(2)^{(2)}$, where $SU(2)^{(2)}$ is gauged.
The spinor decomposes as ${\bf 16} = ({\bf 4},{\bf 2},{\bf 1}) + (\bar{\bf 4},{\bf 1},{\bf 2})$.
The $(0^{k},1,0^{N-k-3})$ instanton sees effectively 4 flavors, and the analogous decomposition
is ${\bf 8}_s = ({\bf 2},{\bf 1},{\bf 1},{\bf 2}) + ({\bf 1},{\bf 2},{\bf 2},{\bf 1})$
under $SO(8)\supset SU(2)^{(k)}\times SU(2)_{BF_{k}}\times  SU(2)_{BF_{k+1}}\times SU(2)^{(k+2)}$.}
More generally one can show that all the extra conserved currents arise from instantons
with topological charges $(0^k,1^l,0^{N-k-l-1})$, where $0\leq k \leq N-3$ and $1\leq l \leq N-k-2$.
Consider for example the $N=4$ case, in which the dual quiver theory is $3+SU(2)^3+1$,
whose global symmetry is 
$SU(4)_F \times SU(2)_{BF_1}\times SU(2)_{BF_2} \times U(1)_{I_1}\times U(1)_{I_2}\times U(1)_{I_3}\times U(1)_F$.
In this case the $SU(8)$ current decomposes as 
(we include also the trivial $U(1)_{I_3}$ charges, but not the trivial $U(1)_F$ charges)
\bea
{\bf 63} &=&  ({\bf 15},{\bf 1},{\bf 1})_{(0,0,0)} + ({\bf 1},{\bf 3},{\bf 1})_{(0,0,0)} + ({\bf 1},{\bf 1},{\bf 3})_{(0,0,0)} 
+ 2\, ({\bf 1},{\bf 1},{\bf 1})_{(0,0,0)} \nonumber \\
&+& \left[({\bf 4},{\bf 2},{\bf 1})_{(1,0,0)} +({\bf 1},{\bf 2},{\bf 2})_{(0,1,0)} + ({\bf 4},{\bf 1},{\bf 2})_{(1,1,0)} + \mbox{c.c.}\right] .
\eea

We can likewise relate the two remaining $U(1)$ symmetries of the two gauge theories.
The basic baryonic operator of the $SU(N)+2N$ theory carries $N$ units of charge under $U(1)_B$, 
and transforms in the $N$-index antisymmetric representation of $SU(2N)_F$.
This operator contributes to the superconformal index at ${\cal O}(x^N)$.
The dual operator in the $3+SU(2)^{N-1}+1$ theory includes both perturbative and instantonic contributions.
The perturbative part is simply the product of all the matter fields through the quiver.
This carries one unit of charge under $U(1)_F$, and
transforms in the $({\bf 6}, {\bf 2}^{N-2})_{(0^{N-1})}$ of 
$SU(4)_F\times SU(2)^{N-2}_{BF}\times U(1)_I^{N-1}$. 
Non-perturbative contributions are obtained, as in the $N=3$ case, by replacing some matter 
fields in the above chain (not including the flavor of the last $SU(2)$) by instantons. 
Consider for simplicity the $N=4$ case.
The decomposition of the 4-index antisymmetric representation of $SU(8)$ is given by
\be
{\bf 70} = ({\bf 6},{\bf 2},{\bf 2})_{(0,0,0)} + \left[(\bar{\bf 4},{\bf 1},{\bf 2})_{(1,0,0)} + ({\bf 6},{\bf 1},{\bf 1})_{(0,1,0)} 
 + (\bar{\bf 4},{\bf 2},{\bf 1})_{(1,1,0)} + ({\bf 1},{\bf 1},{\bf 1})_{(2,1,0)} + \mbox{c.c}\right] .
\ee
The $(\bar{\bf 4},{\bf 1},{\bf 2})_{(1,0,0)}$ state corresponds to the gauge invariant combination of the $(1,0,0)$
instanton, the second bi-fundamental field and the flavor of $SU(2)^{(3)}$,
and the $({\bf 6},{\bf 1},{\bf 1})_{(0,1,0)}$ state to the combination of the $(0,1,0)$ instanton,
the flavor of $SU(2)^{(1)}$ and the flavor of $SU(2)^{(3)}$.
The $(\bar{\bf 4},{\bf 2},{\bf 1})_{(1,1,0)}$ and $({\bf 1},{\bf 1},{\bf 1})_{(2,1,0)}$ states are 
likewise gauge-invariant combinations of the $SU(2)^{(3)}$ flavor and the $(1,1)$ and $(2,1)$ instantons, respectively.
Since all of these states carry one unit of $U(1)_F$ charge,
we conclude that $U(1)_B\times U(1)_I$ maps to $U(1)_F\times U(1)_{I_2}$, with $B= NF$
(generalizing the $N=3$ case).

Verifying all of this explicitly requires an index calculation, which quickly becomes technically difficult as we increase $N$.
We shall not presently pursue it.
We do however expect a further enhancement of the global symmetry at the 5d fixed point to $SU(2N+1)$ due to the
instanton of the last $SU(2)$ factor.

\subsection{$SU(N) + 4 + 2\, \asymm$}

Another possible higher-rank generalization of $SU(3)+6$ is to $SU(N) + 2\,\asymm + 4$.
This too is an exact SCFT in four dimensions with one marginal parameter.
In this case the conjectured dual theory is different for odd and even $N$ \cite{Distler}.
For $N=2n+1$ the dual theory is given by $1+USp(2n)\subset R_{2,2n+1}$, and
for $N=2n$ by $3+USp(2n)\subset R_{2,2n-1}$.
Since $R_{2,2n+1}$ has a global symmetry $SO(4n+6)ֿ\times U(1)$, 
the global symmetry in both cases on both sides of the duality (except for $N=3$) is $SU(4)\times SU(2) \times U(1)^2$.

In the $SU(N)$ theory, all the symmetries come from the matter fields.
In particular one $U(1)$ factor is the baryonic symmetry associated
to the fundamentals, $U(1)_{B_F}$, and the other one to the antisymmetrics, $U(1)_{B_A}$.
In the dual theory for $N=2n$, the $SU(4)$ factor is associated to the three flavors,
one of the $U(1)$'s is intrinsic to the $R_{2,2n-1}$ theory,
and the $SU(2)$ together with the other $U(1)$ come from the embedding 
$SO(4n+2)\supset USp(2n) \times SU(2) \times U(1)$, once the $USp(2n)$ has been gauged.
In the dual theory for $N=2n+1$, the $SU(4)\times SU(2)$ comes from the embedding
$SO(4n+6)\supset USp(2n) \times SU(4) \times SU(2)$, one $U(1)$ is intrinsic to $R_{2,2n+1}$,
and the other is associated with the one flavor.

We will now discuss the 5d lifts of the two cases separately.

\subsubsection{$N=2n$}

\begin{figure}[h]
\center
\includegraphics[width=0.35\textwidth]{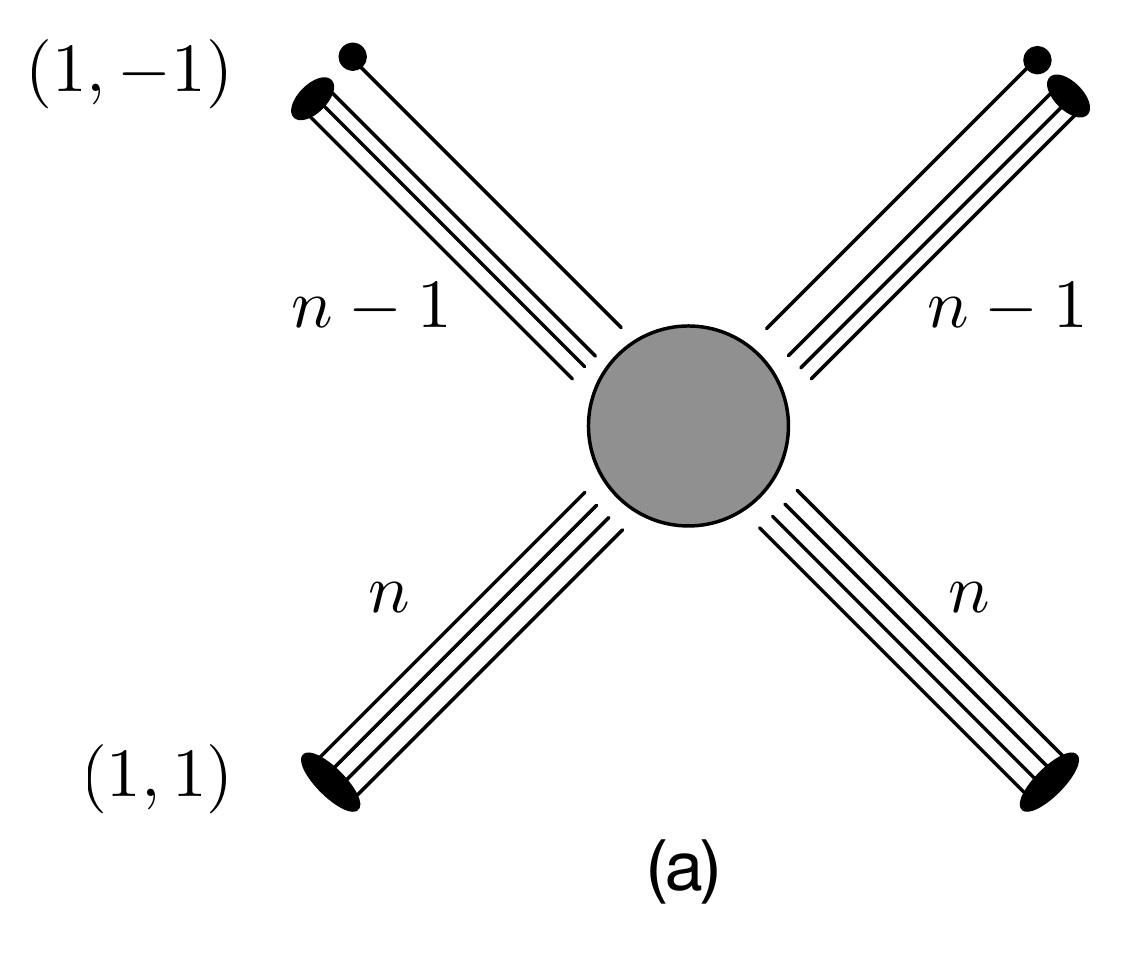} 
\hspace{1cm}
\includegraphics[width=0.25\textwidth]{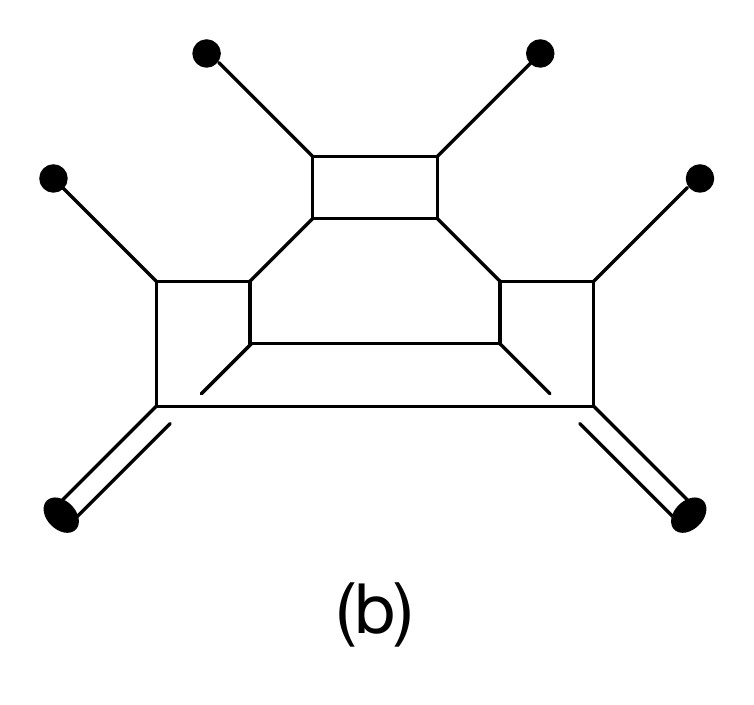} 
\caption{5-brane web for $SU(2n) + 2\, \asymm$}
\label{SU(2n)+2ASjunctions}
\end{figure}

Let us begin with the 5-brane junction for a 5d $SU(2n)$ gauge theory with two antisymmetric hypermultiplets shown in 
Fig.~\ref{SU(2n)+2ASjunctions}a. 
Fig.~\ref{SU(2n)+2ASjunctions}b shows the deformed web for $n=2$ (previously described in \cite{Bergman:2013aca}).
This corresponds to a CS level $\kappa =2$, which will be the one relevant for us.\footnote{One can also generalize 
to other CS levels, as shown in Appendix~\ref{sec:antisymmetric} for the theory with a single antisymmetric field.}

\begin{figure}[h]
\center
\includegraphics[width=0.27\textwidth]{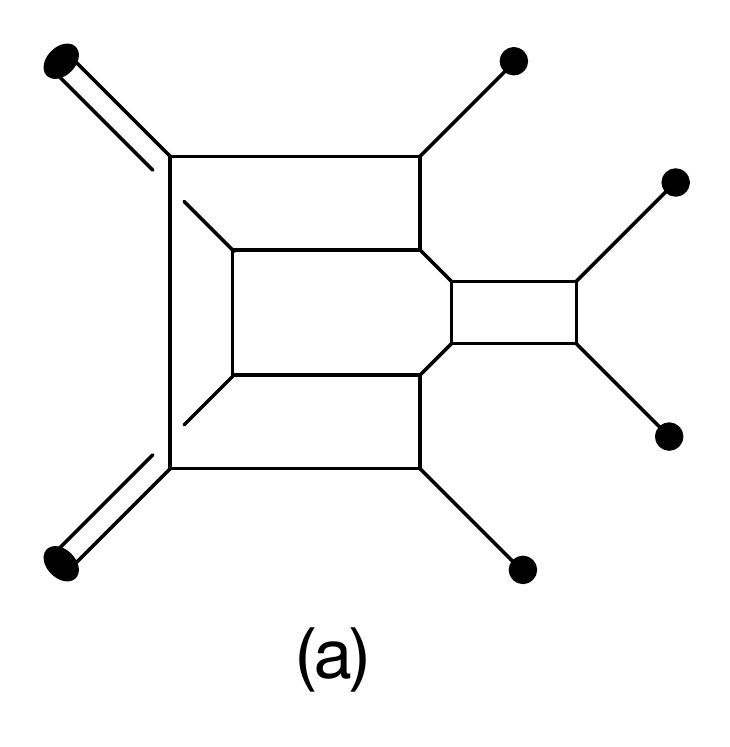} 
\hspace{1cm}
\includegraphics[width=0.45\textwidth]{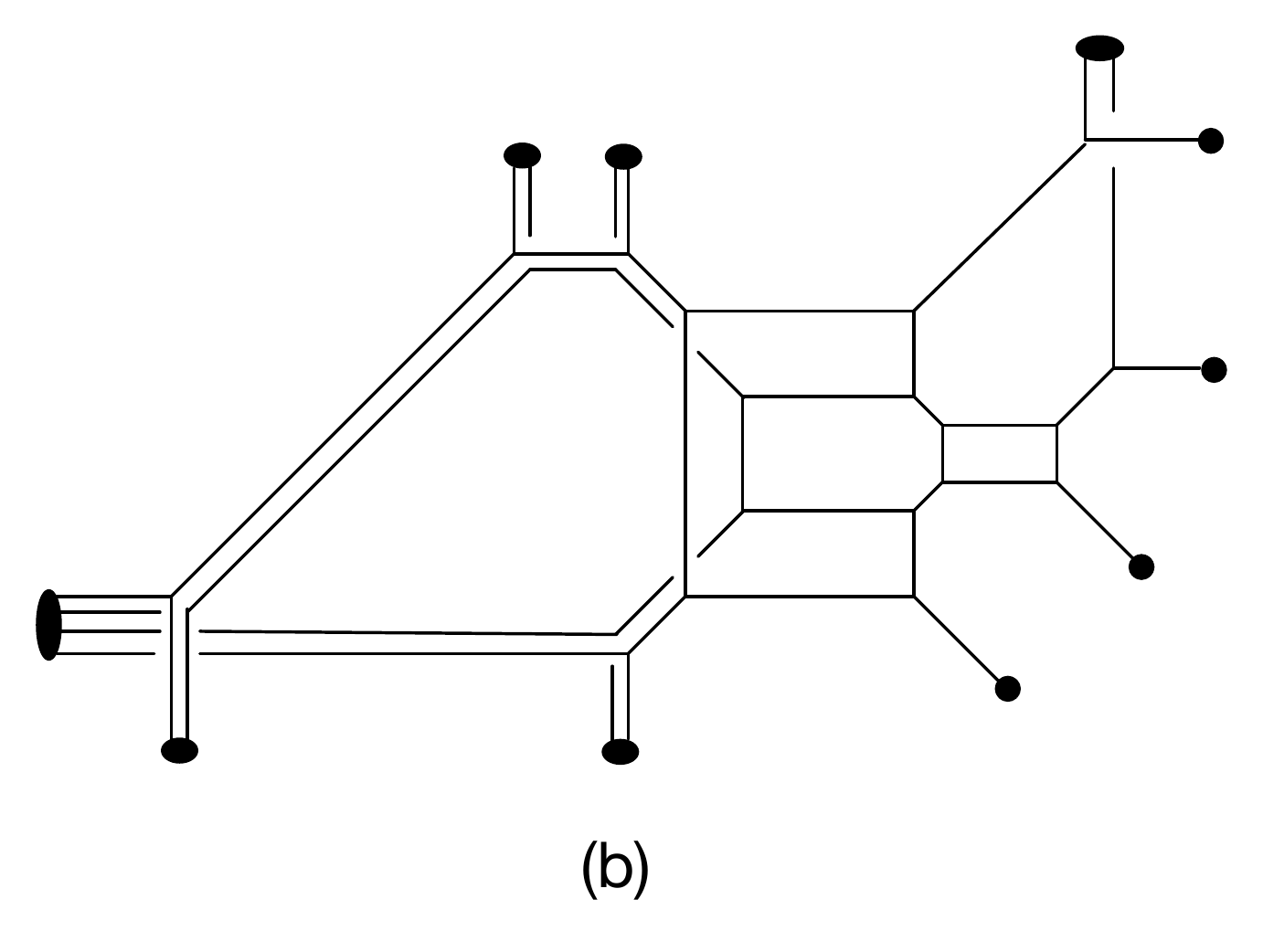}
\caption{5-brane webs for (a) $USp(4)\times USp(2)$ and (b) $3+USp(4)\times USp(2)+1$.}
\label{USpxUSpwebs1}
\end{figure}

To lift the 4d SCFT we need to add four flavors to this.
In terms of the 5-brane web there are several possibilities, resulting in theories with different bare CS levels.
To motivate the correct choice, let us consider the S-dual web, Fig.~\ref{USpxUSpwebs1}a.
This describes a quiver gauge theory with $USp(2n)\times USp(2n-2)$ (shown for $n=2$).
We know that the 5d $R_{2,2n-1}$ theory has a gauge theory deformation described by $USp(2n-2) + 2n+1$,
and a global symmetry $SO(4n+2)$, fully realized by the gauge theory.
Gauging a $USp(2n)$ subgroup of this gives a quiver gauge theory with $USp(2n)\times USp(2n-2) + 1$.
The lift of the proposed 4d dual of $SU(2n) + 2\, \asymm +4$ would therefore seem to be 
the quiver theory with $3+USp(2n)\times USp(2n-2) + 1$.
Therefore on the quiver side of the duality the flavors should be added as shown in Fig.~\ref{USpxUSpwebs1}b.
This web can be obtained by adding D7-branes in the appropriate places and following the procedures described in
Appendix~\ref{sec:flavors}.

S-dualizing back (rotating the web back by 90 degrees) we get $SU(2n)_{\pm 1} + 2\,\asymm + 4$,
namely the bare CS level 
is $\pm 1$.
Thus the 4d duality
\be
SU(2n) + 2\,\asymm +4  \leftrightarrow  3+USp(2n)\subset R_{2,2n-1} 
\ee
lifts to the 5d duality
\be
SU(2n)_{\pm 1} + 2\,\asymm +4 \leftrightarrow  3+USp(2n)\times USp(2n-2) + 1 \,.
\ee
The global symmetries of the proposed 5d duals agree. 
On the $SU(2n)$ side the symmetry is $SU(4)_F \times SU(2)_A \times U(1)_{B_F} \times U(1)_{B_A} \times U(1)_I$,
where $B_F$ is the baryon number associated to the flavors, and $B_A$ is the baryon number associated to the 
antisymmetric fields.
On the $USp\times USp$ side it is $SU(4)_F \times SU(2)_{BF} \times U(1)_{F} \times U(1)_{I_1} \times U(1)_{I_2}$.
There is no enhancement in the 4d reduction (although there may be enhancement at the 5d fixed point).

We can again derive the explicit map of the $U(1)$ charges by finding dual descriptions
of various charged operators.
The simplest baryonic operator on the $SU(2n)$ side is given by the gauge invariant product
of two fundamentals and the conjugate of the antisymmetric field, $q \bar{A} q$.
This transforms in 
$({\bf 6},{\bf 2})_{(2,-1,0)}$ of $SU(4)_F \times SU(2)_A \times U(1)_{B_F} \times U(1)_{B_A} \times U(1)_I$.
The dual operator on the quiver theory side is given by the gauge invariant product of 
a $USp(2n)$ fundmental, the bi-fundamental and the $USp(2n-2)$ fundamental,
which transforms in the $({\bf 6},{\bf 2})_{(1,0,0)}$ of $SU(4)_F \times SU(2)_{BF} \times U(1)_{F}\times U(1)_{I_1}\times U(1)_{I_2}$.

There are also three baryonic operators involving the Levi-Civita symbol,
\be
\epsilon_{i_1 \cdots i_{2n}} A^{i_1 i_2} \cdots A^{i_{2k-1} i_{2k}} q^{i_{2k+1}} \cdots q ^{i_{2n}} \,,
\ee
with $k =n,n-1,n-2$.
For $k=n$ this is the Pfaffian operator $\mbox{Pf}(A)$, 
which transforms as $({\bf 1},({\bf 2})^n_{sym})_{(0,n,0)}=({\bf 1},{\bf n+1})_{(0,n,0)}$ of
$SU(4)_F \times SU(2)_A \times U(1)_{B_F} \times U(1)_{B_A}\times U(1)_I$.
The $k=n-1$ and $k=n-2$ operators transform as $({\bf 6},{\bf n})_{(2,n-1,0)}$
and $({\bf 1},{\bf n-1})_{(4,n-2,0)}$, respectively.
In the dual quiver theory these operators involve the $(0,1)$ instanton.
The Pfaffian corresponds to the gauge-invariant component of
the $(0,1)$ instanton, which transforms as $({\bf 1},{\bf n+1})_{(-\frac{1}{2},0,1)}$ of 
$SU(4)_F\times SU(2)_{BF}\times U(1)_F \times U(1)_{I_1}\times U(1)_{I_2}$.\footnote{The $(0,1)$ and $(1,0)$
instantons both transform as spinors of $SO(4n+2)$.
Under $SO(4n+2) \supset USp(2n)\times SU(2)_{BF} \times U(1)_F$ this decomposes as
${\bf 2}^{2n} = ({\bf 1},{\bf n+1})_{-\frac{1}{2}} + (\, \funda \,,{\bf n})_{\frac{1}{2}} 
+ (\,\asymm \,,{\bf n-1})_{-\frac{1}{2}} + \cdots$, 
where the $USp(2n)$ antisymmetric representations are traceless,
and under $SO(4n+2) \supset USp(2n-2)\times SU(2)_{BF} \times SU(4)_F$ it decomposes as
${\bf 2}^{2n} = ({\bf 1},{\bf n},{\bf 4}) + (\, \funda \,,{\bf n-1},\bar{\bf 4}) + (\,\asymm \,,{\bf n-2},{\bf 4}) + \cdots$.}
The dual of the $k=n-1$ operator involves the component of the $(0,1)$ instanton transforming as
$(\funda \,,{\bf n})_{\frac{1}{2}}$ of $USp(2n)\times SU(2)_{BF}\times U(1)_F$, combined with a
$USp(2n)$ flavor, resulting in a gauge-invariant operator in the 
$({\bf 6},{\bf n})_{(\frac{1}{2},0,1)}$ representation of the global symmetry.
The dual of the $k=n-2$ operator is a little trickier.
If we combine the next component of the $(0,1)$ instanton, $(\asymm \,,{\bf n-1})_{-\frac{1}{2}}$, 
with two $USp(2n)$ flavors we get a gauge-invariant operator in the ${\bf 15}$ of $SU(4)_F$ instead of a singlet.
The correct operator combines the $(\funda \,,{\bf n})_{\frac{1}{2}}$ component of the $(0,1)$ instanton
with the bi-fundamental and the $USp(2n-2)$ flavor into a gauge invariant operator with global symmetry
charges $({\bf 1},{\bf n-1})_{(\frac{3}{2},0,1)}$.

Comparing the $U(1)$ charges of these operators in the two theories we conclude that 
the 5d charge map has the form:
\bea
F &=& \frac{2n-1}{4n} B_F - \frac{1}{2n} B_A + \alpha I \nonumber \\
I_1 &=& \beta I \\
I_2 &=& \frac{1}{2n} B_F + \frac{1}{n} B_A + \gamma I  \,. \nonumber
\eea
This fixes completely the 4d version of the map, given by setting $I=0$.
The two $U(1)$ charges on the $SU(2n)$ side 
are $B_F$ and $B_A$. In the dual theory, $3+USp(2n)\subset R_{2,2n-1}$,
the charge $F$ corresponds to the $U(1)$ in the embedding of $USp(2n)$, and the charge
$I_2$ to the $U(1)$ intrinsic to $R_{2,2n-1}$.
It would be nice to verify this map directly from the 4d point of view.

To determine the coefficients $\alpha, \beta$ and $\gamma$ in the 5d map we need to also consider instantonic 
operators in the $SU(2n)$ theory and their duals in the quiver theory.
We will not do this here.

\subsubsection{$N=2n+1$}

\begin{figure}[h]
\center
\includegraphics[width=0.35\textwidth]{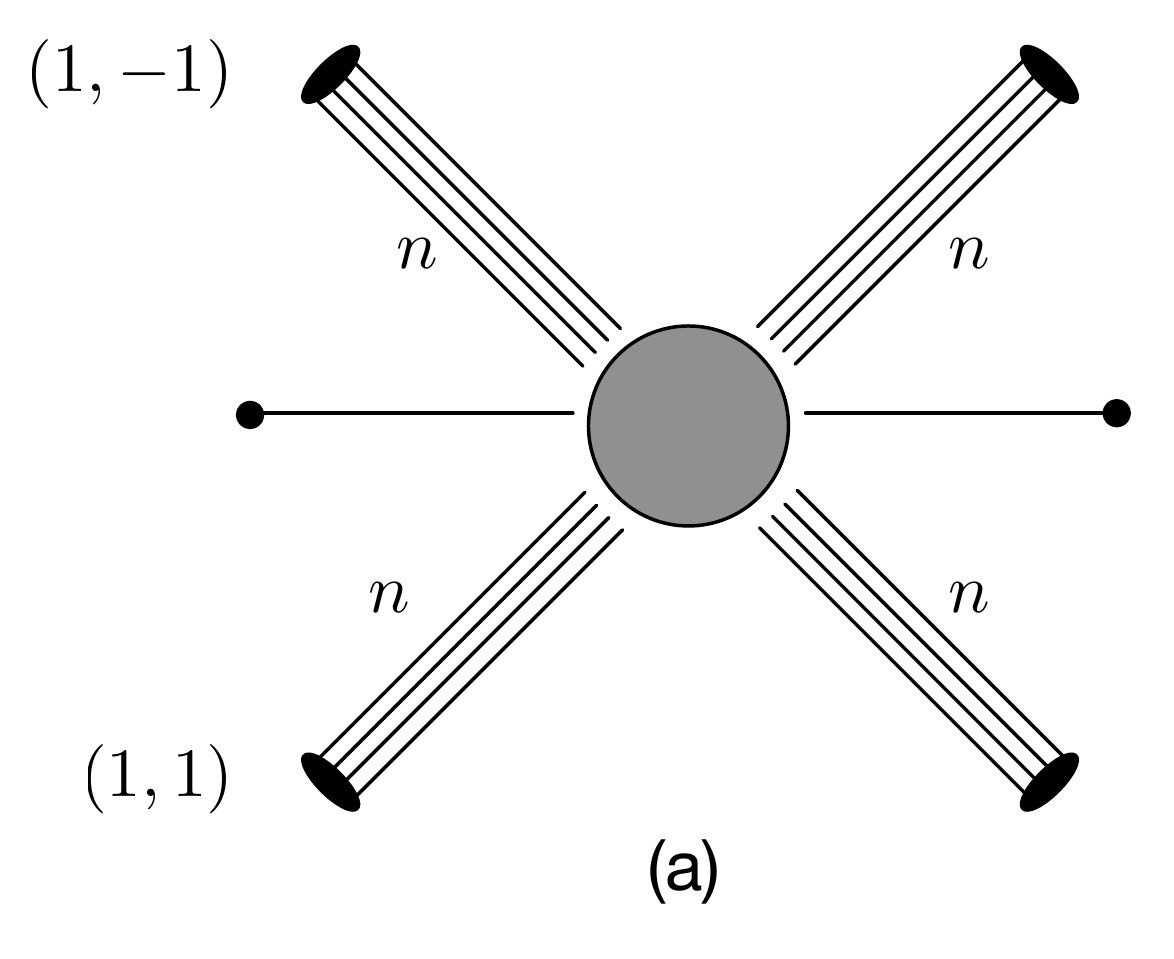} 
\hspace{1cm}
\includegraphics[width=0.27\textwidth]{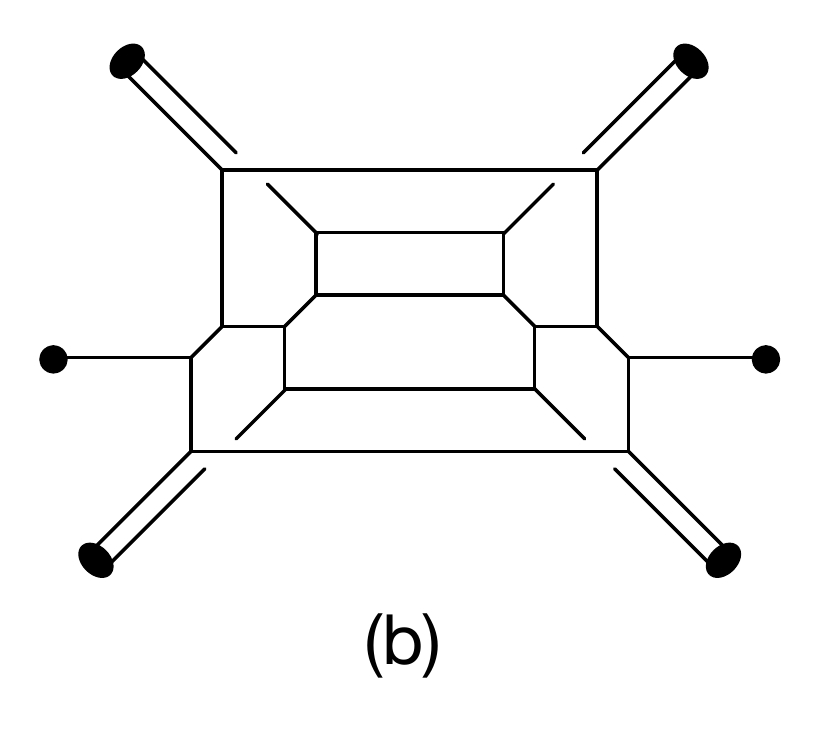} 
\caption{5-brane web of $SU(5)_0 + 2\, \asymm$.}
\label{SU(2n+1)+2ASwebs}
\end{figure}

Most of the analysis here parallels that of the even $N$ case, so we will be somewhat briefer.
The 5-brane junction and mass-deformed web of the gauge theory $SU(2n+1) + 2\, \asymm$ 
(for $n=2$) are shown in 
Fig.~\ref{SU(2n+1)+2ASwebs} (this was also previously described in \cite{Bergman:2013aca}).
The relevant CS level in this case is $\kappa = 0$.

\begin{figure}[h]
\center
\includegraphics[width=0.27\textwidth]{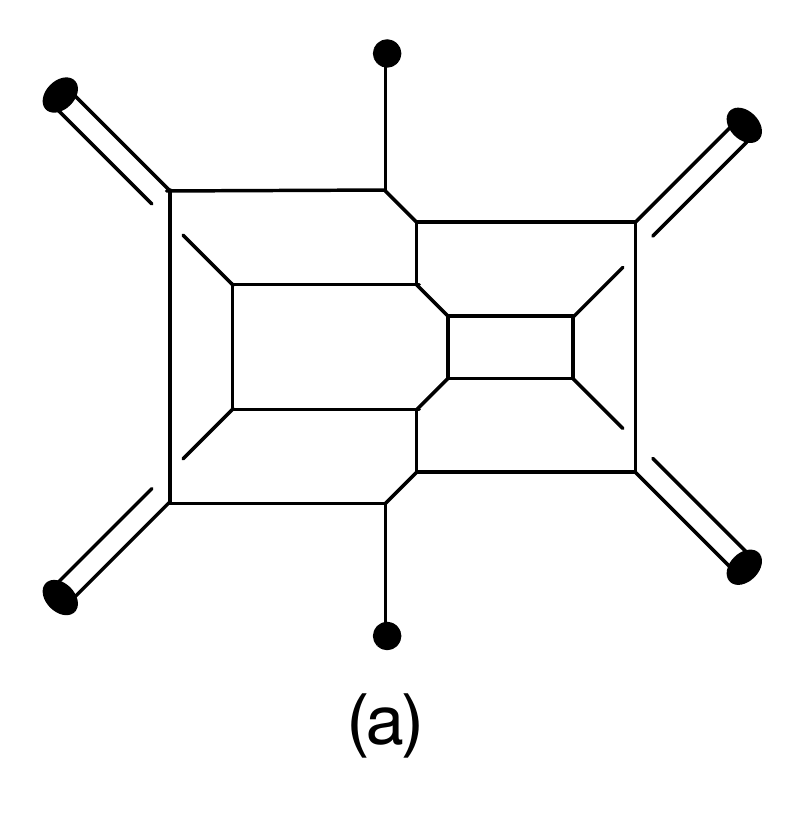} 
\hspace{1cm}
\includegraphics[width=0.45\textwidth]{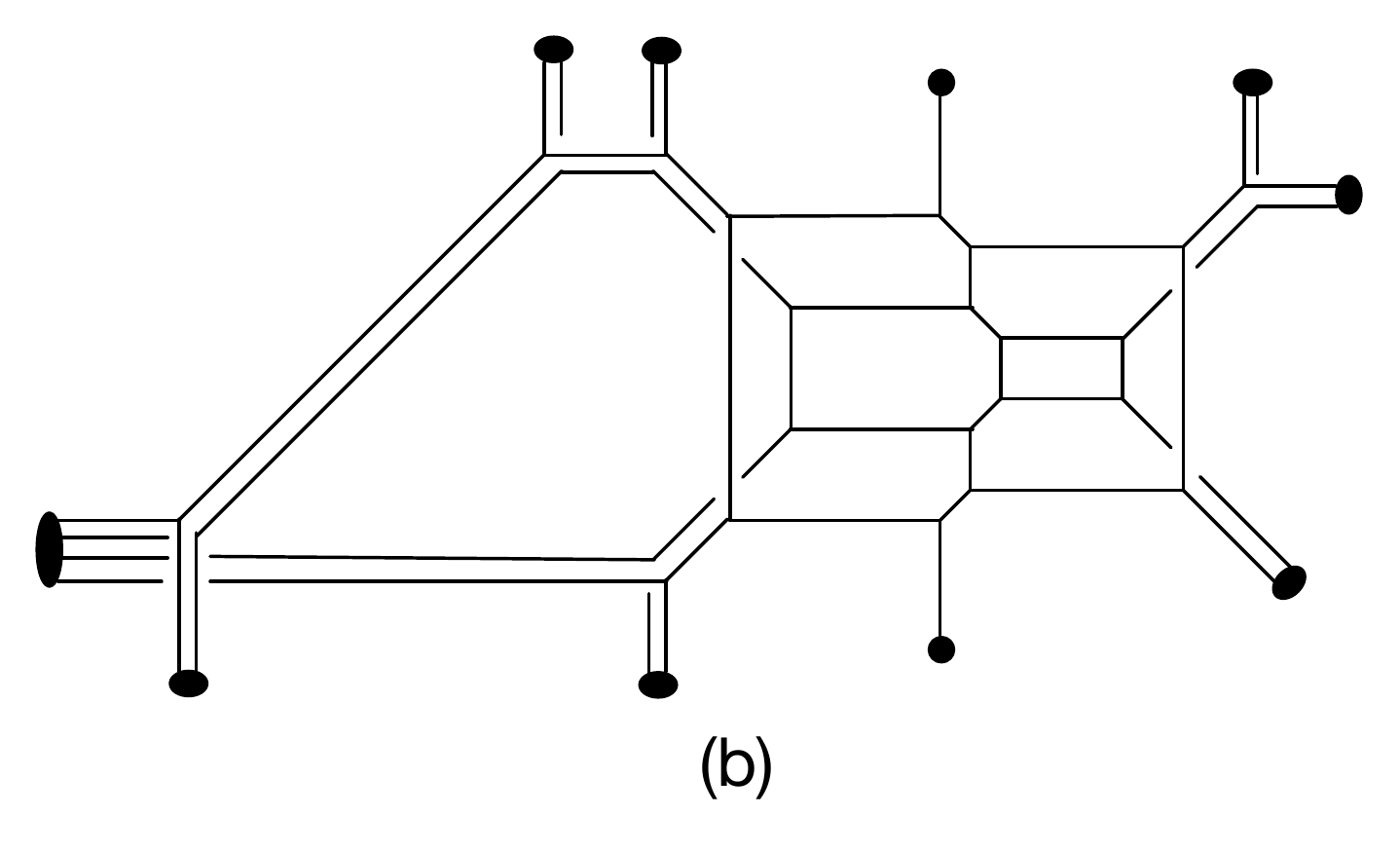} 
\caption{5-brane webs for (a) $USp(4)\times USp(4)$, (b) $3+USp(4)\times USp(4)+1$.}
\label{USpxUSpwebs2}
\end{figure}

Following the same strategy as before, we consider the S-dual web, Fig.~\ref{USpxUSpwebs2}a.
This describes a quiver gauge theory with $USp(2n)\times USp(2n)$.
Making use of the fact that the $R_{2n+1}$ theory is deformable to a $USp(2n) + 2n+3$ gauge theory,
with an $SO(4n+6)$ global symmetry,
and that gauging a $USp(2n)$ subgroup of this leaves $3+ USp(2n)\times USp(2n)$, 
we conclude that we want to consider the quiver theory $3+ USp(2n)\times USp(2n) +1$.
The web for this theory is shown in Fig.~\ref{USpxUSpwebs2}b.
S-dualizing back we then get $SU(2n+1)_{\pm 1} + 2\,\asymm + 4$.
In other words, the 4d duality
\be
SU(2n+1) + 2\,\asymm +4  \leftrightarrow  1+USp(2n)\subset R_{2,2n+1}
\ee
lifts to the 5d duality
\be
SU(2n+1)_{\pm 1} + 2\,\asymm +4  \leftrightarrow  3+USp(2n) \times USp(2n)+1 \,.
\ee
The global symmetries on both sides are the same as in the even $N$ case.

The mapping of the three $U(1)$ charges follows in a very similar manner.
The $q\bar{A}q$ operator is the same as before.
The $SU(2n+1)$ theory has two additional baryonic operators:
\be
\epsilon_{i_1 \cdots i_{2n+1}} A^{i_1 i_2} \cdots  A^{i_{2k-1}i_{2k}} q^{i_{2k+1}} \cdots q^{i_{2n+1}} \,,
\ee
with $k=n, n-1$. The first transforms as $({\bf 4},{\bf n+1})_{(1,n,0)}$
of $SU(4)_F\times SU(2)_A\times U(1)_{B_F}\times U(1)_{B_A}\times U(1)_I$,
and the second as $(\bar{\bf 4},{\bf n})_{(3,n-1,0)}$.
In the dual quiver theory, the former is just the gauge-invariant component of the $(1,0)$ instanton,
which transforms as $({\bf 4},{\bf n+1})_{(0,1,0)}$.
The dual of the latter is given by the gauge-invariant combination of the $(\funda\, ,{\bf n})$ component of the $(1,0)$ instanton
and the flavor of the second $USp(2n)$, which gives an operator in $(\bar{\bf 4},{\bf n})_{(1,1,0)}$.
The 5d charge map then takes the form
\bea
F &=& \frac{n}{2n+1} B_F - \frac{1}{2n+1} B_A + \alpha I \nonumber \\
I_1 &=& \frac{1}{2n+1} B_F + \frac{2}{2n+1} B_A +\beta I \\
I_2 &=& \gamma I \,. \nonumber
\eea
The coefficients of $I$ can again be determined by including
$SU(2n+1)$ instantons, which we will not do here.
The 4d map is given by setting $I=0$.
Now the charge $F$ corresponds to the one flavor in the dual theory $1+USp(2n)\subset R_{2,2n+1}$,
and the charge $I_1$ to the $U(1)$ intrinsic to $R_{2,2n+1}$.

\subsection{$N + SU(N)^k + N$}

As a further generalization of $SU(N)+2N$, let us consider the linear quiver gauge theory with $N+SU(N)^k +N$ ($N\geq 3$), 
which is a superconformal theory with $k$ marginal couplings in 4d.
The interesting limit to consider is when all the couplings become large.
When only some of them become large the problem reduces to finding the dual of smaller quivers.
For $k=2$ the S-dual theory was identified in \cite{Distler} as
$1+SU(2)\times [SU(3)\subset U_N]$. Namely it is a quiver theory with gauge group $SU(2)\times SU(3)$,
a fundamental of $SU(2)$, and a bi-fundamental,
where the $SU(3)$ results from weakly gauging the corresponding part of the global symmetry of $U_N$,
$SU(N)^2\times SU(3) \times U(1)$.
This theory has two marginal couplings and a global symmetry $SU(N)^2\times U(1)^3$,
the same as the $N+SU(N)^2+N$ quiver theory.
There are two more special cases for which the S-dual theories have been identified.
For $k=N-1$ the S-dual theory is $1+SU(2)\times SU(3)\times \cdots \times SU(N-1)\times [1+SU(N)\subset T_N]$
\cite{Gaiotto:2009gz}, and for $k=N-2$ it is
$1+SU(2)\times SU(3)\times \cdots \times [SU(N-1)\subset T_N]$ \cite{Tachikawa:2013kta}.

As far as we know, the dual theories for $2<k<N-2$ and $k>N-1$ have not been identified yet.
We will argue that for $k\geq N-1$ the dual theory is given by
\be
\label{TNdual}
1+SU(2)\times SU(3)\times \cdots \times SU(N-1)\times [SU(N)+1] \times SU(N)^{k-N}\times [SU(N)\subset T_N] \,,
\ee
and that for $k< N-1$ it is given by
\be
\label{chidual}
1+SU(2)\times SU(3)\times \cdots \times SU(k)\times [SU(k+1)\subset \chi_N^k] \,,
\ee
where $\chi_N^k$ is the class of isolated SCFT's that we introduced in section~\ref{sec:chi}.
The former reduces to the example in \cite{Gaiotto:2009gz} for $k=N-1$.
The latter reduces to the example in \cite{Distler} for $k=2$, since $\chi_N^2 = U_N$,
and to the example in \cite{Tachikawa:2013kta} for $k=N-2$, since $\chi_N^{N-2} = T_N$.

We will motivate these dualities by relating them to 5d dualities between gauge theories.
But first let us do some 4d consistency checks.
The global symmetry on both sides is $SU(N)^2\times U(1)^{k+1}$.
In the dual for 
$k\geq N-1$ the $U(1)$'s originate from the two fundamental fields and the $k-1$ bi-fundamental fields,
and in the dual for 
$k<N-1$ they originate from the one fundamental field, the $k-1$ bi-fundamental fields, 
and the $U(1)$ intrinsic to the $\chi_N^k$ theory.
The dimension of the Coulomb branch on both sides is $k(N-1)$.
For the dual theories this comes about by summing the dimension of the Coulomb branch of the    
isolated SCFT and those of the gauge group factors in the product.
For example, the $\chi_N^k$ theory has a $k (N-1)-k(k+1)/2$ dimensional Coulomb branch.
Together with the $1+2+\cdots + k = k(k+1)/2$ dimensional Coulomb branch of the product this gives $k(N-1)$.

\begin{figure}[h]
\center
\includegraphics[width=0.26\textwidth]{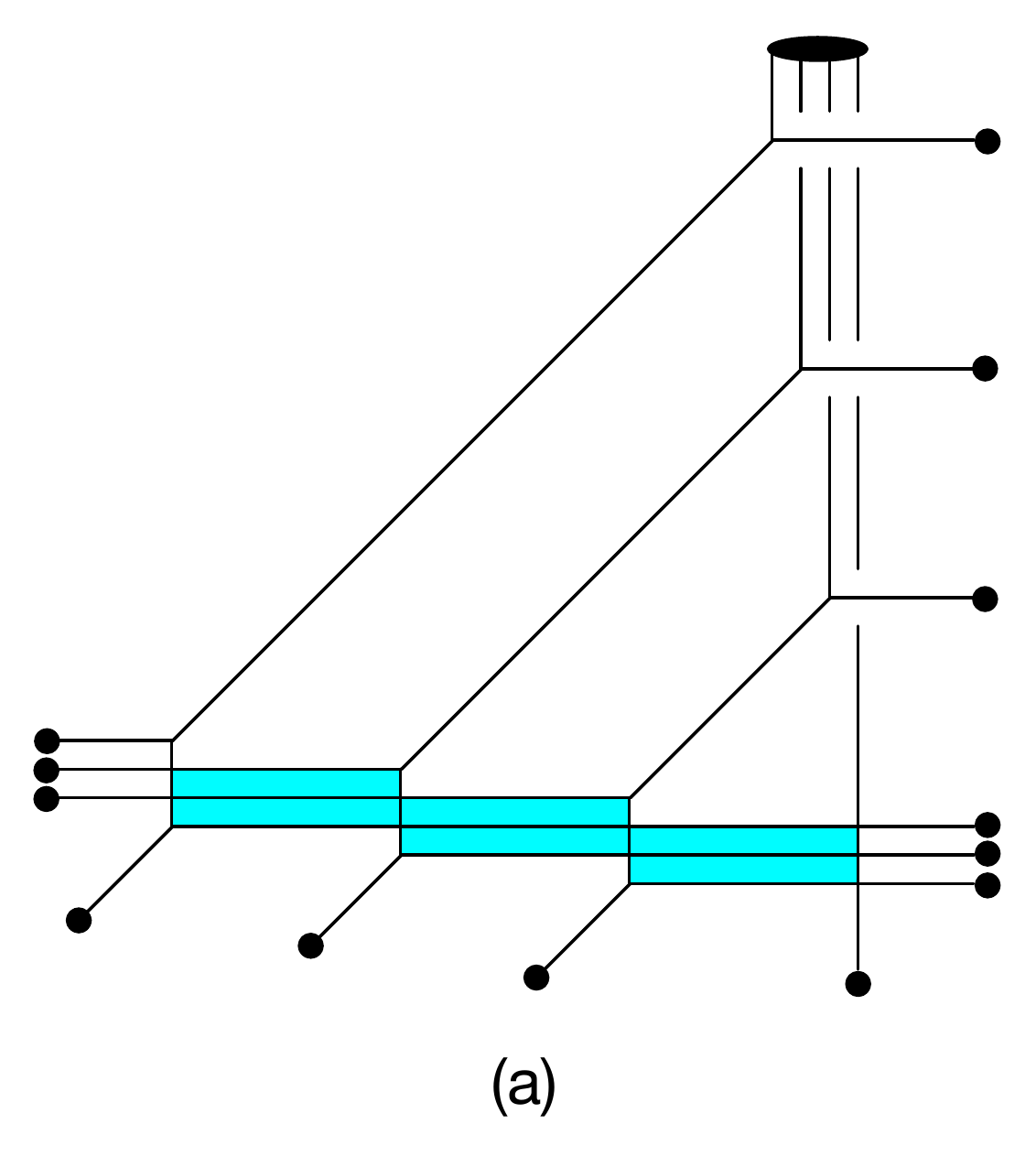} 
\hspace{0.5cm}
\includegraphics[width=0.31\textwidth]{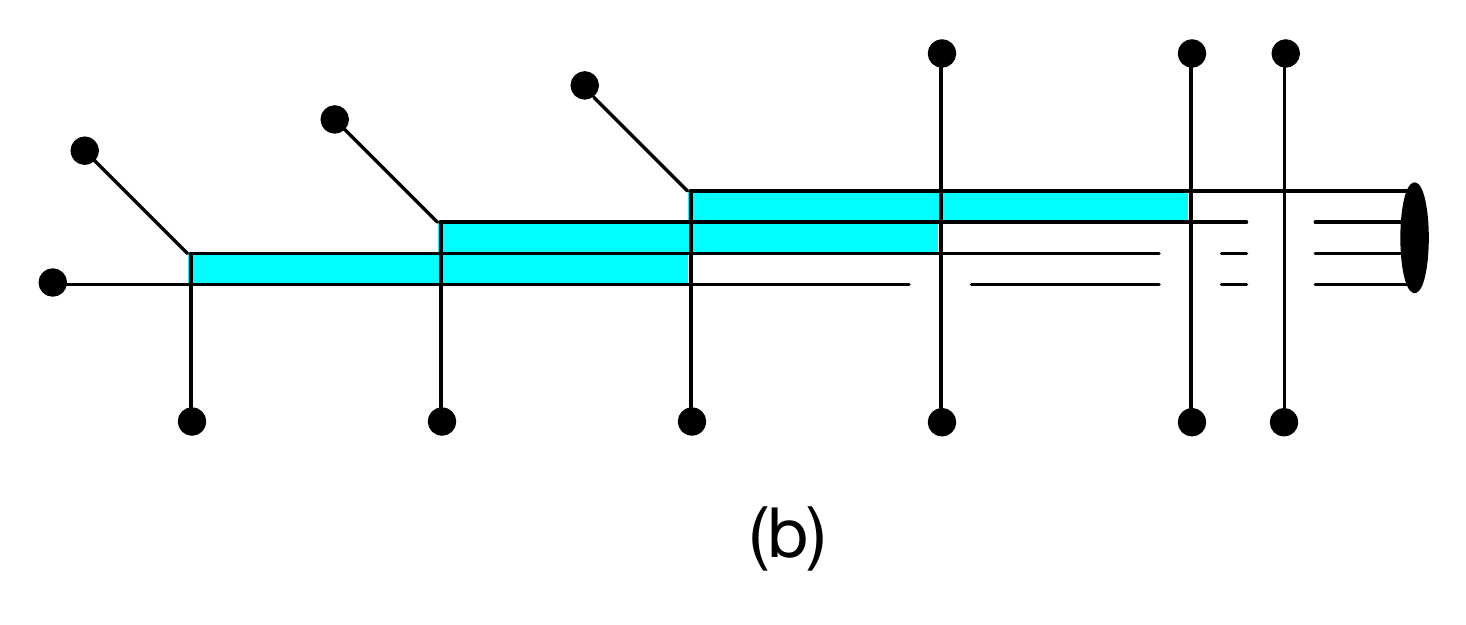} 
\hspace{0.5cm}
\includegraphics[width=0.31\textwidth]{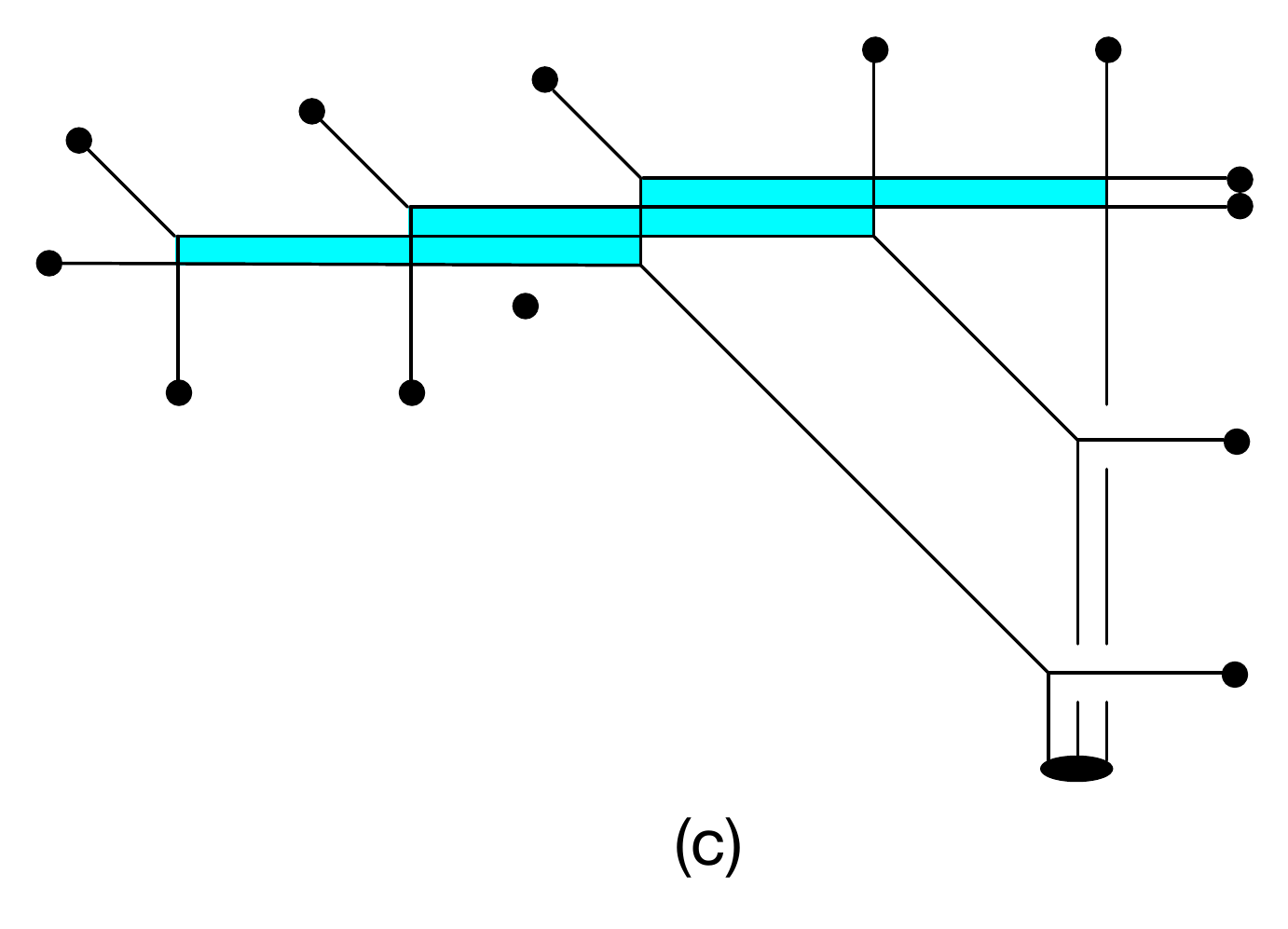} 
\caption{5-brane web of $3+SU(3)_0\times SU(3)_0\times SU(3)_1+3$ and its S-dual.}
\label{SU(3)^3webs}
\end{figure}

\begin{figure}[h]
\center
\includegraphics[width=0.4\textwidth]{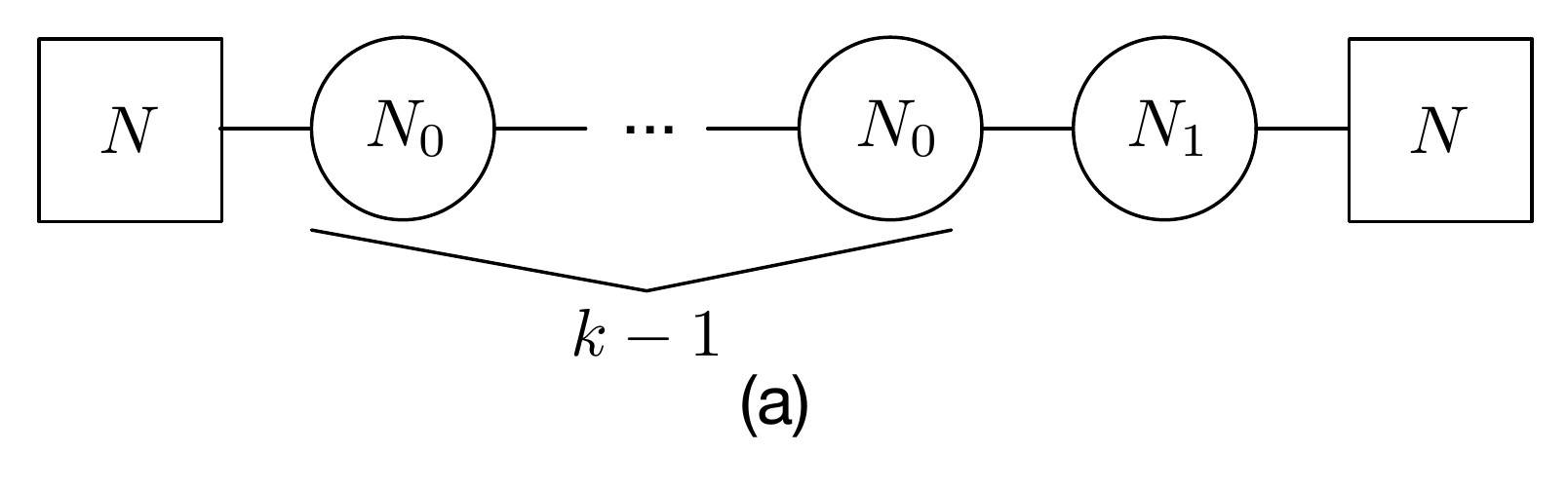}
\hspace{1cm}
 \includegraphics[width=0.5\textwidth]{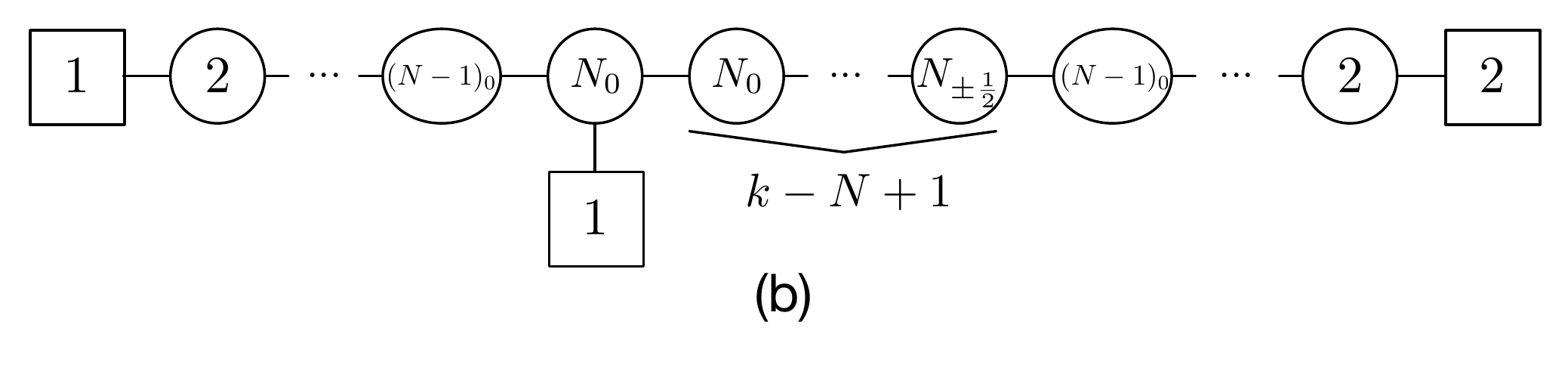}
\caption{The dual quiver gauge theories for $k\geq N-1$.}
\label{SU(N)^kquiver}
\end{figure}

Now let us consider the lift to five dimensions.
We again have to make a choice regarding the CS levels in the $SU(N)^k$ theory.
It turns out that the correct choice for the duality is $N+SU(N)^{k-1}_0 \times SU(N)_1 +N$.
Let us begin with the case of $k\geq N-1$.
As a simple representative, we will take $N=3$ and $k=3$.
The 5-brane web for this theory is shown in Fig.~\ref{SU(3)^3webs}a.
S-duality turns this into the web shown in Fig.~\ref{SU(3)^3webs}b, which after some simple brane manipulations
becomes the web shown in Fig.~\ref{SU(3)^3webs}c.
This describes the theory $1+SU(2)\times [SU(3)_0+1]\times SU(3)_{\pm\frac{1}{2}}\times SU(2)+2$.
More generally the dual pair of 5d quiver gauge theories is shown in Fig.~\ref{SU(N)^kquiver}.
The part on the RHS of the second quiver diagram beginning with the last $SU(N)$ node corresponds precisely to
the gauging of an $SU(N)$ subgroup of the global symmetry of the $T_N$ theory.
This reduces to the 4d dual in (\ref{TNdual}).

\begin{figure}[h]
\center
\includegraphics[width=0.3\textwidth]{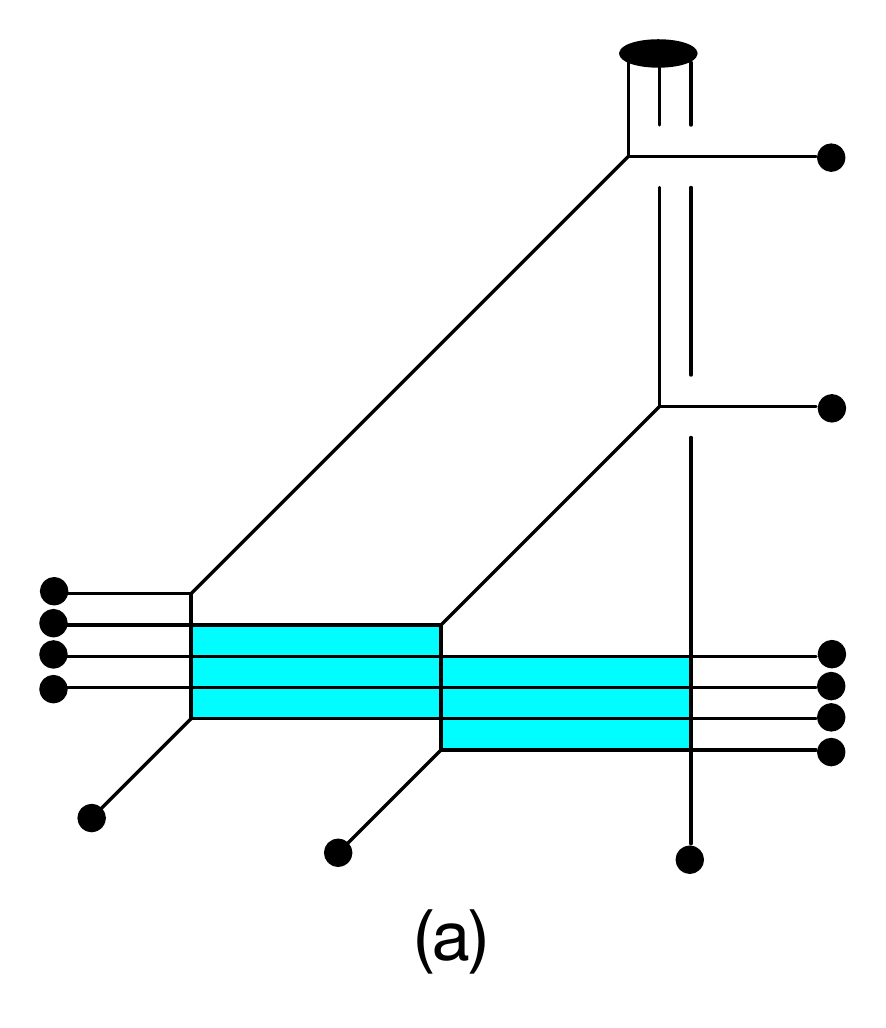} 
\hspace{1cm}
\includegraphics[width=0.45\textwidth]{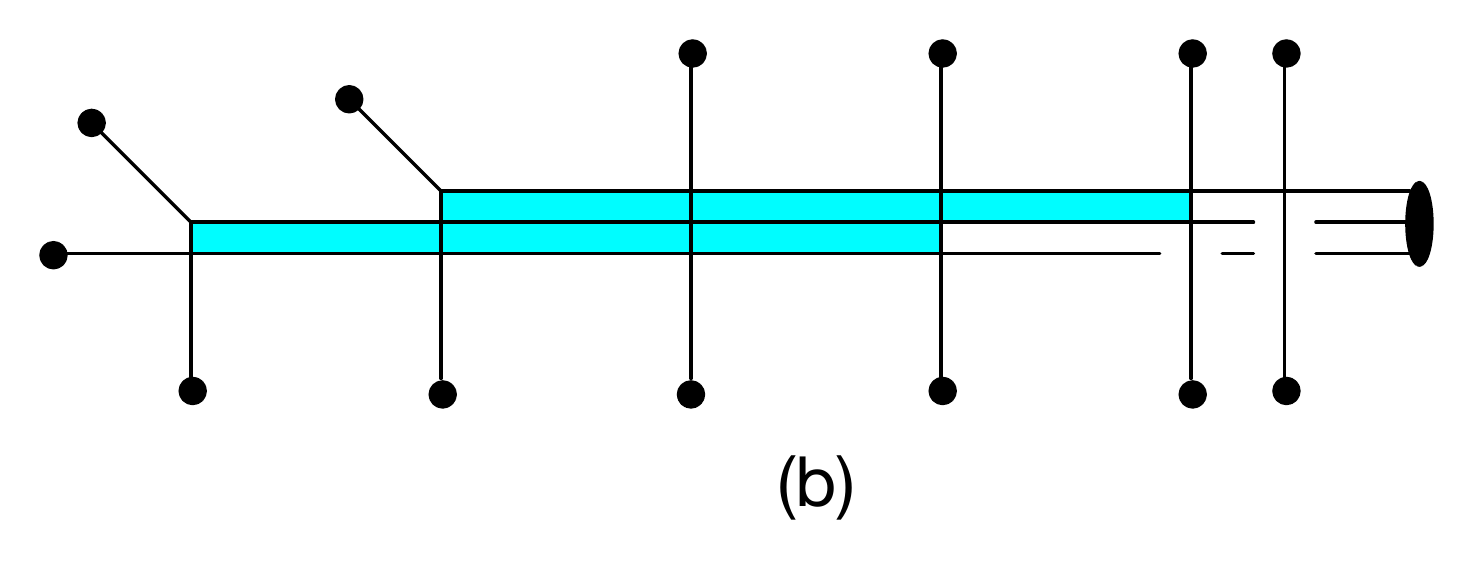} 
\caption{5-brane web of $4+SU(4)_0\times SU(4)_1 +4$ and its S-dual.}
\label{SU(4)^2webs}
\end{figure}

\begin{figure}[h]
\center
\includegraphics[width=0.4\textwidth]{SUNkquiver1}
\hspace{1cm}
 \includegraphics[width=0.5\textwidth]{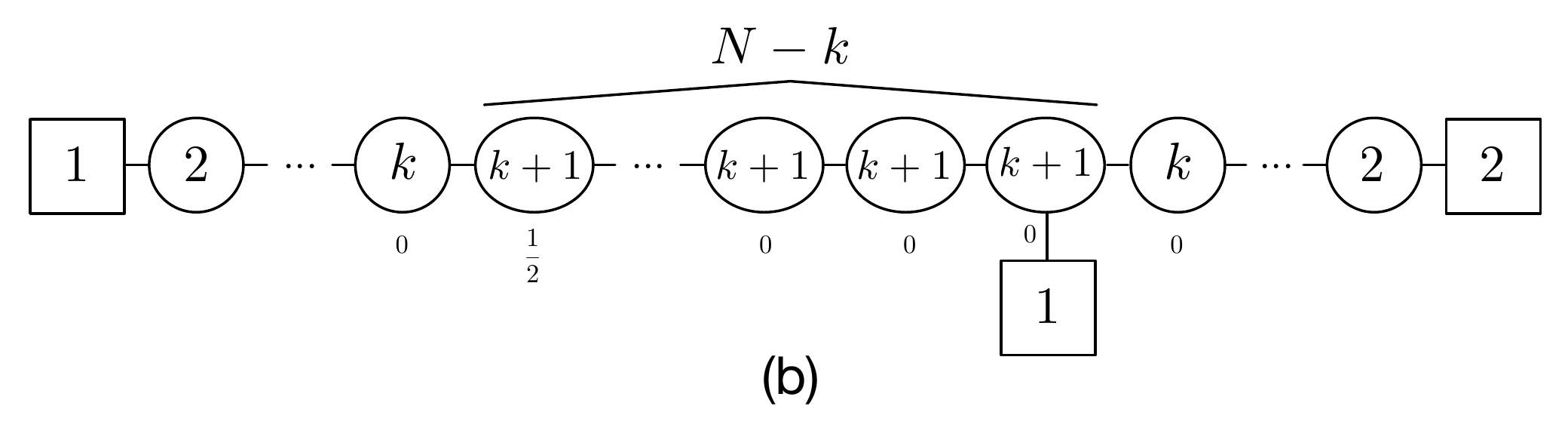}
\caption{The dual quiver gauge theories for $k<N-1$.}
\label{SU(N)^kquiver2}
\end{figure}

For $k<N-1$ we take the example of $N=4$ and $k=2$,
{\em i.e.}, $4+ SU(4)_0 \times SU(4)_1 +4$.
The web and its S-dual are shown in Fig.~\ref{SU(4)^2webs}.
We read-off the dual gauge theory as $1 + SU(2)\times SU(3)_{\frac{1}{2}} \times [SU(3)_0+1]  \times SU(2)+2$.
The dual pair for general $N$ and $k$ with $k<N-1$ is shown in Fig.~\ref{SU(N)^kquiver2}.
One recognizes the RHS of the second quiver starting with the first $SU(k+1)$ node as
the gauging of an $SU(k+1)$ global symmetry of the $\chi_N^k$ theory.
This reduces to the 4d dual claimed in (\ref{chidual}).

Obviously, the classical global symmetries of the two gauge theories (in both cases) are different,
and we expect non-perturbative enhancement, at least in the theories described by the S-dual webs.
The enhancement of symmetry, and more generally the operator map between the two theories,
become increasingly harder to see for larger values of $k$ due to the multitude of
topological $U(1)$ symmetries involved.

To get a flavor for this 
let us consider 
just the simplest case of $k=2$.
On one side we have $N+SU(N)\times SU(N)+N$, which has a classical global symmetry
$SU(N)^2\times U(1)_B^2\times U(1)_{BF}\times U(1)_I^2$. 
The proposed dual theory is $1+SU(2)\times SU(3)^{N-3}\times [SU(3)+1]\times SU(2)+2$.
The classical global symmetry of this theory is $SO(4)_F\times U(1)_F^2\times U(1)_{BF}^{N-1}\times U(1)_I^N$.
We claim that the enhancement to $SU(N)^2$ involves 
all the $U(1)$ symmetries except those associated with the $SU(2)\times SU(3)$ factor on the LHS,
namely $U(1)_{F_1}\times U(1)_{BF_1}\times U(1)_{I_1} \times U(1)_{I_2}$.
For example, the instanton of the last $SU(2)$ factor on the RHS gives a partial enhancement
$SO(4)_F\times U(1)_{BF_{N-1}}\times U(1)_{I_N} \rightarrow SU(3)^2$, which can  be understood as
the usual enhancement of the global symmetry of the $SU(2)+5$ theory to $E_6$, with an $SU(3)$ subgroup gauged.
To get the next level of enhancement we include the next gauge group factor, $SU(3)$.
The $(1,0)$ and $(1,1)$ instantons of $SU(3)\times SU(2)$ then lead to a further enhancement 
to $SU(4)^2$, as we showed for the $T_4$ theory in section~\ref{TN_symmetry}.
This suggests a pattern leading to $SU(N)^2$ once all the gauge group factors except the $SU(2)\times SU(3)$ on the LHS
have been included.

This leaves the five $U(1)$ symmetries, which in the dual theory are the four associated to the first
$SU(2)\times SU(3)$ factor plus one combination of the others.

\section{Conclusions}

In this paper we addressed two related aspects of the connection between 4d ${\cal N}=2$ theories and 5d 
${\cal N}=1$ theories.
In the first, building on the idea of \cite{BBT},
we obtained Lagrangian descriptions for several classes of isolated
4d ${\cal N}=2$ SCFT's in terms of 5d ${\cal N}=1$ gauge theories.
These gauge theories correspond to relevant deformations of 5d SCFT's described by 5-brane webs in
Type IIB string theory, which are in turn related by dimensional reduction to the 4d SCFT's.
The theories considered in this paper are all of $A_N$ type, including the $T_N$ theories
and theories obtained in limits on the Higgs branch of the $T_N$ theories.
It would be interesting to generalize this to the $D_N$ type theories.

In the second aspect, we showed that S-dualities in four dimensions,
relating the strong coupling limit of one ${\cal N}=2$ superconformal gauge theory to the weak coupling limit of another,
lift to 5d dualities between different gauge theory deformations of the same 5d SCFT.
It is important to stress that here it is the gauge theories themselves, not the 5d SCFT,
that are dimensionally reduced to four dimensions.
The examples studied in this paper include the Argyres-Seiberg duality involving $SU(3)+6$,
and the generalizations to $SU(N)+2N$, $SU(N)+2\,\asymm +4$ and $N+SU(N)^k+N$.
In the latter case the dual theories for $2<k<N-2$ have not been identified previously.
This procedure can, in principle, be extended to any 4d ${\cal N}=2$ superconformal gauge theory,
if one can identify the 5-brane web corresponding to its 5d lift.

An important question that our analysis raises is under what conditions does a duality 
between 5d supersymmetric gauge theories reduce to an S-duality between 4d superconformal gauge theories?
We do not have a complete answer to this question.
Clearly we need the reduction to produce a conformal theory, which will only happen
for specific gauge groups and matter content, and in specific scaling limits.
For example, the 4d superconformal gauge theory with $SU(3)+6$ corresponds to 
the 5d gauge theory with same content, compactified on a circle in the limit $R\rightarrow 0$, with 
$g_4^2 = g_5^2 /R$ held fixed.
On the other hand the reduction of $SU(3)+5$ with the same scaling limit gives $SU(3)+5$ in
4d, which is not conformal.
Also if we take the same content, $SU(3)+6$, but a different scaling limit, say fixing 
$m R$, where $m$ is the mass of one of the hypermultiplets, it is not clear what we get in four dimensions.
A sufficient condition to get a 4d duality, other than obtaining a conformal theory in four dimensions, 
is that the 5d duality maps a YM coupling on one side to a YM coupling on the other. 
This is the coupling that is scaled in the reduction on
both sides of the 5d duality, leading to an S-duality relative to the corresponding marginal coupling in 4d.

Another important question that deserves further study is to determine how the duality maps the parameters of the 
dual theories in 5d. 
In particular, the relation between the scaled YM couplings should reduce to the S-duality map of the 4d couplings.
The main quantitative tool used to study the 5d duality, the superconformal index, is insensitive
to the values of the mass parameters, and in particular to the YM couplings. 
We must therefore look for a different approach.

\medskip

\noindent {\bf Note added:} While this paper was being finalized, the paper \cite{McGrane:2014pma} appeared,
containing some overlap with section 3.4 of our paper, which discusses the dual of $N+SU(N)^k+N$.
The theories we call $\chi_N^k$ are called $T_{N,k}$ in \cite{McGrane:2014pma}.

\section*{Acknowledgements}

We thank Diego Rodriguez-Gomez for useful conversations.
G.Z. is supported in part by the Israel Science Foundation under grant no.~352/13 and by the German-Israeli Foundation
for Scientific Research and Development under grant no.~1156-124.7/2011.
O.B. is supported in part by the Israel Science Foundation under grant no. 352/13,
the German-Israeli Foundation for Scientific Research and Development under grant no.~1156-124.7/2011,
and the US-Israel Binational Science Foundation under grant no. 2012-041.

\appendix

\section{The 5d superconformal index}
\label{appendix:index}

The superconformal index is a characteristic of superconformal field theories\cite{KMMS}. It counts BPS operators 
modulo the merging of pairs to form non-BPS operators.
As such it is a rigid quantity, invariant under all continuous deformations of the theory that preserve the supersymmetry.
It is also given by 
the functional integral of the theory on $S^{D-1} \times S^1$.

In $D=5$ dimensions
the representations of the superconformal group are labeled by the highest weights of its $SO(5) \times SU(2)_R$ subgroup:
$j_1, j_2$ and $R$. The superconformal index is defined as \cite{KKL}:
\be
\mathcal{I}={\rm Tr}\,(-1)^F\,x^{2\,(j_1+R)}\,y^{2\,j_2}\,\mathfrak{q}^{\mathfrak{Q}}\,,\label{eq:ind}
\ee
where $x,\,y$ are the fugacities associated with the superconformal group, and $\mathfrak{q}$ denotes
collectively fugacities associated to other charges $\mathfrak{Q}$ that commute with the chosen supercharge.
These can include symmetries associated to matter fields, as well as topological (instantonic) $U(1)_I$ symmetries.  

If a Lagrangian description is available, the index can in principle be evaluated from the path integral via supersymmetric localization.
This was done in a number of examples with $SU(N)$ and $USp(2N)$ gauge groups in \cite{KKL}.
The full index is the product of the perturbative index, corresponding to the one-loop determinant of the field theory, 
and a sum of instantonic contributions, integrated over the gauge group.
For the perturbative part, each vector multiplet contributes
\be
f_{vector}(x,y,\alpha) = - \frac{x (y + \frac{1}{y})}{(1 - x y)(1 - \frac{x}{y})}\sum_{\bold{R}}e^{-i\bold{R}\cdot \alpha} \,,
\label{eq:vec}
\ee
where $e^{i\alpha_i}$ are the gauge fugacities and the sum runs over the root lattice,
and each matter hypermultiplet contributes
\be
f_{matter}(x,y,\alpha) = \frac{x}{(1 - x y)(1 - \frac{x}{y})}\sum_{\bold{w}\in \bold{W}}\sum^{N_f}_{i=1} (e^{i\bold{w}\cdot\alpha+im_i}
+e^{-i\bold{w}\cdot\alpha-i m_i}) \,,
\label{eq:mat}
\ee
where $e^{i m_i}$ are the fugacities associated with the matter degrees of freedom,
and the first sum is over the weights of the matter representations.
These give the one-particle index.
In order to evaluate the full perturbative contribution one needs to put this in a plethystic exponent, defined as
\be
PE[f(\cdot)] \equiv \exp\left[\sum^{\infty}_{n=1} \frac{1}{n} f(\cdot^n)\right] \,,
\label{eq:plesh}
\ee 
where the $\cdot$ represents all the variables in $f$. 

The instantonic contributions come from instantons localized at the north pole of the $S^4$ and anti-instantons
localized at the south pole, which also satisfy the supersymmetric localization conditions.
These are computed by integrating over the full instanton partition function. 
The result can be expressed  as a power series in the instanton number,
\be
\mathcal{Z}^{inst} = 1 + q Z_1 + q^2 Z_2 + \cdots \,, 
\label{insum}
\ee 
where $q$ is the instanton, {\em i.e.}, $U(1)_I$, fugacity.
The $k$-instanton partition function $Z_{k}$ is the 5d version of the Nekrasov partition function for $k$ instantons 
\cite{Nekrasov:2004vw}, expressed as an integral over the ADHM dual gauge group, where the integrand
has contributions from both the gauge field degrees of freedom and the charged matter degrees of freedom.
The gauge field contributions generically introduce poles, which must be dealt with 
by giving an appropriate prescription \cite{KKL,Hwang:2014uwa,Bergman:2013aca}.
Matter fields in representations other than the fundamental introduce additional poles,
and the correct prescription for dealing with them can be found in \cite{Hwang:2014uwa}. 
In some cases there are also poles at the origin or infinity. 
The prescription for these will be mentioned shortly.

There are a number of subtle issues related to the instanton computation.
Specifically for $SU(N)$, there are two issues related the fact that the computation really uses $U(N)$.
Naively the diagonal $U(1)$ part decouples, and one would assume that the result for $SU(N)$ can be obtained
simply by setting $\sum\alpha_i=0$. However this is not generally true, and there are ``$U(1)$ remnants" that
must be removed by hand.
The first is just a sign given by $(-1)^{\kappa_0 +  N_f/2}$, where $\kappa_0$ is the bare Chern-Simons level \cite{Bergman:2013aca}.
The second type of remnants are instantonic states that do not belong to the $SU(N)$ theory,
but whose contribution nevertheless remains in the instanton partition function after removing the diagonal $U(1)$.
In some cases these states can be identified in the 5-brane web construction, where they correspond to D1-branes
suspended between parallel external NS5-branes.
They are removed by multiplying the partition function by the appropriate factor   
\cite{Bao:2013pwa,Hayashi:2013qwa,Hwang:2014uwa,Bergman:2013aca}, for example
\be
\mathcal{Z}_{c} = PE\left[\frac{x^2  q f^F}{(1-x y)(1-\frac{x}{y})}\right] \mathcal{Z} \,,
\label{eq:cpf}
\ee 
for a remnant instantonic state with $F$ units of flavor charge.

However a simple brane description is not always available.
More generally, the presence of remnant states manifests itself as the lack of invariance under $x\rightarrow 1/x$,
which is part of the superconformal group, and in some cases, the lack of invariance under the full classical global symmetry.
The correction factor can then be determined by the requirement that the corrected partition function respect these symmetries. 
We have also checked that all the partition functions we used where invariant under these symmetries.
The violation of $x\rightarrow 1/x$ is intimately connected to the presence of poles at the origin or infinity.
The prescription for such poles in the contour integral, namely the choice of whether or not to include them,
is absorbed in the correction factor.
In particular, in (\ref{eq:cpf}) all the poles enclosed by the contour are included.

For $SU(2)$ there is another check that one can perform, since $SU(2)=USp(2)$.
The formulas for instanton partition functions of $USp(2N)$ are different, and in particular
do not exhibit the $U(1)$ remnants that the $SU(N)$ partition functions do.

Explicit formulas for the instanton partition functions in many cases have appeared elsewhere, and we will not reproduce them here.
We refer the reader to \cite{KKL,Bergman:2013aca}.
The only expression we will need, which, as far as we know, has not appeared in the literature, is 
the contribution of a bifundamental hypermultiplet in $SU(N_1) \times USp(2N_2)$.
We evaluated this using the methods of \cite{Shadchin:2005cc}.
The dual gauge group for a $(k_1,k_2)$ instanton is $U(k_1)\times O(k_2)$.
The $O(k_2)$ part has two contributions, $O_+$ and $O_-$, corresponding to the two disconnected components.
Using $z$ for the fugacity of the bi-fundamental symmetry $U(1)_{BF}$, $a,b$ for the instanton fugacities of 
$SU(N_1)$ and $USp(2N_2)$, respectively, $u_i,v_j$ for the fugacities of the dual gauge groups 
$U(k_1)$ and $O(k_2)$, respectively, and setting $k_2 = 2n_2 +\chi_2$, where $\chi_2=0$ or $1$, we find: 
\bea
& & Z^{U\times USp}_{BF+} =  \left[\prod^{N_1}_{i=1} (\sqrt{z a_i}-\frac{1}{\sqrt{z a_i}}) \prod_{m=1}^{k_1} \frac{(z u_m+\frac{1}{z u_m}- y-\frac{1}{y})}{(z u_m+\frac{1}{z u_m}- x-\frac{1}{x})} \right]^{\chi_2} \\ \nonumber & & \prod_{i,j=1}^{N_1,n_2} (z a_i+\frac{1}{z a_i}- v_j-\frac{1}{v_j}) \prod_{n,m=1}^{N_2,k_1} (z u_m+\frac{1}{z u_m}-b_n-\frac{1}{b_n}) \\ \nonumber & & \prod_{m,j=1}^{k_1,n_2} \frac{(z u_m+\frac{1}{z u_m}- v_j y-\frac{1}{v_j y})(z u_m+\frac{1}{z u_m}-\frac{v_j}{y}-\frac{y}{v_j})}{(z u_m+\frac{1}{z u_m}- v_j x-\frac{1}{v_j x})(z u_m+\frac{1}{z u_m}-\frac{v_j}{x}-\frac{x}{v_j})} \label{uus1}
\eea
for the $O^+$ part,
\bea
& & Z^{U\times USp}_{BF-O} =  \prod^{N_1}_{i=1} (\sqrt{z a_i}+\frac{1}{\sqrt{z a_i}}) \prod_{m=1}^{k_1} \frac{(z u_m+\frac{1}{z u_m}+ y+\frac{1}{y})}{(z u_m+\frac{1}{z u_m}+ x+\frac{1}{x})} \\ \nonumber & & \prod_{i,j=1}^{N_1,n_2} (z a_i+\frac{1}{z a_i}- v_j-\frac{1}{v_j}) \prod_{n,m=1}^{N_2,k_1} (z u_m+\frac{1}{z u_m}-b_n-\frac{1}{b_n}) \\ \nonumber & & \prod_{m,j=1}^{k_1,n_2} \frac{(z u_m+\frac{1}{z u_m}- v_j y-\frac{1}{v_j y})(z u_m+\frac{1}{z u_m}-\frac{v_j}{y}-\frac{y}{v_j})}{(z u_m+\frac{1}{z u_m}- v_j x-\frac{1}{v_j x})(z u_m+\frac{1}{z u_m}-\frac{v_j}{x}-\frac{x}{v_j})} \label{uus2}
\eea
for the $O^-$ part and odd $k_2$, and
\bea
& & Z^{U\times USp}_{BF-E} =  \prod^{N_1}_{i=1} (z a_i -\frac{1}{z a_i}) \prod_{m=1}^{k_1} \frac{(z^2 u^2_m+\frac{1}{z^2 u^2_m}- y^2 -\frac{1}{y^2})}{(z^2 u^2_m+\frac{1}{z^2 u^2_m}- x^2-\frac{1}{x^2})} \\ \nonumber & & \prod_{i,j=1}^{N_1,n_2-1} (z a_i+\frac{1}{z a_i}- v_j-\frac{1}{v_j}) \prod_{n,m=1}^{N_2,k_1} (z u_m+\frac{1}{z u_m}-b_n-\frac{1}{b_n}) \\ \nonumber & & \prod_{m,j=1}^{k_1,n_2-1} \frac{(z u_m+\frac{1}{z u_m}- v_j y-\frac{1}{v_j y})(z u_m+\frac{1}{z u_m}-\frac{v_j}{y}-\frac{y}{v_j})}{(z u_m+\frac{1}{z u_m}- v_j x-\frac{1}{v_j x})(z u_m+\frac{1}{z u_m}-\frac{v_j}{x}-\frac{x}{v_j})} \label{uus2}
\eea
for the $O^-$ part and even $k_2$.

These contributions also add poles to the integral. The prescription for dealing with them follows from the results of \cite{Hwang:2014uwa}. Specifically, one defines $p=\frac{1}{z x}$ and $d=\frac{z}{x}$, calculates the integral assuming $x, p, d << 1$, 
and only at the end returns to the original variables.

\section{Flavors in webs}
\label{sec:flavors}

Matter in the fundamental representation of the gauge group is usually referred to as ``flavors".
There are different, but equivalent, ways to represent flavor degrees of freedom in 5-brane webs.
These are related by brane-creation, or Hanany-Witten (HW), transitions.
In some representations, the counting of flavors is not obvious, so it is useful to be able to
map to other representations in which the counting is obvious.
We will illustrate this with four examples, which should make the general process clear.
In the main body of the paper we will refer to this idea whenever we have a web in which the counting is unclear.

\begin{figure}[h]
\center
\includegraphics[width=0.17\textwidth]{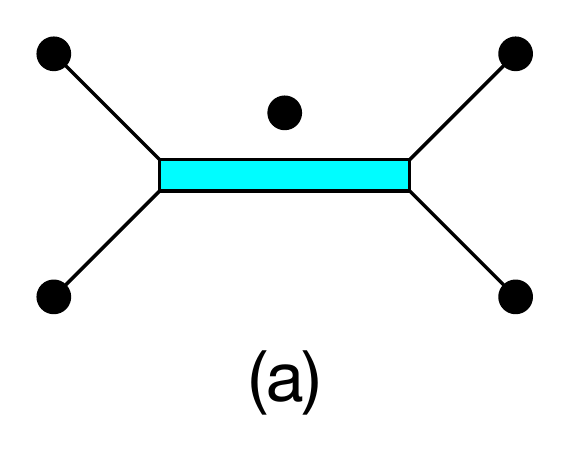} 
\hspace{1cm}
\includegraphics[width=0.17\textwidth]{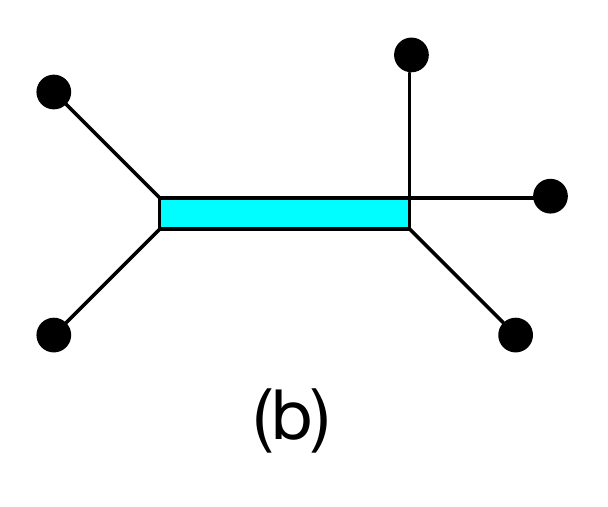} 
\caption{Two representations of $SU(2)+1$}
\label{SU(2)+1}
\end{figure}

In the first example, we add a single flavor to the pure $SU(2)$ theory by adding a D7-brane, Fig.~\ref{SU(2)+1}a.
Moving the D7-brane to the right across the $(1,1)$ 5-brane we get an external D5-brane,  Fig.~\ref{SU(2)+1}b.
These are two equivalent representation of a hypermultiplet in the fundamental representation of $SU(2)$.
The first corresponds to the D5-D7 strings, and the second to D5-D5 strings across the NS5-brane.

\begin{figure}[h]
\center
\includegraphics[width=0.2\textwidth]{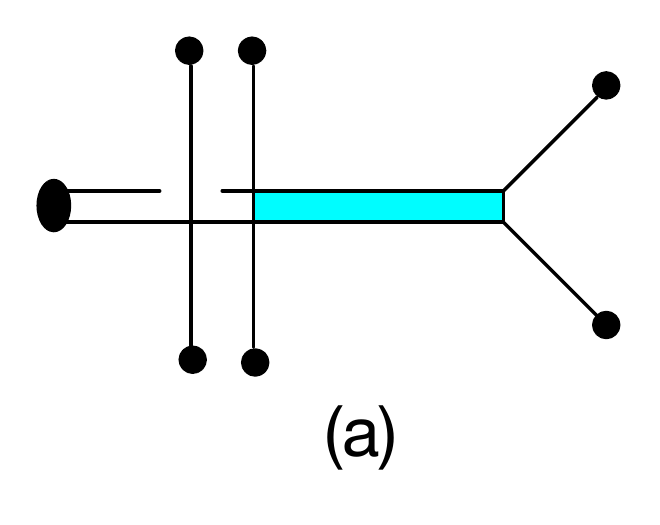} 
\hspace{0.5cm}
\includegraphics[width=0.2\textwidth]{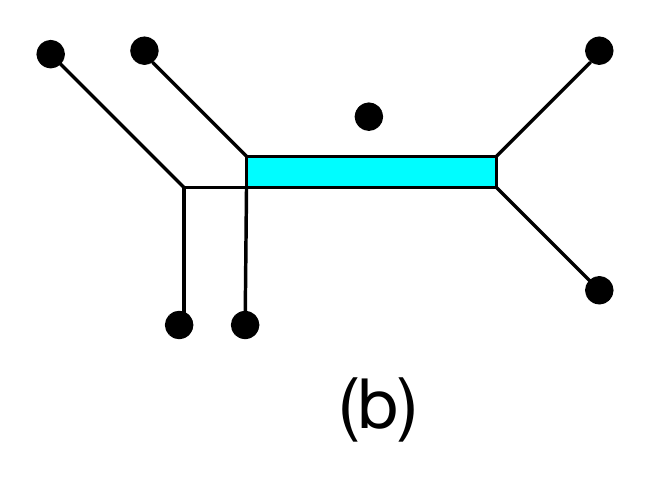} 
\hspace{0.5cm}
\includegraphics[width=0.2\textwidth]{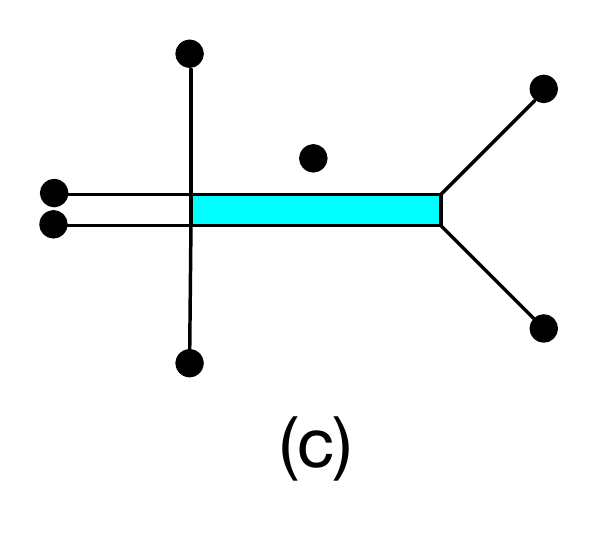} 
\hspace{0.5cm}
\includegraphics[width=0.2\textwidth]{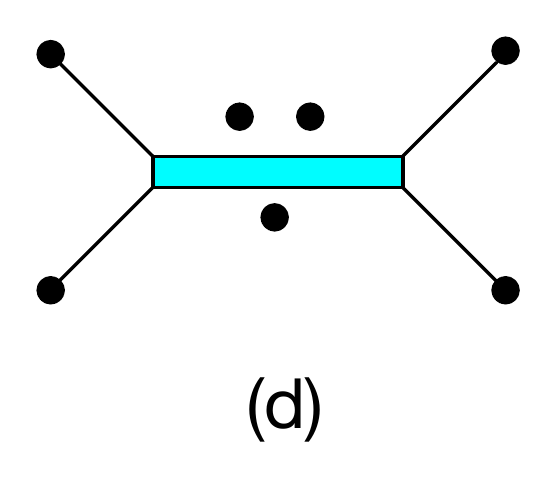} 
\caption{Representations of $SU(2)+3$}
\label{SU(2)+3}
\end{figure}

The second example is a little more interesting.
Now we begin with the 5-brane web shown in Fig.~\ref{SU(2)+3}a.
This also describes an $SU(2)$ theory, but with how many flavors?
Note that there is one avoided intersection due to the ``s-rule".
There seem to be three independent external D5-branes, which leads us to conclude that there are three flavors.
Let us verify this by going to a simpler representation.
First we move the one D7-brane across both NS5-branes, leading to the configuration in Fig.~\ref{SU(2)+3}b.
Then we move the left-most $(0,1)$ 7-brane across the two $(1,-1)$ 5-branes, giving Fig.~\ref{SU(2)+3}c.
The three flavors are now clearly visible. We can also move the two D7-branes on the left across the NS5-brane
to get the representation in Fig.~\ref{SU(2)+3}d.

\begin{figure}[h]
\center
\includegraphics[width=0.17\textwidth]{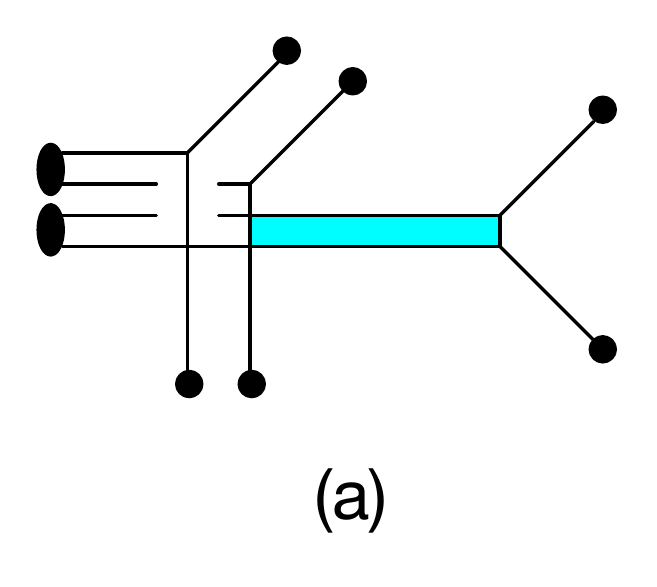} 
\hspace{0.3cm}
\includegraphics[width=0.17\textwidth]{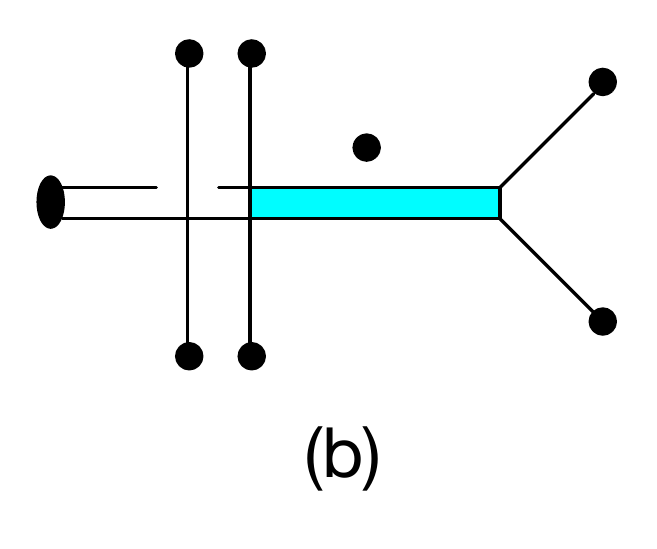} 
\hspace{0.3cm}
\includegraphics[width=0.17\textwidth]{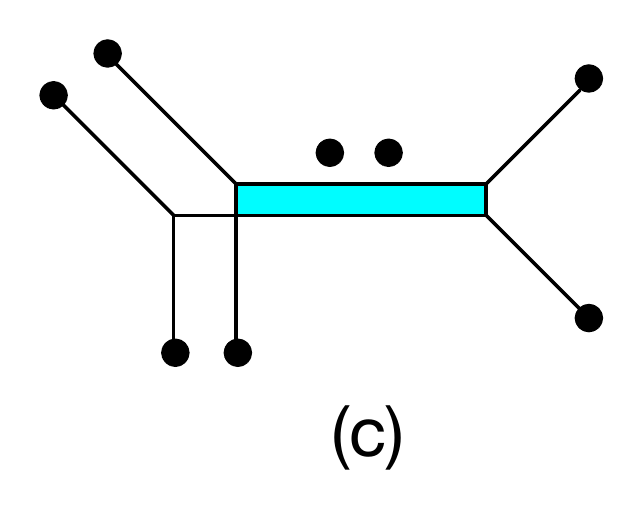} 
\hspace{0.3cm}
\includegraphics[width=0.17\textwidth]{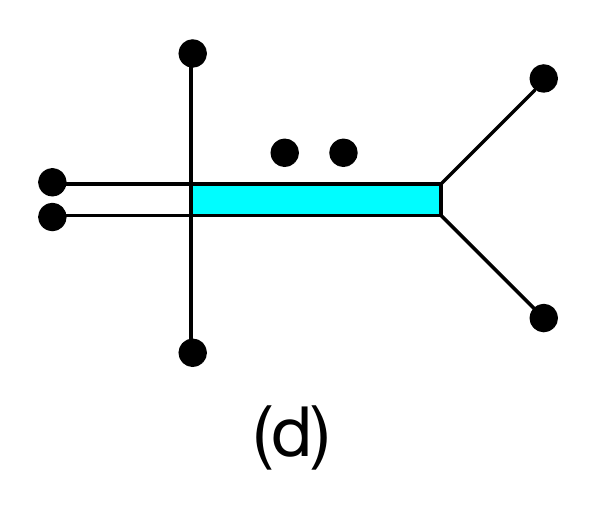} 
\hspace{0.3cm}
\includegraphics[width=0.17\textwidth]{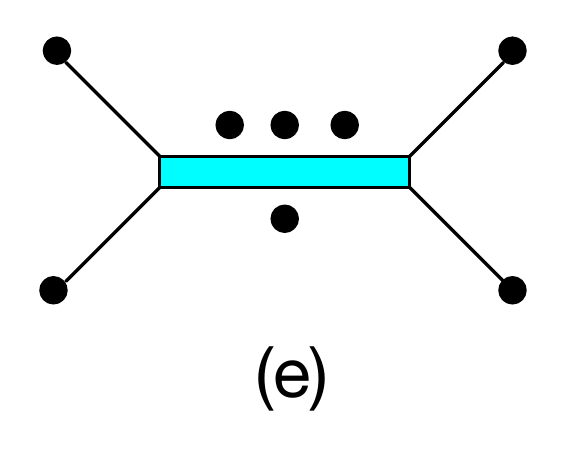} 
\caption{Representations of $SU(2)+4$}
\label{SU(2)+4}
\end{figure}

Our third example, Fig.~\ref{SU(2)+4}, is an elaboration of the previous example.
The steps are basically the same, showing that the original web describes an $SU(2)$ gauge theory with four flavors.

\begin{figure}[h]
\center
\includegraphics[width=0.25\textwidth]{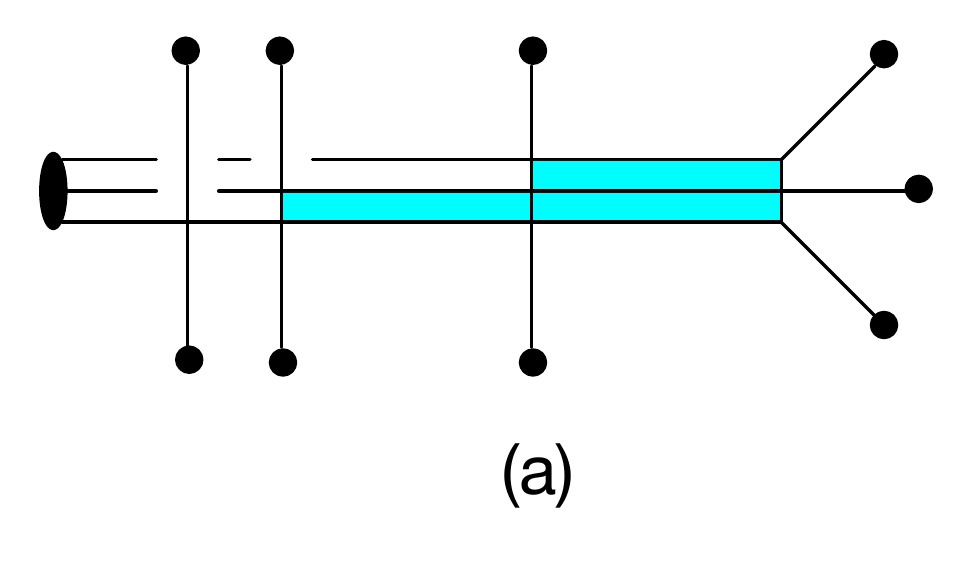} 
\hspace{1cm}
\includegraphics[width=0.25\textwidth]{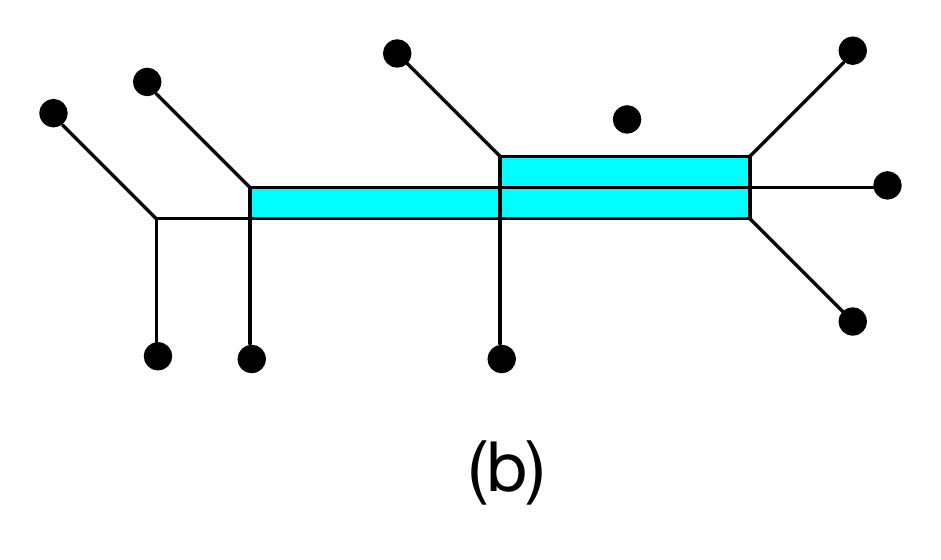} 
\hspace{1cm}
\includegraphics[width=0.25\textwidth]{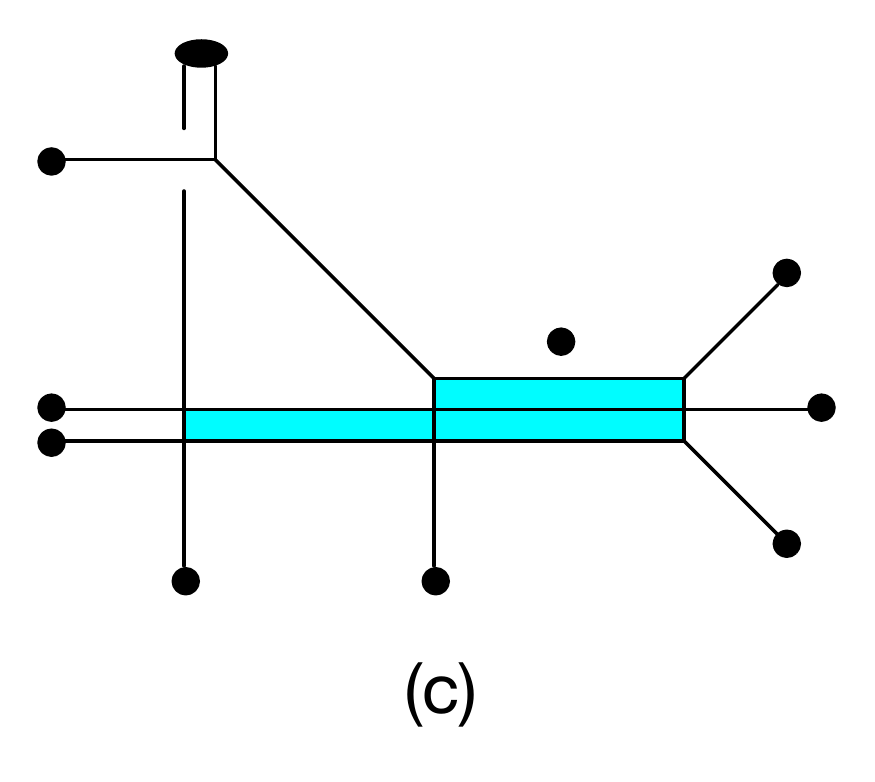} 
\caption{Representations of $2+SU(2)\times SU(3)+2$}
\label{SU(2)xSU(3)}
\end{figure}

Our final example involves a product group and a 7-brane with three attached 5-branes, Fig.~\ref{SU(2)xSU(3)}.
The steps are similar.

\section{Webs for antisymmetric matter}
\label{sec:antisymmetric}

There is no general prescription for including matter in 5-brane webs in representations other than fundamental or bi-fundamental.
Nevertheless it is possible to incorporate some other representations in some cases.
The most common cases are $SU(N)$ or $USp(2N)$ and matter in the rank 2 antisymmetric representation.
The latter and some examples of the former were previously discussed in \cite{Bergman:2013aca}.
We do not have a precise (microscopic) understanding of how these fields arise (in terms of open strings),
but it is possible to argue for their existence indirectly by going on the Higgs branch associated to them,
and confirming the pattern of gauge symmetry breaking.
Here we will consider $SU(N)$ with one antisymmetric plus fundamentals.

\begin{figure}[h]
\center
\includegraphics[width=0.4\textwidth]{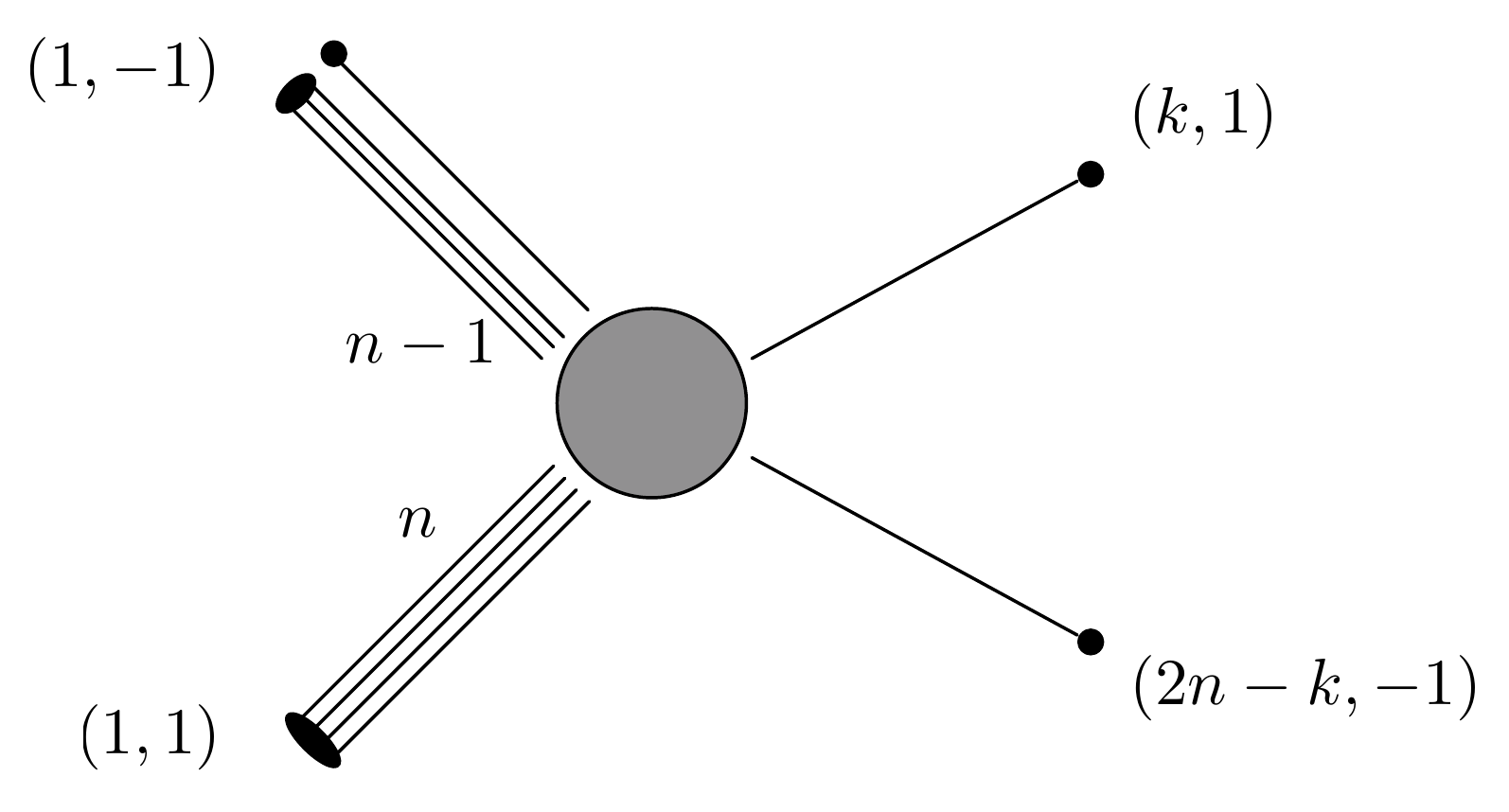} 
\caption{5-brane junction for $SU(2n)_{n+1-k}$ with an antisymmetric.}
\label{ASjunction1}
\end{figure}

We begin with the claim that the 5-brane junction shown in Fig.~\ref{ASjunction1} corresponds
to the UV fixed point of an $SU(2n)$ theory with CS level $\kappa_0 = n+1-k$ and one antisymmetric matter field.
Roughly speaking, the matter multiplet is a degree of freedom associated with the two $(1,-1)$ 7-branes.

\begin{figure}[h]
\center
\includegraphics[width=0.27\textwidth]{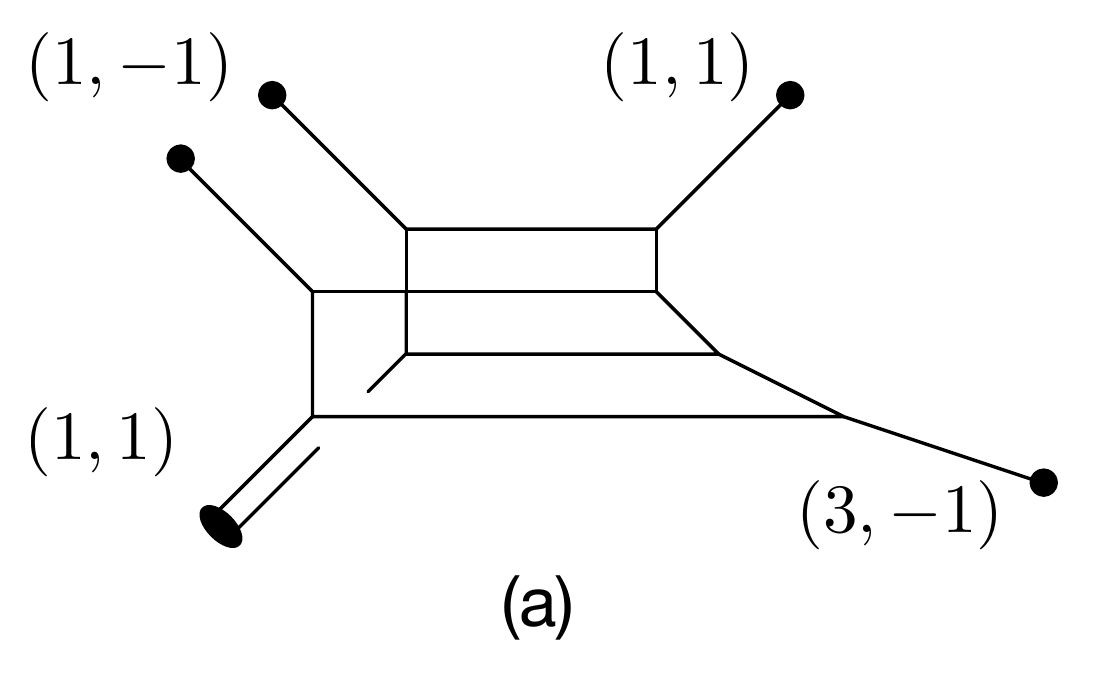} 
\hspace{0.5cm}
\includegraphics[width=0.24\textwidth]{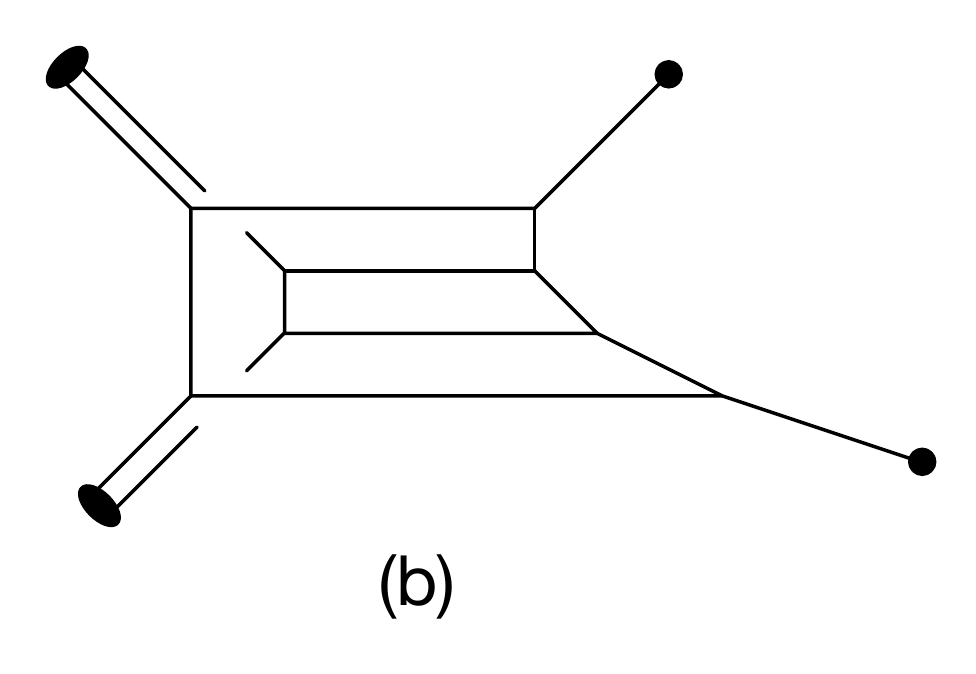} 
\hspace{0.5cm}
\includegraphics[width=0.33\textwidth]{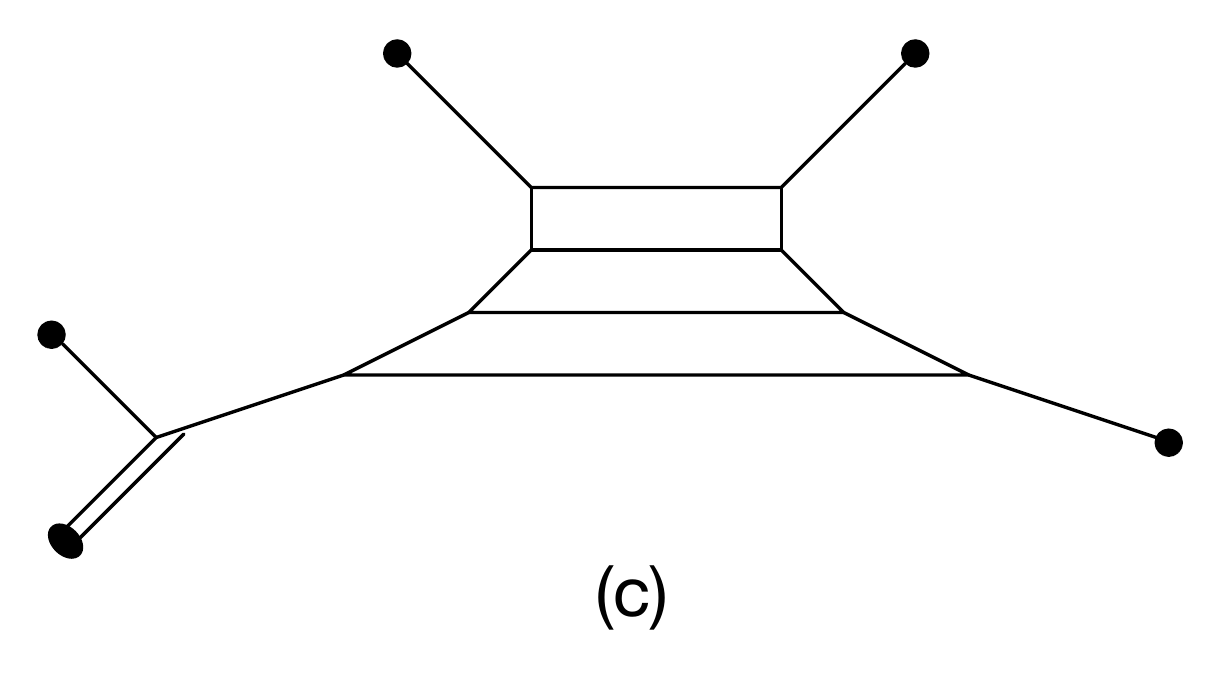} 
\caption{Web for $SU(4)_2 + \asymm$ : (a) Coulomb branch, (b) Higgs branch $USp(4)$, (c) a mass deformation.}
\label{ASweb}
\end{figure}

This can be confirmed by performing various deformations of the web.
For simplicity we will demonstrate this for $n=2$ and $k=1$, namely $SU(4)_2$.
The deformation corresponding to a finite Yang-Mills coupling is shown in Fig.~\ref{ASweb}a.
Then going on the Higgs branch corresponds to the web in Fig.~\ref{ASweb}b, where one of the $(1,-1)$ 5-branes breaks,
and the broken piece is removed along the two $(1,-1)$ 7-branes.
The remaining web describes a pure $USp(4)$ theory (see \cite{Bergman:2013aca}), which is consistent with 
a VEV for a single matter field in the antisymmetric representation of $SU(4)$.
The CS level can be determined by turning on a mass for the antisymmetric field, described by the web in Fig.~\ref{ASweb}c.
The remaining pure $SU(4)$ web shows a renormalized CS level $\kappa = 2$.
On the other hand, for $SU(N)$ with antisymmetric matter $\kappa = \kappa_0 + N -4$ (the cubic Casimir of the antisymmetric
representation of $SU(N)$ is $N-4$), so in this case $\kappa_0=2$ as well.
One can easily generalize this argument for the web in Fig.~\ref{ASjunction1}.

We can add flavors, {\em i.e.}, hypermultiplets in the fundamental representation,
by attaching D5-branes (ending on D7-branes) on the RHS of the web.
Note that it makes a difference whether we attach a D5-brane to the top or the bottom part of the web.
This determines the sign of the mass, and therefore affects the value of the bare CS level for the massless theory.
Also, there is a limit to the number of flavors we can add.
Beyond some number, external 5-branes will intersect, which in principle means that the corresponding 
fixed point theory does not exist.
Some amount of intersection is however ``resolvable" via HW transitions (see for example the webs for the $T_N$ theories).

\begin{figure}[h]
\center
\includegraphics[width=0.2\textwidth]{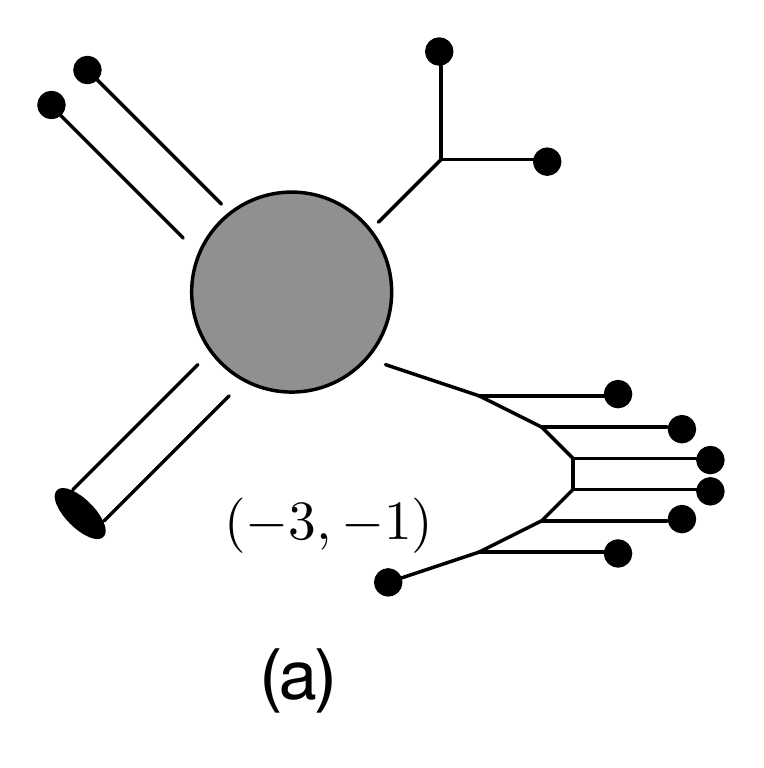} 
\hspace{1cm}
\includegraphics[width=0.2\textwidth]{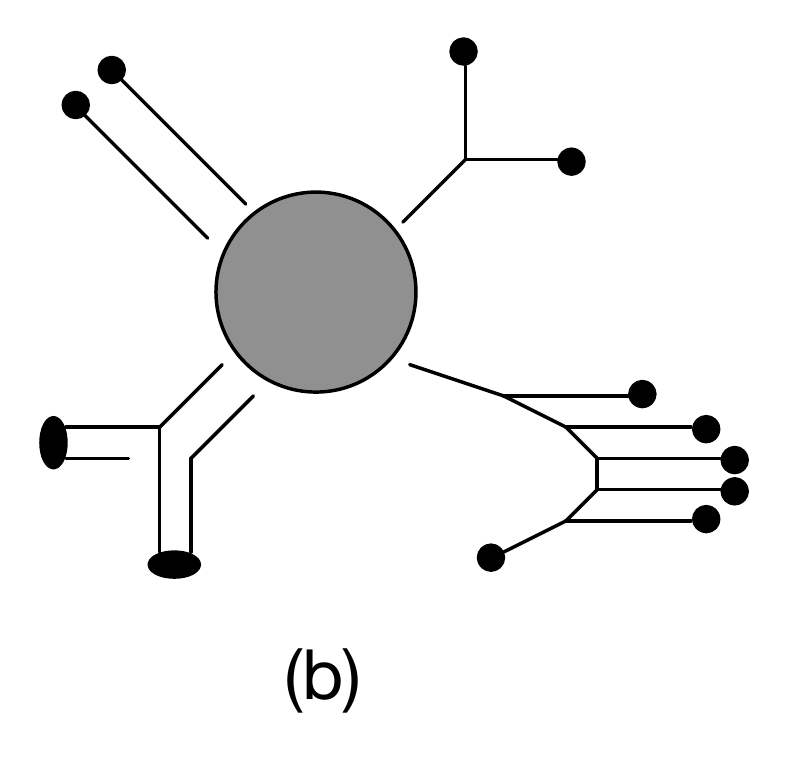} 
\hspace{1cm}
\includegraphics[width=0.2\textwidth]{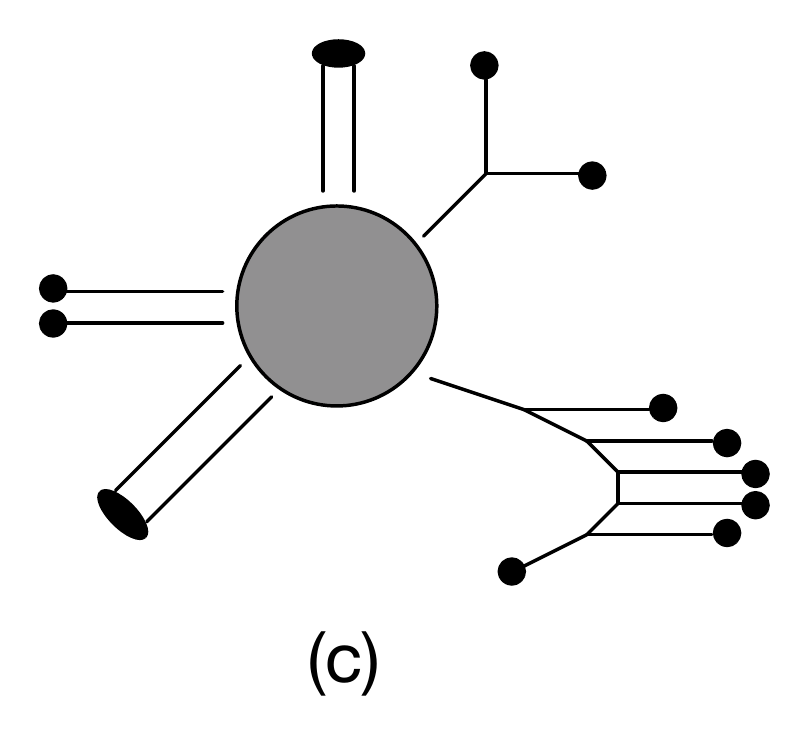} 
\hspace{1cm}
\includegraphics[width=0.15\textwidth]{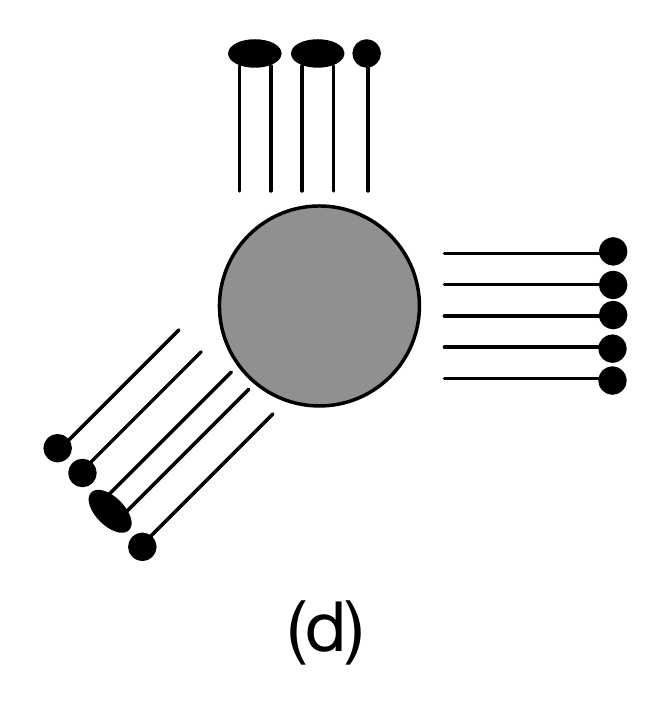} 
\caption{Web for $SU(4)_{-\frac{1}{2}} + \asymm + 7 \cdot \funda$}
\label{AS+7web}
\end{figure}

As a concrete example, let us find a 5-brane web for $SU(4)_{-\frac{1}{2}}$ with an antisymmetric and $N_f=7$ flavors.
(This will play a role in section \ref{section:R1N}.)
Starting with the web for $SU(4)_2 + \asymm$, we need to add one D5-brane at the top and six at the bottom (Fig.~\ref{AS+7web}a).
Then $\kappa_0 = \kappa + \frac{1}{2} - \frac{6}{2} = - \frac{1}{2}$.
Two HW transitions involving the lowest D7-brane lead to the web in Fig.~\ref{AS+7web}b,
and then a couple involving the $(0,1)$ 7-brane at the bottom lead to the web in Fig.~\ref{AS+7web}c.
Repeating these steps for the next D7-brane leads to the web in Fig.~\ref{AS+7web}d.
The latter is related by $SL(2,\mathbb{Z})$ to the $R_{1,5}$ web.

\end{document}